\documentclass[lettersize,journal]{IEEEtran}
\usepackage{amsmath,amsfonts}
\usepackage{algorithmic}
\usepackage{algorithm}
\usepackage{array}
\usepackage[caption=false,font=footnotesize,labelfont=rm,textfont=rm]{subfig}
\usepackage{textcomp}
\usepackage{stfloats}
\usepackage{url}
\usepackage{verbatim}
\usepackage{graphicx}
\usepackage{cite}
\usepackage{subfiles}
\usepackage{multirow}
\usepackage{booktabs}
\usepackage{threeparttable}
\usepackage[pagebackref=true,breaklinks=true,colorlinks,bookmarks=false]{hyperref}

\begin{document}

\title{High-Efficiency Lossy Image Coding Through Adaptive Neighborhood Information Aggregation}

\author{Ming Lu,~\IEEEmembership{Student Member,~IEEE,}
        Fangdong Chen,
        Shiliang Pu,
        and 
        Zhan Ma,~\IEEEmembership{Senior Member,~IEEE}
        % <-this % stops a space
\thanks{M. Lu and Z. Ma are with Nanjing University, Nanjing, Jiangsu, China.
E-mails: luming@smail.nju.edu.cn, mazhan@nju.edu.cn.} \\
\thanks{F. Chen and S. Pu are with Hikvision Inc., Hangzhou, Zhejiang, China. \\
E-mails: chenfangdong@hikvision.com, pushiliang.hri@hikvision.com.}% <-this % stops a space
%\thanks{Manuscript received April 19, 2021; revised August 16, 2021.}
}



\maketitle

\begin{abstract}
Questing for learned lossy image coding (LIC) with superior compression performance and computation throughput is challenging. The vital factor behind it is how to intelligently explore Adaptive Neighborhood Information Aggregation (ANIA) in transform and entropy coding modules. To this end, Integrated Convolution and Self-Attention (ICSA) unit is first proposed to form a content-adaptive transform to characterize and embed neighborhood information dynamically of any input. Then a Multistage Context Model (MCM) is devised to progressively use available neighbors following a pre-arranged spatial-channel order for accurate probability estimation in parallel. ICSA and MCM are stacked under a Variational AutoEncoder (VAE) architecture to derive rate-distortion optimized compact representation of input image via end-to-end learning. Our method reports state-of-the-art compression performance surpassing the VVC Intra and other prevalent LIC approaches across Kodak, CLIC, and Tecnick datasets; More importantly, our method offers  $>$60$\times$ decoding speedup  using a comparable-size model when compared with the most popular LIC method. All materials are made publicly accessible at \url{https://njuvision.github.io/TinyLIC} for reproducible research.
\end{abstract}

\begin{IEEEkeywords}
Learned image coding, adaptive neighborhood information aggregation, convolution, self-attention, multistage context model.
\end{IEEEkeywords}

\section{Introduction}\label{sec:introduction}
\IEEEPARstart{T}{he} pursuit of high-efficiency lossy image coding is ever increasingly critical for vast networked applications such as photo sharing, commercial advertisements, remote medical diagnosis, etc. In principle, lossy image coding searches for the {\it optimal compact representation} of input source in a computationally feasible way that leads to the best rate-distortion (R-D) performance~\cite{Gibson2017RateDF} defined in
\begin{align}
J &= R + \lambda D.
\label{eq:vae_rd}
\end{align} Here, $\lambda$ is the Lagrange multiplier that controls the desired compression trade-off between the rate and distortion. $R$ represents the number of bits to encode the input data, and $D$ can be measured using Mean Square Error (MSE) or Multiscale Structural Similarity (MS-SSIM)~\cite{wang2003multiscale}.

Though conceptually any input source can be represented using vector quantization, it is practically infeasible for a high-dimensional source because of unbearable complexity~\cite{balle2020nonlinear}. In the light of computationally manageable coding solution, it then leads to the {\it Transform Coding}  that divides the image coding problem into three consecutive simple steps, e.g., transform, quantization, and entropy coding, as stated in~\cite{952802}. 

\begin{figure}[!t]
\centering
\includegraphics[width=\linewidth]{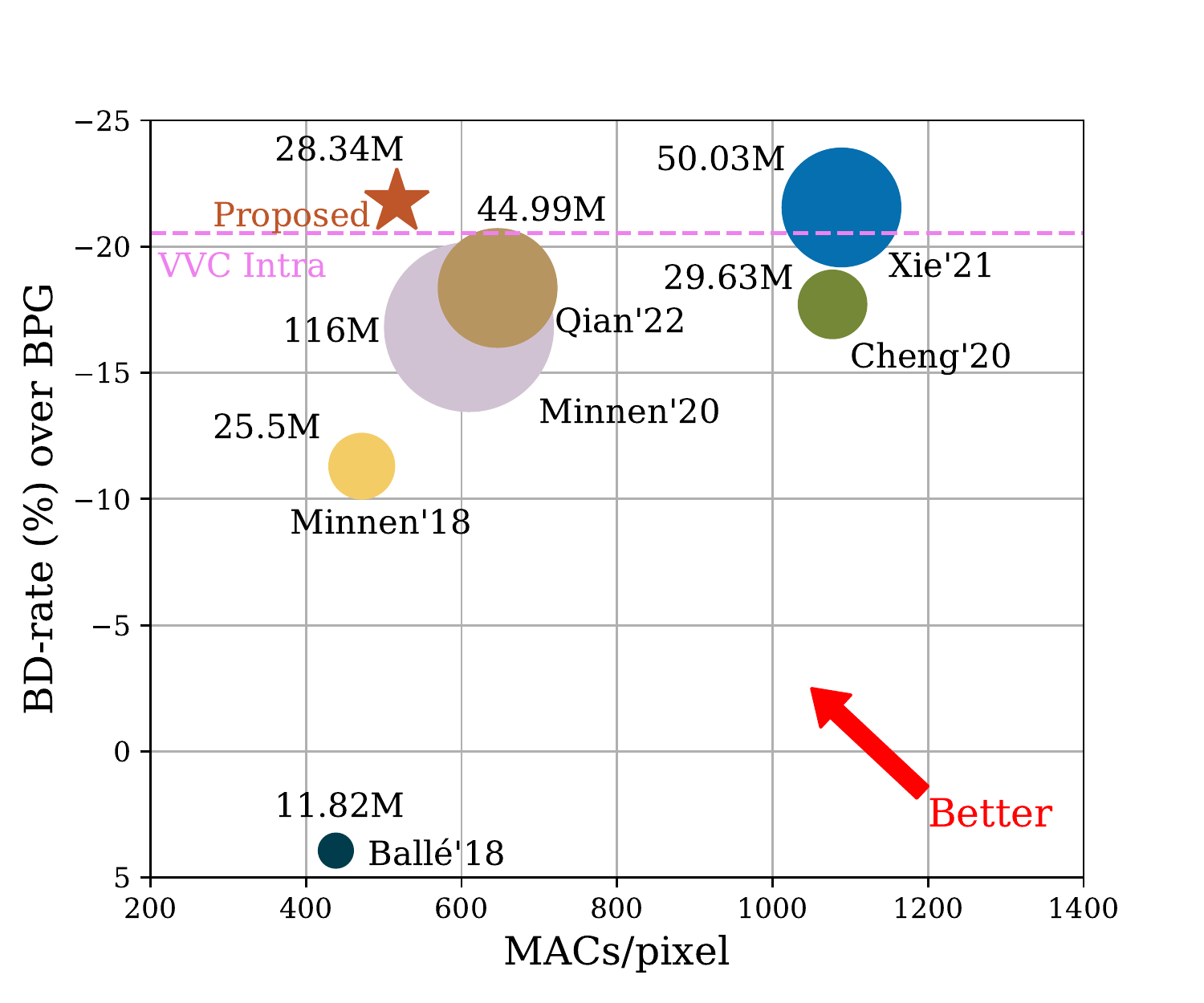} \label{fig:perf_vs_complexity}
\caption{{\bf  Performance versus Complexity.} Performance gain is measured by BD-rate~\cite{BDData} against the HEVC Intra anchor~\cite{HEVC_overview} (a.k.a., BPG), and complexity measures include the Multiply–Accumulate Operations per pixel (MACs/pixel) and the size of model parameters in bytes. Here {the models at the highest quality level with the largest model are used for comparison.} Notable LIC methods like 
Ball\'e'18~\cite{balle2018variational}, Minnen'18~\cite{minnen2018joint}, Cheng'20~\cite{cheng2020learned}, Minnen'20~\cite{minnen2020channel}, Xie'21~\cite{xie2021enhanced}, Qian'22~\cite{Qian2021Entroformer}, and the VVC Intra~\cite{bross2021overview} are evaluated. BD-rate is averaged using all test images in the Kodak dataset. % RGB images are directly encoded under 444 mode~\cite{hevc_intra,vvc_intra}. 
Our method reports the best performance with fewer MACs/pixel and parameters.}
\label{fig:complexity}
\end{figure}

\IEEEpubidadjcol

\subsection{Motivation}

% \begin{figure*}
% \begin{center}
% \medskip
% \begin{minipage}[t]{\linewidth}
% \begin{center}
% \subfloat[Balle \etal]{
% \includegraphics[width=0.2\linewidth]{figs/erf/vis_balle_q3.pdf}
% }
% \subfloat[Minnen \etal]{
% \includegraphics[width=0.2\linewidth]{figs/erf/vis_minnen_q3.pdf}
% }
% \subfloat[Ours w/ CNXB]{
% \includegraphics[width=0.2\linewidth]{figs/erf/vis_conv_q3.pdf}
% }
% \subfloat[Ours]{
% \includegraphics[width=0.2\linewidth]{figs/erf/vis_cvt_q3.pdf}
% }

% \subfloat[Balle \etal]{
% \includegraphics[width=0.2\linewidth]{figs/erf/vis_balle_q7.pdf}
% }
% \subfloat[Minnen \etal]{
% \includegraphics[width=0.2\linewidth]{figs/erf/vis_minnen_q7.pdf}
% }
% \subfloat[Ours w/ CNXB]{
% \includegraphics[width=0.2\linewidth]{figs/erf/vis_conv_q7.pdf}
% }
% \subfloat[Ours]{
% \includegraphics[width=0.2\linewidth]{figs/erf/vis_cvt_q7.pdf}
% }
% \end{center}
% \end{minipage} %\par
% \caption{\textbf{Compactness Visualization of Latent Features.}  Close-ups of the gradient maps averaged over Kodak dataset.}
% \label{fig:visual}
% \end{center}
% \end{figure*}

In general, the ``transform'' module
%~\footnote{Usually a transform module is comprised of a pair of analysis $\mathsf{g}_a(\cdot)$ (or forward) and synthesis $\mathsf{g}_s(\cdot)$ (or backward) functions.} 
converts an image block in the pixel domain to a latent space (e.g., frequency domain), by which less nonzero coefficients are retained to represent the input source~\cite{952802};  Then the ``quantization'' function uses finite  symbols to represent transformed coefficients with the least bitrate desire under a certain distortion target~\cite{GrayQuantization}. Finally, the ``entropy coding'' engine is  devised to further reduce statistical redundancy by accurately modeling the probability distribution of each quantized symbol~\cite{cabac}.  Finetuning the transform, quantization, and entropy coding jointly is enforced for  decades to pursue better image compression as defined in \eqref{eq:vae_rd}~\cite{bross2021overview,sullivan_videoConcepts,JPEG,JPEG2K}. 

Given that the scalar quantizer is widely applied in mainstream image compression solutions, we keep using it and have the main focus of this work on transform and entropy coding.

{\bf Transform Function.} Since the 1970s, a great amount of studies have been devoted to advance the transform module, from the very first Discrete Cosine Transform (DCT)~\cite{DCT}, to variable-size Karhunen-Lo\`eve Transform (KLT)~\cite{zhang2020image}, to Hybrid intra Prediction/Transform (HiPT) that applies spatial intra prediction and residue DCT across variable-size tree blocks~\cite{sullivan_videoConcepts,hevc_intra,vvc_intra}, and to Nonlinear Neural Transform using attention optimized Convolutional Neural Networks (CNN)~\cite{balle2020nonlinear,cheng2020learned,chen2021end}. 
%with the support of attention mechanism\cite{cheng2020learned,chen2021end}. 
All of these endeavors give a clear direction: exploiting redundancy exhaustively with better energy compaction desires content-adaptive transforms that can {\it effectively characterize the neighborhood distribution conditioned on the dynamic input}. 

{\bf Entropy Context Model.} As revealed in a sequence of image/video coding standards, context-adaptive arithmetic coding  demonstrates its superior capacity to model non-stationary and high-order statistics among syntax elements (e.g., quantized coefficients). Unlike transforms that can use block-level parallelism to some extent~\cite{VVC_parallelism}, the context model, especially in the decoding phase,  operates sequentially because of causal dependency~\cite{cabac}. Thus, high-performance and high-throughput entropy modeling is of great importance in practice~\cite{HT-CABAC}. How to {\it leverage neighborhood  dependency to arrange a more appropriate order for context modeling} that not only provides accurate probability estimation but also assures computationally-efficient processing in the entropy coding  engine is crucial.

As seen, the vital factor behind high-efficiency LIC for ensuring high-performance compression and high-throughput computation jointly is highly related to the efficient  use of neighborhood information that is defined as the {\it Adaptive Neighborhood Information Aggregation} (ANIA).

\subsection{Our Method}

This work, therefore,  fulfills the use of ANIA in respective transform and entropy coding modules for high-efficiency LIC.

{\bf Content-Adaptive Transform Through Integrated Convolution \& Self-Attention.} Past explorations have suggested us leveraging neighborhood dependency adaptively for better transformation~\cite{vvc_intra,zhang2020image}. Although deep CNN-based nonlinear transforms have been devised in a collection of LIC approaches shown in Fig.~\ref{fig:complexity} because of their powerful representation capacity to embed neighborhood information of underlying content, they do have limitations~\cite{liang2021swinir}. For example, offline-trained  CNN models are presented with fixed receptive fields and weights  in inference, making them generally inefficient for unseen images that exhibit different content distribution from training samples~\cite{jia2016dynamic}.

To tackle it, we propose the Integrated Convolution and Self-Attention (ICSA) unit that is comprised of a convolutional layer and multiple self-attention layers realized by local window-based Residual Neighborhood Attention Blocks (RNABs)~\cite{hassani2022neighborhood}. 
%to adaptively aggregate neighborhood information (elements) through window based weighting conditioned on the input. 
{The convolutional layer is applied in each ICSA unit to not only reduce the data dimensionality~\cite{chen2021end} but also exploit the hierarchical characteristics of the content~\cite{zhang2020image}. In comparison to fixed-weights convolutions used in pre-trained CNN models, the self-attention mechanism in succeeding RNABs can  weigh and aggregate neighboring elements on-the-fly with which instantaneous content input can be better characterized to some extent.}

{\bf Entropy Coding Using Multistage Context Model.} Adaptive context modeling conditioned on hyperpriors and spatial-channel neighbors jointly that was originally proposed by Minnen et al.~\cite{minnen2018joint} and extended in succeeding followups~\cite{cheng2020learned,chen2021end,lu2021transformer,xie2021enhanced} is able to accurately approximate the probability of latent features following an autoregressive manner. % For simplicity, we call it autoregressive (AR) model in short.
However, the sequential processing of spatial or spatial-channel autoregressive neighbors (in a raster scan order) makes the image decoder extremely impractical, e.g., taking hours to reconstruct a 1080p RGB image due to element-by-element computation as reported in~\cite{chen2021end,Liu_CSTR}\footnote{Image encoding speedup can be easily fulfilled by parallel processing since elements are all available but the casual data dependency enforces the sequential processing strictly in image decoding. Thus image decoding runtime or latency is another vital factor for practical application.}.

\begin{table}[t]
    \centering
    \caption{Notations} \label{tab:notations}
    \begin{tabular}{c|c}
    % \toprule
    \hline 
    % \rowcolor{gray!40} 
    Abbr. & Description\\
    \hline
%    Sax (as Sacs) & Self-Attention and Convolution Stack\\

    % L$^2$IC &  Learned Lossy Image Coding\\
ANIA & Adaptive Neighborhood Information Aggregation\\
ICSA & Integrated Convolution and Self-Attention\\
RNAB & Residual Neighborhood Attention Block\\
MCM & Multi-stage Context Model\\
GCP & Generalized Checkerboard Pattern\\
\hline
%MLP & Multilayer Perceptron\\
   %R-D & Rate-distortion\\
    VAE & Variational Auto-Encoder\\
         LIC    &   Lossy Image Coding \\
    %\hline
    MAC & Multiply–Accumulate Operation\\
     BD-rate~\cite{BDData} & Bj{\o}ntegaard Delta Rate\\
     PSNR & Peak Signal-to-Noise Ratio\\
     MSE & Mean Square Error\\
     MS-SSIM~\cite{wang2003multiscale} & Multiscale Structural Similarity\\
     \hline
     BPG & Better Portable Graphics\\
     HEVC~\cite{HEVC_overview} & High-Efficiency Video Coding\\
     VVC~\cite{bross2021overview} & Versatile Video Coding\\
     \hline
     % \bottomrule
    \end{tabular}
\end{table}

Thus, devising a method that not only best maintains the performance of the autoregressive model but also enables parallel processing for high-throughput computation is of great desire. Apparently, the efficiency of the autoregressive model comes from the utilization of causal neighbors for conditional probability estimation. % (see spatially-ordered elements in Fig.~\ref{fig:seq_proc} as an example). 
Simply enforcing the independent processing of each latent element by completely ignoring the inter dependency across neighbors for  concurrency can improve the throughput but definitely hurt the compression performance. It urgently calls for intelligently exploiting neighborhood dependency using a different conditional manner (or scan order).

% \lm{He \etal~\cite{he2021checkerboard} firstly introduced the checkerboard pattern for spatial arrangement of neighborhood context modeling. By re-organizing the decoding order, it achieved significantly improved computational efficiency. However, half of the latents in their method were predicted only depend on the hyperpriors and the contextual information would not be fully utilized, which might constraint the performance to some extent. Minnen \etal~\cite{minnen2020channel} proposed a channel-wise conditional model to replace the spatial contextual modeling. By slicing the latent features into several groups in channel dimension, the entropy parameters of latent representations within latter groups can be predicted using the hyper priors and the predicted groups before. This method proved better RD performance than the serial autoregressive model but brought in a huge amount of parameters and multiply-accumulation operations which was calculation unfriendly. He \etal~\cite{he2022elic} then introduced a spatial-channel context model by considering both spatial and inter-channel contextual relationships together, which obviously achieved superior performance.}

{As inspired by recent studies in~\cite{he2022elic,minnen2020channel} where they arranged the context prediction across evenly-grouped feature channels, and/or uniformly-grouped spatial neighbors for parallel probability estimation, we propose the Multistage Context Model (MCM) to process nonuniformly-grouped spatial-channel features in a pre-arranged  context modeling order for optimal performance-complexity trade-off.}

{We first slice the latent feature tensor along with the channel dimension into four sub-tensors with variable channels following the Cosine slicing strategy where the number of channels increases gradually from the first to the fourth channel-grouped sub-tensor. Upon each channel-grouped sub-tensor, a Generalized Checkerboard Pattern (GCP) is utilized to group spatial neighbors for multi-step processing where concurrent context prediction is applied for same-group elements using available spatial-channel neighbors previously-processed in preceding steps. The granularity of GCP decreases from one stage to another for the processing of corresponding channel-grouped sub-tensor, e.g., 4-Step GCP at the first stage, 2-Step GCP for both the second and their stage, and direct channel-wise context prediction without spatial GCP for the last stage. As revealed later,  progressively processing such non-uniformly grouped spatial-channel features ensures accurate and high-throughput context modeling simultaneously.}

{\bf End-to-End Architecture.} We stack the ICSA and MCM units upon the prominent VAE structure  to form a novel LIC. We call it {\it TinyLIC} as  in Fig.~\ref{fig:network}. Such VAE architecture has been well generalized in various LICs~\cite{balle2018variational,minnen2018joint,cheng2020learned,chen2021end}. As seen, main and hyper coders are paired with encoding and decoding processes. In the main encoder, it generally performs the analysis transform $g_a(\cdot)$ using four consecutive ICSA units to derive latent features of input image $\bf x$ while the main decoder mirrors the encoding as the synthesis transform $g_s(\cdot)$ to reconstruct $\hat{\bf x}$. To efficiently encode quantized latent features $\hat{\bf y}$, the MCM jointly utilizes  the hyperpriors and spatial-channel neighbors, where hyperpriors are generated by the hyper coder that uses two paired ICSA units and a factorized model-based entropy coding~\cite{balle2018variational}.

\subsection{Contribution}

Our contributions are summarized below:

1) This work shows that high-efficiency LIC with both high-performance compression and high-throughput computation can be successfully fulfilled by  adaptive neighborhood information aggregation (ANIA) to best exploit neighborhood characteristics in  transform and entropy coding; As for transform function, the ANIA dynamically adapts itself to the input to best  embed neighborhood information; while for entropy coding, it carefully arranges the order of context modeling upon non-uniformly grouped spatial-channel features, which  not only retains the efficiency as the autoregressive model but also enables high-throughput parallel processing.

2) This work exemplifies the design of ANIA by using the Integrated Convolution and Self-Attention unit for content-adaptive transform, and  the Multistage Context Model in entropy coding, respectively, to form the proposed {\it TinyLIC}; Extensive comparisons report the superior compression efficiency of the {\it TinyLIC}, outperforming the VVC Intra and other notable LICs for three popular datasets; More importantly, the {\it TinyLIC} offers the best complexity-performance tradeoff, reporting $>10$ absolute percentage BD-rate  points improvement against the same HEVC Intra, $>60 \times$ decoding speedup with a comparable-size model to the Minnen'18~\cite{minnen2018joint} - the seminal foundation for other LIC approaches.

3) The proposed {\it TinyLIC} further reports its generalization by thoroughly examining a variety of settings in modular components such as the backbone structure (e.g., feature embedding, self-attention method), entropy context modeling (e.g., conditional estimation method) in ablation studies. {Additional experiments are also carried out to report the efficiency of the {\it TinyLIC} in a companion supplementary material.}

\section{Related Work} \label{sec:related_work}
This section briefs the developments in transform coding for image compression including classical rules-based approaches and recently-emerged learning-based solutions.

\subsection{Rules-based Transform Coding} \label{sec:related_transform_coding}

%\subsubsection{Transform Function}

{\bf Fixed-Weights Transforms.} Prominent transforms like DCT (Discrete Cosine Transform)~\cite{DCT} and Wavelet~\cite{J2K_Wavelet} use linear transformations that are generally comprised of a set of linear and orthogonal bases.
They have been used in famous image coding standards like JPEG~\cite{wallace1992jpeg} and JPEG2000~\cite{rabbani2002jpeg2000}.
Later, DCT alike Integer transforms~\cite{DCT_drift} are adopted in intra profile of respective H.264/AVC~\cite{AVC_transform}, HEVC~\cite{HEVC_overview}, and VVC~\cite{bross2021overview} to process predictive residues. 

Apparently, linear transformation with fixed bases can not best exploit the redundancy because the content of the underlying image block is non-stationary and does not strictly follow the mathematical distribution as assumed (e.g., Gaussian source~\cite{952802}). Therefore, devising transformation with data-driven bases to better exploit non-stationary content distribution attracts intensive attention. Notable approaches include the dictionary learning~\cite{yao_selfDict,Xu_CompressDict,RLS-DLA}, KLT~\cite{zhang2020image} and recently-emerged CNN transforms~\cite{balle2020nonlinear} (see Sec.~\ref{sec:related_LLIC} for more details). %Note that transforms with data-driven bases are also with fixed weights after training.

{\bf Content-Adaptive Transforms.} Although data-driven transforms have improved energy compaction~\cite{zhang2020image} to some extent compared with fixed-basis DCT or wavelet, the model generalization is still a challenging problem due to fixed weights after training. For example, if the distribution of test data is different from the training samples, energy compaction is largely suffered with poor coding performance~\cite{zhang2020image}.

Given that neighborhood pixels often presented high coherency, adaptively weighting local spatial neighbors through an autoregressive predictive means~\cite{li2001edge} or predefined directional patterns~\cite{vvc_intra,hevc_intra} had been proposed and extensively studied over the past decades. Since the late 1990s,  spatial intra prediction was integrated with the aforementioned fixed-basis transforms (e.g., DCT), forming the normative toolset in mainstream intra profiles of video coding standards like H.264/AVC Intra, HEVC Intra, and VVC Intra, because of the superior performance on redundancy removal and energy compaction~\cite{zhang2020image}.  

Such Hybrid intra Prediction/Transform (HiPT) dynamically characterizes and embeds spatial neighbors, making it content adaptive. Then after, variable-size HiPT has been extended along with the recursive tree structures, by which the non-stationary image characteristics in different regions can be well and adaptively captured and modeled. 

The use of reconstructed neighbors in HiPT leverages the neighborhood coherency through handcrafted rules to best reflect the  dynamics of the input content, which motivates us to develop the content-adaptive transformation from a learning perspective. 

% \begin{table}[t]
% \caption{An overview of the transform coding methods utilized in notable LICs.}
% \label{tbl:bd-rate}
% \centering
% \begin{tabular}{cccc}
% \toprule
% \multirow{2}{*}{Method} & \multirow{2}{*}{Transform Function} & \multicolumn{2}{c}{Entropy Coding} \\
% & & Hyper & Context \\
% \midrule
% Ball\'e'18~\cite{balle2018variational} & CNN & CNN & H \\
% Minnen'18~\cite{minnen2018joint} & CNN & CNN & H + S \\
% Cheng'20~\cite{cheng2020learned} & CNN + Attention & CNN & H + S \\
% Minnen'20~\cite{minnen2020channel} & CNN & CNN & H + C \\  
% Chen'21~\cite{chen2021end} & CNN + Attention & CNN & H + S \\ 
% Xie'21~\cite{xie2021enhanced} & Flow + CNN & CNN & H + S \\ 
% Zhu'22~\cite{zhu2021transformer} & Attention & Attention &  H + C \\ 
% Qian'22~\cite{Qian2021Entroformer} & CNN & Attention &  H + T \\ 
% He'22~\cite{he2022elic} & CNN & CNN &  H + B \\ 
% \bottomrule
% \end{tabular}
% \begin{tablenotes}
% \item Here, H denotes hyperpriors~\cite{balle2018variational}, S represents serial autoregressive model proposed in \cite{minnen2018joint}, C is the channel-wise conditional model~\cite{minnen2020channel}, T stands for the checkerboard context model~\cite{he2021checkerboard} and B corresponds the spatial-channel context model~\cite{he2022elic}.
% \end{tablenotes}
% \end{table}

{\bf Entropy Model.}
Quantized transform coefficients are subsequently encoded into binary strings for efficient storage or network delivery, by further exploiting  their statistical correlations. Extensive explorations conducted in the past~\cite{cabac} have clearly revealed that an accurate context model conditioned on neighborhood elements plays a vital role in high-efficiency entropy coding. Examples include the context-adaptive variable-length coding (CAVLC) and context-adaptive binary arithmetic coding (CABAC)~\cite{cabac}. And, because of the superior efficiency offered by the arithmetic codes, CABAC, and its variants, are widely deployed in mainstream compression recommendations like HEVC, VVC, JPEG2000, etc, where associated context models are mainly developed following  empirical rules and experimental observations.

Computation throughput limitation incurred by the sequential data dependency in  context modeling was extensively investigated since the standardization of HEVC a decade ago.  High-throughput and high-performance were then jointly evaluated during the development of the entropy coding engine~\cite{HT-CABAC,VVC_Q_Entropy}. Well-known examples include symbol parsing dependency unknitting, bins grouping, etc that more or less rely on the utilization of contextual correlation in a local neighborhood.

\subsection{Learning-based Transform Coding} \label{sec:related_LLIC}

Given that LIC methods jointly optimize transform and entropy coding modules through end-to-end learning, we review them together.

{\bf CNN Models.} As CNNs have shown their remarkable capacity for generating compact representation features from underlying image data in various visual tasks, numerous attempts have been made in recent years to use CNN models for image compression. For instance, in 2017, Ball\'e et al.~\cite{balle2016end} showed that stacking convolutions could replace the traditional transform coding to form an end-to-end trainable image compression method with better efficiency than the JPEG~\cite{JPEG}, in which a CABAC alike entropy coding engine was used. Then, hyperpriors and spatial-channel neighbors were jointly used in~\cite{minnen2018joint,cheng2020learned,chen2021end} for context modeling assuming the Gaussian distribution following an autoregressive manner, which further improved the image compression efficiency. As seen such a context model conditioned on joint hyperprior and autoregressive neighbors mostly utilized the local correlations. Recently, Qian et al.~\cite{qian2020learning} and Kin et al.~\cite{kim2021joint} extended the utilization of only local correlation to the use of both global and local correlation by the inclusion of additional global priors.

In addition to these methods mainly utilizing convolutions to aggregate information locally, our early exploration in~\cite{chen2021end} applied nonlocal attention to optimizing intermediate features generated by the convolutional layer for more effective information embedding. However, the nonlocal computation is expensive since it typically requires a large amount of space to host a correlation matrix with the size of $H_fW_f\times H_fW_f$. Here $H_f$ and $W_f$ are the height and width of the input feature map. A similar convolution-based spatial attention mechanism was also used in~\cite{cheng2020learned} and other related works. 

{\bf ViT Models.}  LIC solutions discussed above mainly leveraged CNNs to formulate nonlinear transform and high-performance entropy coding. With the surge of self-attention-based Vision Transformers (ViT) in various tasks~\cite{ViT_Survey,carion2020end,dosovitskiy2020image,zhou2021deepvit,wang2021uformer,liu2021swin}, a number of attempts had been made to apply Transformer alike schemes to improve transform and entropy coding in a LIC. For example, Zhu et al.~\cite{zhu2021transformer} replaced stacked convolutions with Swin Transformer~\cite{liu2021swin} to form the nonlinear transform and kept using the channel-wise context prediction as in~\cite{minnen2020channel}; while Qian et al.~\cite{Qian2021Entroformer} retained CNN transform   but replaced the convolution-based context modeling with the Transformer. Coincidentally at the same time, our preliminary study in~\cite{lu2021transformer} extended the Transformer architecture to both transform and entropy coding modules. 

{\it Discussion.} Most works have claimed that the ability of  long-range dependency capturing in ViTs improves the  CNN models that operate locally. Yet, we have a different view: we believe that the compression gains are mainly contributed by the self-attention mechanism that can best weigh neighborhood information of  the dynamic input. As reported in~\cite{liu2022convnet}, large-kernel convolutions can also capture relatively long-range dependency as ViTs for various tasks.

\begin{figure*}[t]
\centering
\includegraphics[width=\linewidth]{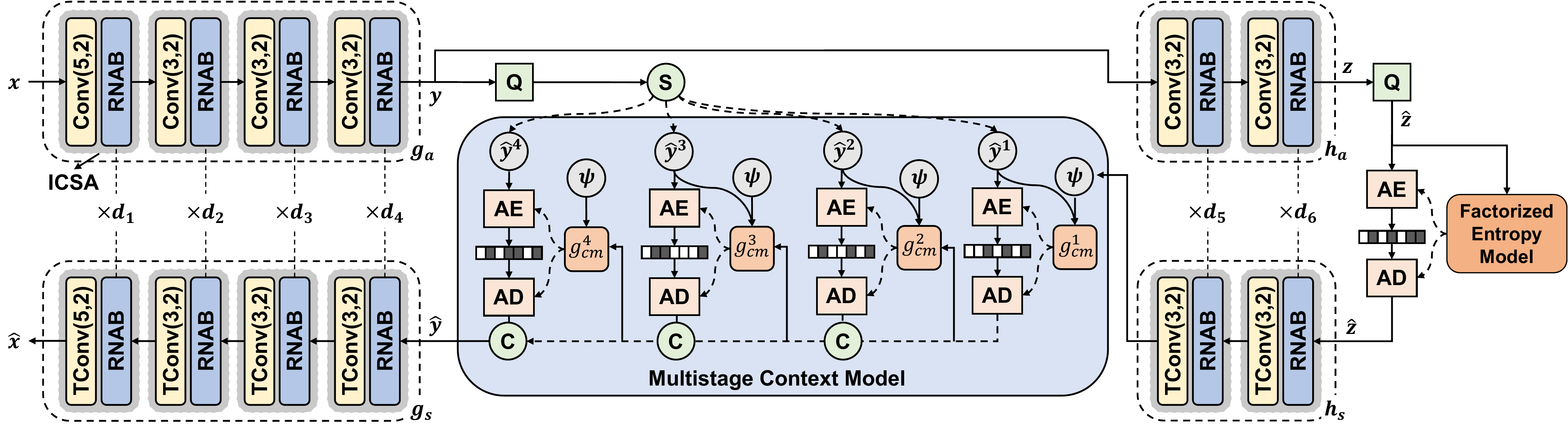}
% where an .eps filename suffix will be assumed under latex, 
% and a .pdf suffix will be assumed for pdflatex; or what has been declared
% via \DeclareGraphicsExtensions.
\caption{{{\bf TinyLIC}.} Prominent VAE structure is used with main and hyper encoder-decoder pairs. Four and two paired ICSA units are used in main and hyper coders. Each ICSA is comprised of a convolutional layer for feature embedding and spatial resampling, and multiple RNABs to adaptively aggregate neighborhood information through attention-based weighting. $d_i, i=1,2,\ldots,6$ is the number of RNABs used at $i$-th stage. Convolution Conv($k$,$s$) and its transposed version TConv($k$,$s$) apply the kernel at a size of $k\times k$ and a stride of $s$. $k$ = 3 or 5, and $s$ = 2. The MCM fully utilizes the hyperpriors $\psi$ and non-uniformly grouped channel and spatial elements for better probability estimation, and a simple factorized entropy model is used to encode hyperpriors. S and C represent the tensor Slicing and Concatenation; Uniform Quantization is used in Q; AE and AD stand for respective Arithmetic Encoding and Decoding.}
\label{fig:network}
\end{figure*}

% \subsubsection{Efficient Learned LIC}
{\bf Efficient Entropy Model.}
Computation efficiency is another key factor determining whether the solution can be used in practice. Existing LIC approaches were rigorously criticized for exhaustive  computing and caching. 
% One direction to develop computationally efficient learned LIC is to enable variable-rate support using few pretrained models since most solutions have trained specific model for individual bitrate~\cite{chen2020variable,Sun21_InterpCA,MMSP_Liu_VariableRate}. This would reduce the model switching for supporting wider bitrate range, and also decrease the storage requirement for caching multiple models. {As shown at our webpage, the proposed {\it TinyLIC} can be easily extended to support variable-rate compression without noticeable performance loss.} In the meantime, to execute learned LIC with less peak memory in inference, native floating-point model could be quantized using integer network~\cite{balle2018integer} and fixed-point network~\cite{hong2020efficient}.
The most computationally exhaustive subsystem is the sequential processing of syntax elements in entropy coding. For instance, the decoding runtime of a popular context model conditioned on joint hyperprior and spatial autoregressive neighbors is a function of $\mathcal{O}(H\times W)$, which is unbearable for practical image applications. Massively-parallel context modeling was then developed by exploring channel-wise concurrency like channel-wise grouping~\cite{minnen2020channel}, and spatial concurrency like checkerboard patterning~\cite{he2021checkerboard} or column-wise/row-wise parallelism~\cite{chen2021end} to improve the throughput with reasonable performance compromise.

{ He et al.~\cite{he2021checkerboard} solely relied on the 2-step checkerboard pattern to perform context modeling across grouped spatial neighbors. Although it significantly improved computational efficiency, the context probability estimation of half of the latent features used the hyperpriors only, which led to a noticeable performance loss to the default autoregressive model (see reproduced results of ``P'' model in Fig.~\ref{fig:entropy_model}). Minnen et al.~\cite{minnen2020channel} then proposed a channel-wise conditional model by slicing the latent feature tensor into ten equal-channel groups to avoid the use of spatial autoregressive neighbors for context modeling. The probability of latent features in the latter grouped channels can be predicted using the hyperpriors and the previously-processed groups. This method showed better R-D performance than the serial autoregressive model by additionally costing a huge amount of parameters and multiply-accumulation operations. Later, the combination of non-uniform channel grouping and uniform 2-Step spatial checkerboard grouping in each grouped channels was given in~\cite{he2022elic} with improved efficiency. 
The aforementioned methods performed the uniform feature grouping either spatially or channel-wisely to do context modeling following a pre-arranged order.

%He \etal~\cite{he2022elic} then tried to utilize both spatial-wise and channel-wise relationships together, which obviously achieved 
%superior performance.
}

{The proposed MCM groups the latent features non-uniformly from both spatial and channel dimensions. In this way, our method extends existing methods in~\cite{minnen2020channel,he2021checkerboard,he2022elic} with a generalized solution which offers the best performance-complexity tradeoff as reported in subsequent studies.}

 %through a 2$\times$2 spatial patterning for context prediction using valid neighbors and stage-wise massive parallelism. Contrast with the checkerboard context modeling in~\cite{he2021checkerboard}, e.g., a typical example of the two-stage spatial patterning shown in Fig.~\ref{fig:para_2CC}, the proposed MCM provides better compression performance by gradually aggregating valid neighbors from one stage to another but only needs to use small-kernel convolutions.

\section{Proposed Method} \label{sec:method}
{This section first overviews the proposed {\it TinyLIC}. More details are given subsequently for transform and entropy coding. For better comprehension, notations are given in  Table~\ref{tab:notations}.}

\subsection{Overview}

Figure~\ref{fig:network} depicts the {\it TinyLIC} which follows the end-to-end VAE architecture~\cite{minnen2018joint} to construct main and hyper encoder-decoder pairs to layer-wisely analyze and aggregate neighborhood information for R-D optimized compact representation.

{Given an input image ${\bf x} \in \mathbb{R}^{H \times W \times 3}$, the analysis transform $g_a(\cdot)$ is first applied to derive the latent representation $\bf y$, which is further processed to generate hyper features $\bf z$ through $h_a(\cdot)$. Quantization is used to discretize both $\bf y$ and $\bf z$, producing $\hat{\bf y}$ and $\hat{\bf z}$ for entropy coding. A simple factorized entropy model is applied for  $\hat{\bf z}$~\cite{balle2018variational}. %assuming only a very small percentage of the total file size is consumed by the hyper priors. 
The Gaussian conditional model is used to characterize $\hat{\bf y}$ for probability estimation where its mean $\mu$ and scale $\sigma$ parameters are predicted using decoded hyperpriors $\psi$ (e.g., via $h_s(\cdot)$) and available spatial-channel neighbors~\cite{minnen2018joint} following a pre-arranged order. The final $\hat{\bf x}$ is reconstructed using the synthesis transform $g_s(\cdot)$.}

Thus, rate-distortion optimization of the VAE in Fig.~\ref{fig:network} can be extended from \eqref{eq:vae_rd} as
\begin{align}
   J = \mathbb{E}_{{\bf x} \sim p_{\bf x}}[-\log_{2}p_{\mathbf{\hat{\bf y}}}(\hat{\bf y} | \mathbf{\hat{z}})] &+ \mathbb{E}_{{\bf x} \sim p_{\bf x}}[-\log_2p_{\mathbf{\hat{\bf z}}}(\hat{\bf z})] \nonumber \\ 
   + \lambda \cdot \mathbb{E}_{{\bf x} \sim p_{\bf x}}[\mathsf{d}({\bf x}, \hat{\bf x})], \label{eq:J_cost}
\end{align} 
where $p_{\bf x}$ is the distribution of input source image, $p_{\hat{{\bf y}}}$ and $p_{\hat{{\bf z}}}$ are the probability distribution of respective $\hat{\bf y}$ and $\hat{\bf z}$ at bottleneck layer for entropy coding. 

Next, the ICSA based content-adaptive transform and the MCM based context modeling are given in detail.

\subsection{Content-Adaptive Transform via Stacked ICSAs} \label{sec:nonlinear_transform}

\subsubsection{Convolutional Feature Embedding \& Resampling}
To ensure the spatial coherency as suggested in Vision Transformer studies~\cite{dosovitskiy2020image,carion2020end}, we perform the convolutional feature embedding to tokenize the input image ${\bf x}$ for the first ICSA unit in Fig.~\ref{fig:network} into a latent space. The same tokenizations are enforced for subsequent ICSA units to process the output from the RNAB module of proceeding ICSA for hierarchical feature embedding. It is worth pointing out that we also apply the spatial resampling at the convolution layer in each ICSA. This is because convolutions can aggregate spatial neighbors within the receptive field to some extent, by which we can down-sample the resolution to reduce data dimensionality spatially with negligible information loss. For simplicity, we apply uniform sampling at each dimension with a stride of 2.

Since convolutional feature embedding can implicitly encode the position information~\cite{islam2019much} to capture the spatial relationship between tokenized latents, it does not require the explicit position signaling in the tokenization phase as reported in~\cite{dosovitskiy2020image,carion2020end}. Additionally, compared with non-overlapping pixel patches used in~\cite{zhu2021transformer}, convolutional features used for succeeding window-based self-attention computation can avoid blocky artifacts, and is beneficial to early visual processing and stable training as reported in~\cite{gulati2020conformer,xiao2021early} and our simulations in Sec.~\ref{sec:backbone}.

\subsubsection{Window-based Self-Attention via RNAB}

Years ago, despite the great success of Transformers in high-level vision tasks (e.g., classification)~\cite{dosovitskiy2020image,carion2020end}, it is difficult to directly migrate the self-attention layer in Transformers to low-level vision tasks (e.g., compression) because of the quadruple computation complexity of input image size. Recently, the emergence of window-based self-attention~\cite{liu2021swin, hassani2022neighborhood}  demonstrates outstanding efficiency with much less computation consumption. 

\begin{figure}[t]
\centering
\includegraphics[scale=0.4]{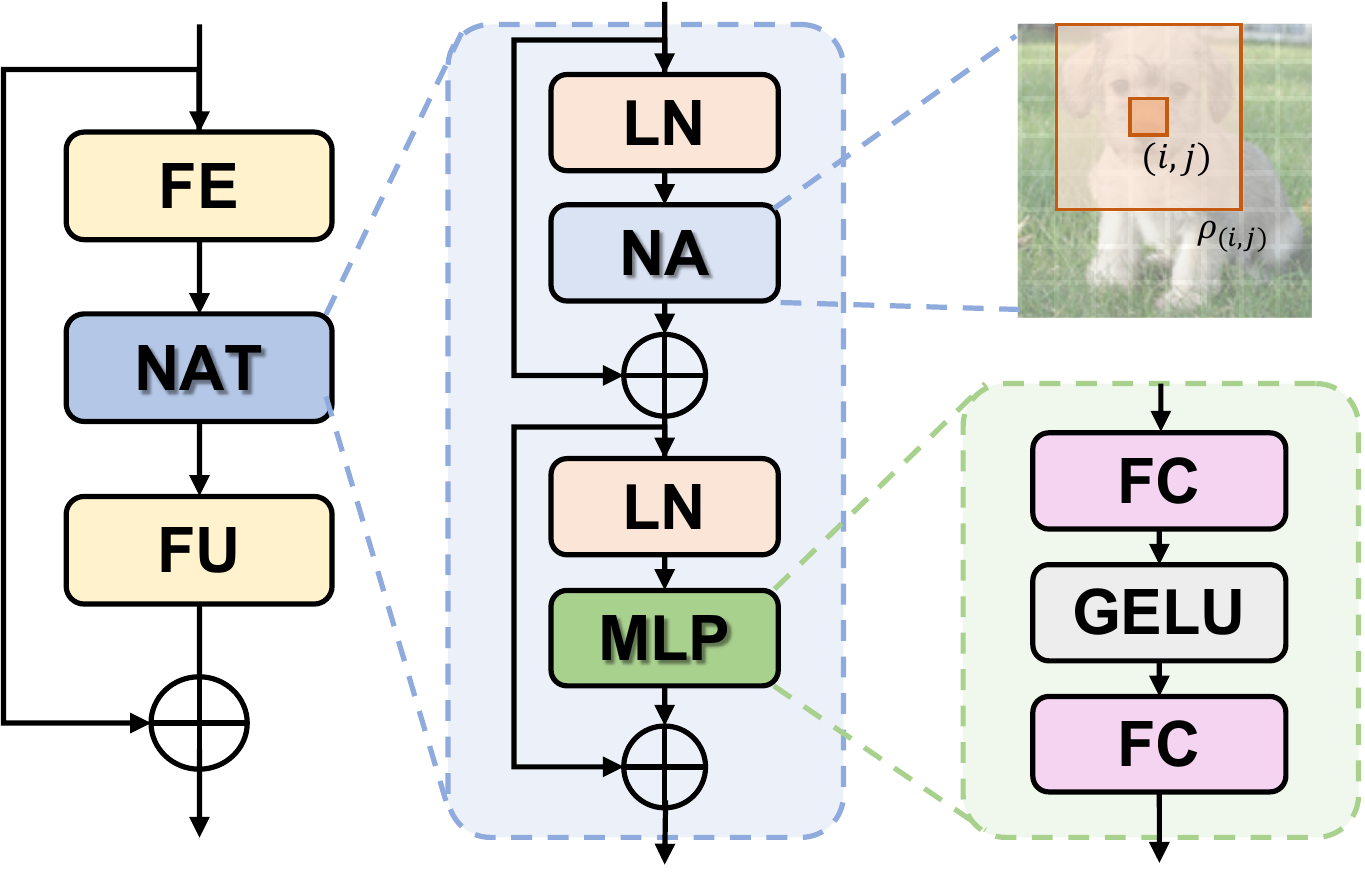}
\caption{{\bf Residual Neighborhood Attention Block.} FE and FU denote the feature embedding and unembedding layer. NAT is the Neighborhood Attention Transformer consisting of layer normalization (LN), Neighborhood Attention (NA), and multi-layer perceptron (MLP) layers. Two fully connected (FC) layers are interleaved with a GELU activation layer to form the MLP layer. Residual skip connections are applied.}
\label{fig:rnab}
\end{figure}

% \begin{figure}[t]
% \centering
% \subfloat[]{\includegraphics[scale=0.45]{Figures/swin.pdf}\label{fig:twoswin}}
% \hspace{0.5cm}
% \subfloat[]{\includegraphics[scale=0.45]{Figures/neighborhood.pdf}}
% \caption{{\bf Two Successive Transformer Blocks.} (a) Swin Transformers apply a shifted window to enlarge the receptive field with the attention coefficients reweighted by the second block. (b) Neighborhood Attention Transformers will weight twice the attention coefficients during the same window with neighbor pixels.}
% \end{figure}

{As a result, we use the Neighborhood Attention Transformer (NAT) proposed in~\cite{hassani2022neighborhood} as an example to form the Residual Neighborhood Attention Block (RNAB) for window-based self-attention computation. Other window-based self-attention mechanisms like Swin Transformer~\cite{liu2021swin} can be applied as well (see Sec.~\ref{sec:backbone}). Multiple RNABs are often stacked and connected with a convolutional layer to form an ICSA unit as in Fig.~\ref{fig:network}.} 
%we propose to stack multiple Residual Neighborhood Attention Blocks (RNABs) right after the convolutional layer to form the ICSA unit. 

{Figure~\ref{fig:rnab} briefly sketches the processing flow of an RNAB.
\begin{itemize}
    \item First, the feature embedding (FE) layer projects input feature tensor at a size of $ H_f \times W_f \times C$ to a dimension of $H_fW_f \times C$ for processing;
    \item Subsequently, the NAT aggregates neighborhood information by stacking the layers for respectively processing the neighborhood attention (NA), multi-layer perception (MLP) and layer normalization (LN);
    \item Finally, a feature unembedding (FU) layer remaps attention-weighted features back to the original resolution at $ H_f \times W_f \times C$.
\end{itemize} Following the convention, residual skip connections are used for better information aggregation and model training \cite{He_2016_CVPR}.}

%where the first , the following NAT aggregates neighborhood information by stacking the neighborhood attention (NA) and multi-layer perception (MLP) layers.
%, both of them are with the layer normalization (LN) pre-processing.
% \begin{align}
% y^l &= \mathbf{NA}(\mathbf{LN}(y^{l-1})) + y^{l-1} \nonumber \\ 
% y^{l+1} &= \mathbf{MLP}(\mathbf{LN}(y^l)) + y^l.
% \end{align}
{The feature aggregation in NA layer can be formulated as 
\begin{align}
NA_{(i,j)}=softmax(\frac{Q_{i,j}K_{\rho(i,j)}^T+B_{i,j}}{\sqrt{d}})V_{\rho(i,j)}, \label{eq:na_features}
\end{align} where  $(i,j)$ is the element location and  $\rho(i,j)$ defines a local neighborhood centered at  $(i,j)$.
$Q_{i,j}$, $K_{\rho(i,j)}$ and $V_{\rho(i,j)}$ (i.e., query, key, and value) are linearly-transformed features.
 $B_{i,j}$ denotes the relative positional bias. $d$ is the query/key dimension. The MLP layer consists of two fully-connected (FC) layers and an activation layer GELU~\cite{hendrycks2016gaussian} in between.}

{As shown in \eqref{eq:na_features},  $Q_{i,j}$, $K_{\rho(i,j)}$ and $V_{\rho(i,j)}$ are computed on-the-fly to weigh local information which well reflects the content characteristics of any input. }
%And finally a feature unembedding (FU) layer remaps attention-weighted features back to the original size of $ H \times W \times C$. }

% \begin{figure}[!t]
% \centering
% \includegraphics[width=\linewidth]{Figures/mcm.pdf} 
% \caption{Multistage Context Model.}
% \label{fig:mcm}
% \end{figure}

\subsection{Multistage Context Model} \label{sec:entropy_mcc}

Using hyperpriors and spatial-channel neighbors jointly for context modeling brings significant R-D performance gain~\cite{minnen2018joint,cheng2020learned,chen2021end}. However, the sharp increase of decoding runtime due to the sequential processing of autoregressive neighbor is unacceptable where each latent feature element is calculated serially using causal neighbors through a masked convolution. 

As the strong correlation exists among  spatial-channel neighbors, instead of the sequential raster scan order used in the autoregressive model, we perform the MCM in a pre-arranged order by non-uniformly grouping spatial-channel features for accurate probability estimation in parallel.

\subsubsection{Four-Stage Processing Pipeline}
% A similar Two-stage Checkerboard Context Model was suggested by He \etal~in~\cite{he2021checkerboard}, further confirming the advantages of the use of alternative scan order in context modeling of entropy coding for the pursuit of compression performance and computation throughput jointly. Compared with the checkerboard patterning exemplified in Fig.~\ref{fig:para_2CC}, our MCM method offers better compression performance with the use of fine-grained local neighbors (see Table~\ref{table:latency} and Fig.~\ref{fig:entropy_model}).

% Next, we offer a detailed sketch of how to implement the Multistage Context Model  (MCM) in entropy coding. Stacked convolutions used for entropy parameter determination and context prediction are the same for encoding and decoding as specified in Sec.~\ref{sec:ablation_entropy_context}.

\begin{figure}[t]
\centering
{\includegraphics[width=\linewidth]{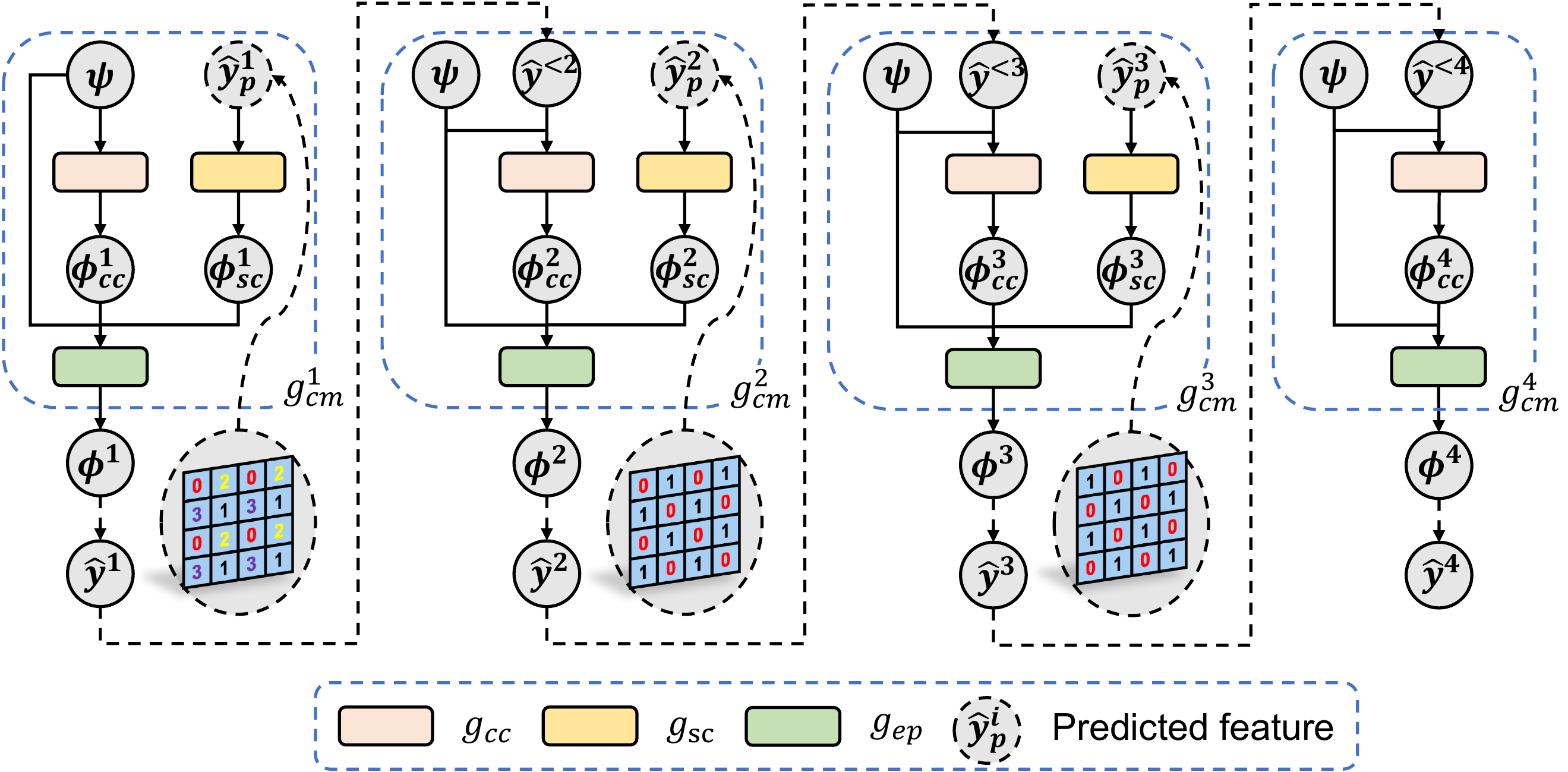}}
\caption{{\bf Multistage Context Model.} The channel conditional model $g_{cc}$ and spatial conditional model $g_{sc}$ are applied for contextual prediction. $\hat{\bf y}_{p}^{i}$ denotes the discrete latent features which have been already predicted.}
\label{fig:mcm}
\end{figure}

Figure~\ref{fig:mcm} pictures the processing pipeline of the proposed 4-Stage MCM, for which quantized latent feature tensor $\bf \hat{y}$ at the bottleneck is first sliced into four groups  along with the channel dimension, e.g., $\hat{\bf y}^i, (i=1, 2, 3, 4)$. 
The Cosine slicing is applied to generate groups with variable channels in learning, where we increase the number of channels gradually from $\hat{\bf y}^1$ to $\hat{\bf y}^4$.
On the contrary, existing methods either slice the tensor to equal-size groups uniformly~\cite{minnen2020channel} (e.g., Linear slicing) or handcraft the  channel slicing with variable-size groups~\cite{he2022elic}.
\begin{itemize}
    \item For the first stage, the entropy parameter $\phi^1$ used for deriving the feature probabilities in $\hat{\bf y}^1$ are computed by processing the concatenation of the hyperpriors $\psi$, channel-wisely aggregated neighbors (channel-wise neighbors) \begin{align}\phi_{cc}^1 = g_{cc}^1(\psi),\end{align} and spatially-aggregated neighbors (spatial neighbors) \begin{align}\phi_{sc}^1 = g_{sc}^1(\hat{\bf y}_p^1 ),\end{align} having $\hat{\bf y}_p^1$ as available spatial neighbors obtained step-wisely, e.g.,
    \begin{align}
\phi^1 = g_{ep}^1(\mathcal{C}\{\psi, \phi_{cc}^1, \phi_{sc}^1\}). \label{eq:stage_1_ep}
\end{align}
    Channel-wise aggregation  $g_{cc}(\cdot)$ stacks convolutional layers for computation while spatial aggregation $g_{sc}(\cdot)$ utilizes the Generalized Checkerboard Prediction (GCP). At this first stage, 4-Step GCP is used for fine-grained spatial information utilization. $g_{ep}(\cdot)$ stacks simple $1\times1$ convolutions to derive entropy parameters, i.e., mean $\mu$ and scale $\sigma$ assuming the Gaussian distribution. $\mathcal{C}\{\cdot\}$ processes the tensor concatenation. 
    \item For the second stage, the entropy parameters $\phi^2$  are computed using $\phi^2 = g_{ep}^2(\mathcal{C}\{(\psi, \phi_{cc}^2, \phi_{sc}^2)\})$ where $\phi_{cc}^2 = g_{cc}^2(\mathcal{C}\{\psi,\hat{\bf y}^1\})$ and $\phi_{sc}^2 = g_{sc}^2(\hat{\bf y}_{p}^2)$. Note the the processing at the third stage is almost the same as it of the second stage, but applies different channel-wise neighbors $\phi_{cc}^3 = g_{cc}^3(\mathcal{C}\{\psi,\hat{\bf y}^1,\hat{\bf y}^2\})$. Also, instead of using 4-step GCP, simpler 2-Step GCP is used at the second and third stages, where a slight difference is setting complementary checkerboard arrangement for spatial prediction.
    \item Finally, the entropy parameters $\phi^4$ are derived using $\phi^4 = g_{ep}^4(\mathcal{C}\{(\psi, \phi_{cc}^4)\})$ with $\phi_{cc}^4 = g_{cc}^4(\mathcal{C}\{(\psi, \hat{\bf y}^1, \hat{\bf y}^2, \hat{\bf y}^3)\})$. As seen, only hypepriors and channel-wise neighbors are used at this stage to simplify the computations greatly.

    %by processing the concatenation of the hyperpriors $\psi$, channel-wisely aggregated priors $\phi_{cc}^2 = g_{cc}^2(\mathcal{C}{\psi,\hat{\bf y}^1})$, and spatially-aggregated priors $\phi_{sc}^1 = g_{sc}^1(\hat{\bf y}^1 )$, e.g.,
    
\end{itemize}

These operators, e.g., $g_{cc}(\cdot)$, $g_{sc}(\cdot)$, and $g_{ep}(\cdot)$ share with the same architecture across the proposed four stages. Their implementation details are shown in the supplemental material. We use the superscript with them to specifically identify the stage-wise computation. Next, we exemplify the 4-Step GCP in detail while the simpler 2-Step GCP can be easily extended.

\subsubsection{4-Step GCP}
%\lm{Here, we offer a detailed sketch of the proposed MCM for entropy coding. Given the discrete latent features $\hat{\bf y}^i (i=1,2,3,4)$ grouped according to the ``Cosine'' function, we follow the re-defined mechanism to progressively predict the entropy probabilities shown in Fig.~\ref{fig:mcm}. In this work, we suggest to encode a small number of channels firstly can help to predict the remaining channels more precisely, which will be discussed in \ref{sec:channel_split}. As a result, for the first stage, the channel distributions can be estimated only depend on the hyperpriors $\psi$. By using the channel conditional model $g_{cc}$ which comprises several convolutional layers, we can get the inter-channel context information
%\begin{align}
%\phi_{cc}^1 &= g_{cc}^1(\psi). \label{eq:5}
%\end{align}
%On the other hand, a local autoregressive spatial context model is introduced for precise spatial de-redundancy, by which the former predicted latents (denoted as $\hat{\bf y}_p^1$) can be used to estimate the distribution of the elements at the next position to obtain the final results $\phi_{sc}^1$. It takes totally four steps for such spatial estimation in which the positions with the same number (denoted in dashed box) will be predicted at the same time. The encoding and decoding processes for such spatial rearrangement can be found below:}

\begin{figure}[t]
\centering
\subfloat[]{\includegraphics[scale=0.35]{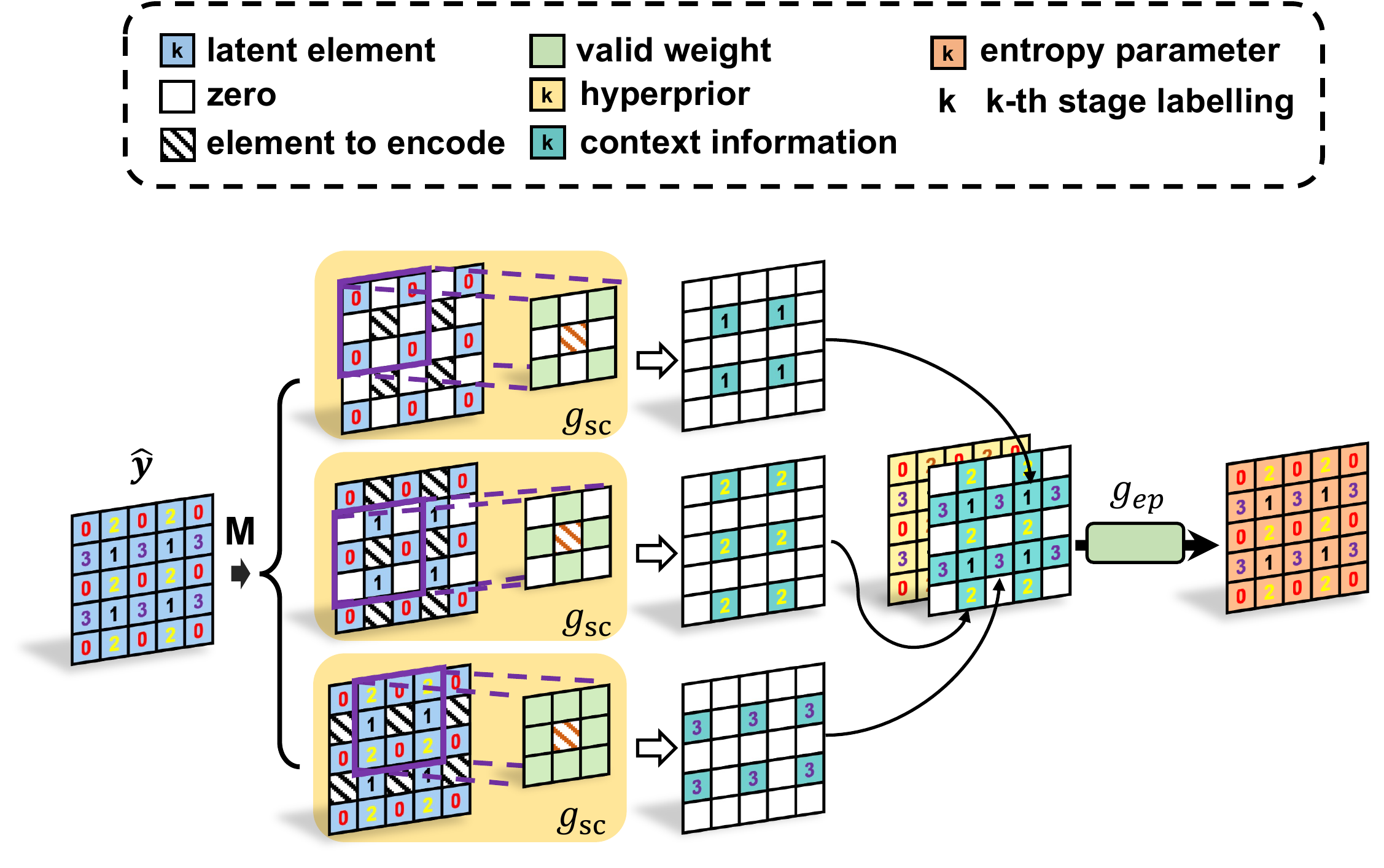}\label{fig:mcc_encoder}} \\
\subfloat[]{\includegraphics[scale=0.35]{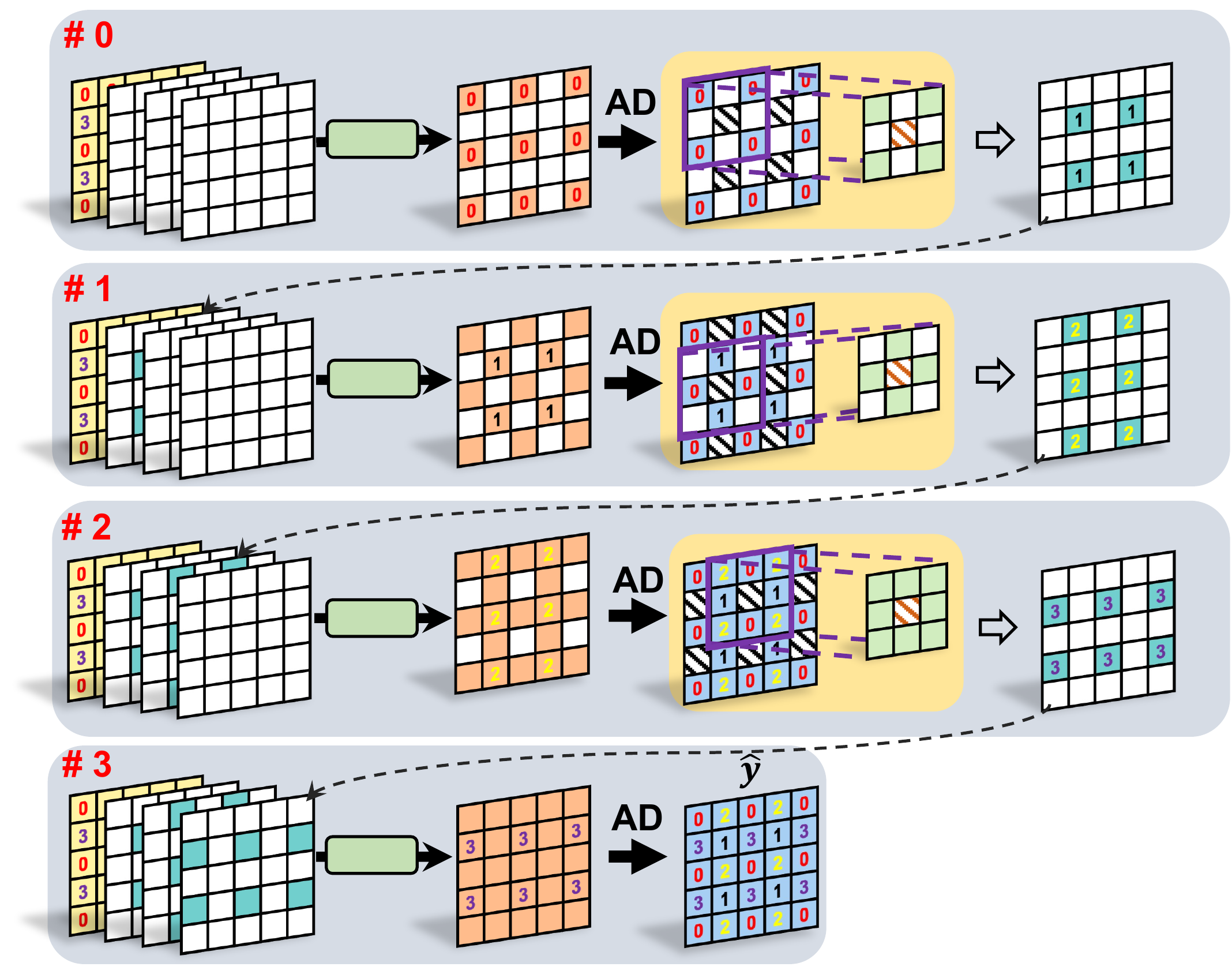}\label{fig:mcc_decoder}}
% where an .eps filename suffix will be assumed under latex, 
% and a .pdf suffix will be assumed for pdflatex; or what has been declared
% via \DeclareGraphicsExtensions.
\caption{{\bf 4-Step GCP.} (a) One-stage Parallel Encoding (b) 4-step Parallel Decoding. \textbf{M} represents binary masks, \textbf{AD} denotes the Arithmetic Decoding to derive latent elements. At the decoder, massive parallelism can be achieved by processing latent features at the same step concurrently. Note that we perform spatial context modeling for  grouped channels  at one time, thus the channel dimension is omitted using 2D illustration for simplicity.}
\label{fig:lar}
\end{figure}

\paragraph{One-shot Parallel Encoding} \label{sec:parallel_encoding}

%To efficiently encode latents $\bf \hat{y}$, the probability of each element in $\bf \hat{y}$ shall be accurately estimated. 
Following the Gaussian distribution used in~\cite{minnen2018joint,chen2021end},  the probability estimation is reformulated as the derivation of entropy parameters for all latent features. 
Apparently, latent features in $\bf \hat{y}$ are all available for encoding.  This section shows how to derive entropy parameters concurrently  as plotted in  Fig.~\ref{fig:mcc_encoder}.

Recalling the spatial tiling using $2 \times 2$ element block in Fig.~\ref{fig:mcc_encoder}, entropy parameters for upper-left latent elements marked with ``0'' (e.g., step\#0 latents) are generated using corresponding co-located hyperpriors only through stacked 1$\times$1 convolutions.  Simultaneously, the entropy parameters of bottom-right step\#1 latents are generated  by applying $3 \times 3$ masked convolutions upon four neighboring step\#0 latents and corresponding hyperpriors. Similarly, the entropy parameters of step\#2 and step\#3 latents are generated using the same stage hyperpriors and  available spatial neighbors in a local 3$\times$3 window, respectively.

% Note that the determination of entropy parameters, a.k.a., {\it epd} module, at different stages shares the same stacked $1\times1$ convolutions to process concatenated hyperpriors and context features, and the context feature generation, i.e., {\it ctx} module, applies 3$\times$3 masked convolutions to aggregate valid neighbors with stage-specific mask accordingly. Detailed settings for entropy parameters and context prediction are given at our project page. As will be unfolded in Sec.~\ref{sec:ablation_entropy_context}, we report that 3$\times$3 convolution is sufficient for cross-stage context prediction which is due to the fact that the proposed content-adaptive transform offers more compact representation of input images with better redundancy removal.  In practice, deploying 3$\times$3 convolutions is more lightweight than existing works using 5$\times$5 masked convolution~\cite{he2021checkerboard,chen2021end,minnen2018joint}.

\paragraph{Step-wise Parallel Decoding}

As depicted in Fig.~\ref{fig:mcc_decoder}, four consecutive steps are involved in decoder to progressively reconstruct the latents $\hat{\bf y}$, which basically mirrors the encoding operations. However, because of the casual dependency in decoding, it can only offer the parallel processing at the same step.

\begin{enumerate}
    \item At the first step, only hyperpriors are used to generate the entropy parameters of step\#0 latents for entropy decoding and reconstruction; and then decoded step\#0 latents are processed with masked 3 $\times$ 3 convolutions to produce step\#1 context features for the second stage;
    \item At the second step, co-located hyperpriors, and step\#1 context features are processed to generate proper entropy parameters to reconstruct step\#1 latents that are subsequently convoluted to derive step\#2 context features;
    \item At the third step, both hyperpriors {at step\#0} and context features at {step\#1 and \#2} are used to derive the entropy parameters to properly decode step\#2 latents; similarly,  step\#2 latents  are then convoluted to derive step\#3 context features for the fourth step;
    \item In the end (at the fourth step), we follow the same way in previous steps to reconstruct step\#3 latents to complete the $\hat{\bf y}$.
\end{enumerate}

\section{Experimental Evaluations} \label{sec:experiment}
This section conducts comparative studies  to understand the compression performance and complexity of the proposed {\it TinyLIC}.

\subsection{Experimental Setup}

{\bf Training.} We choose the Flicker2W~\cite{liu2020unified} as the training dataset in which image samples are randomly cropped into fixed patches at a size of $256 \times 256\times 3$. {More than 20k random patches are generated where 99\% of them, i.e., 20k in total, are used for training, and the rest few hundred of patches are retained for  quick model validation.} Adam is used as the optimizer with default parameters provided in~\cite{kingma2014adam} and the batch size is 8 for each iteration. All training threads run on a single RTX 3090 GPU for 400 epochs in total, having the learning rate at $10^{-4}$ initially, and then at $10^{-5}$ after 300 epochs. 

% \begin{figure*}[htbp]
% \begin{center}
% \begin{minipage}[t]{0.8\textwidth}
% \begin{center}
% \subfloat[]{
% \label{fig:mse_rd}
% \includegraphics[width=0.5\linewidth]{Figures/rd_curve_kodak_mse.pdf}
% }
% % \hspace{0.8mm}
% \subfloat[]{
% \label{fig:msssim_rd}
% \includegraphics[width=0.5\linewidth]{Figures/rd_curve_kodak_msssim.pdf}
% }
% \caption{R-D performance averaged on Kodak dataset: (a) MSE optimized; (b) MS-SSIM optimized.}
% \label{fig:rd_kodak}
% \end{center}
% \end{minipage} \\
% \begin{minipage}[t]{0.4\textwidth}
% \begin{center}
% \includegraphics[width=1\linewidth]{Figures/rd_curve_tecnick.pdf}
% \caption{R-D performance averaged on Tecnick dataset.}
% \label{fig:rd_tecnick}
% \end{center}
% \end{minipage}
% % \hspace{1.2cm}
% \begin{minipage}[t]{0.4\textwidth}
% \begin{center}
% \includegraphics[width=1\linewidth]{Figures/rd_curve_clic.pdf}
% \caption{R-D performance averaged on Tecnick dataset.}
% \label{fig:rd_clic}
% \end{center}
% \end{minipage}
% \end{center}
% \end{figure*}

\begin{figure*}[htbp]
\begin{center}
\subfloat[]{
\label{fig:rd_kodak}
\includegraphics[width=0.33\linewidth]{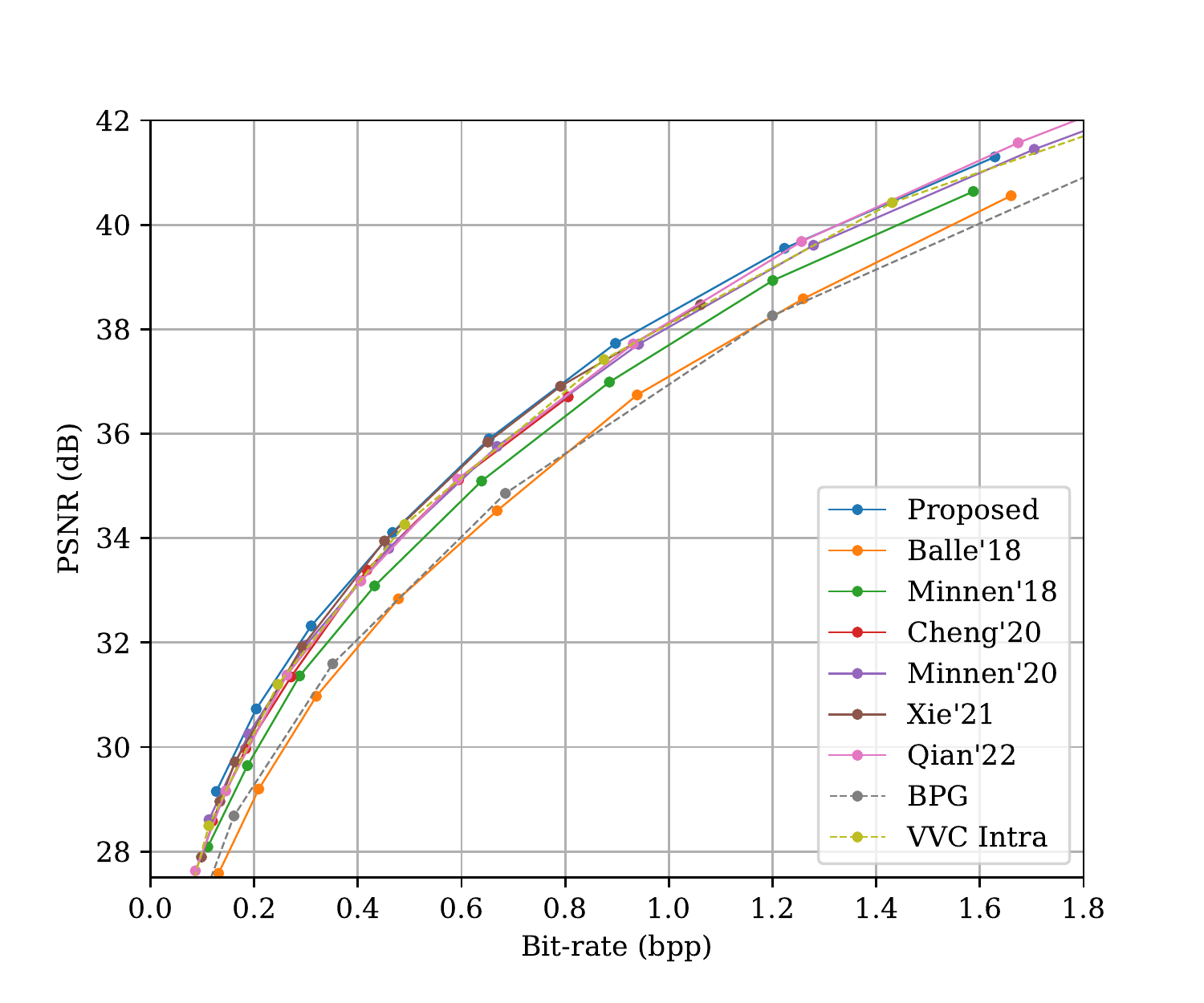}
}
\subfloat[]{
\label{fig:rd_tecnick}
\includegraphics[width=0.33\linewidth]{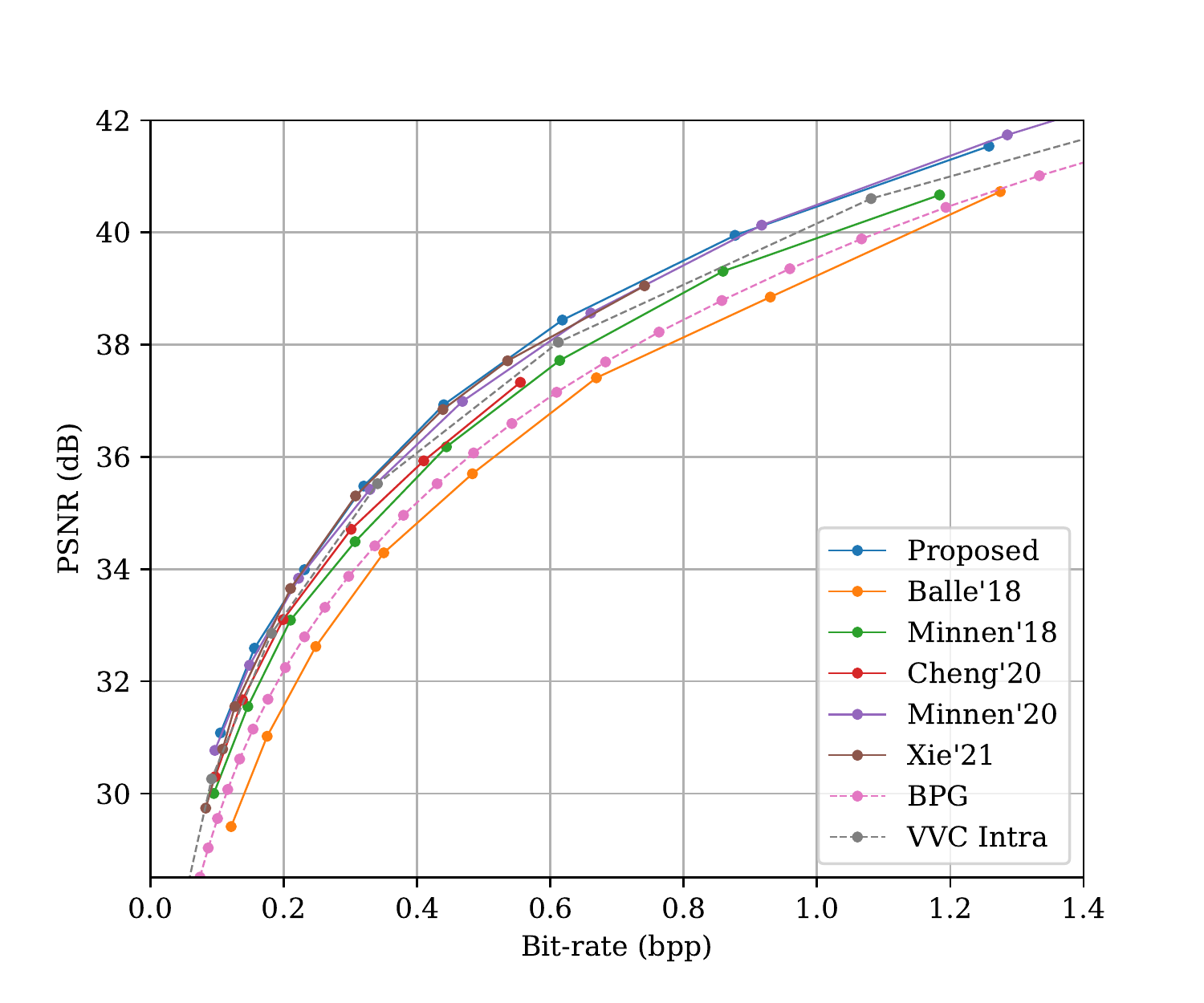}
}
\subfloat[]{
\label{fig:rd_clic}
\includegraphics[width=0.33\linewidth]{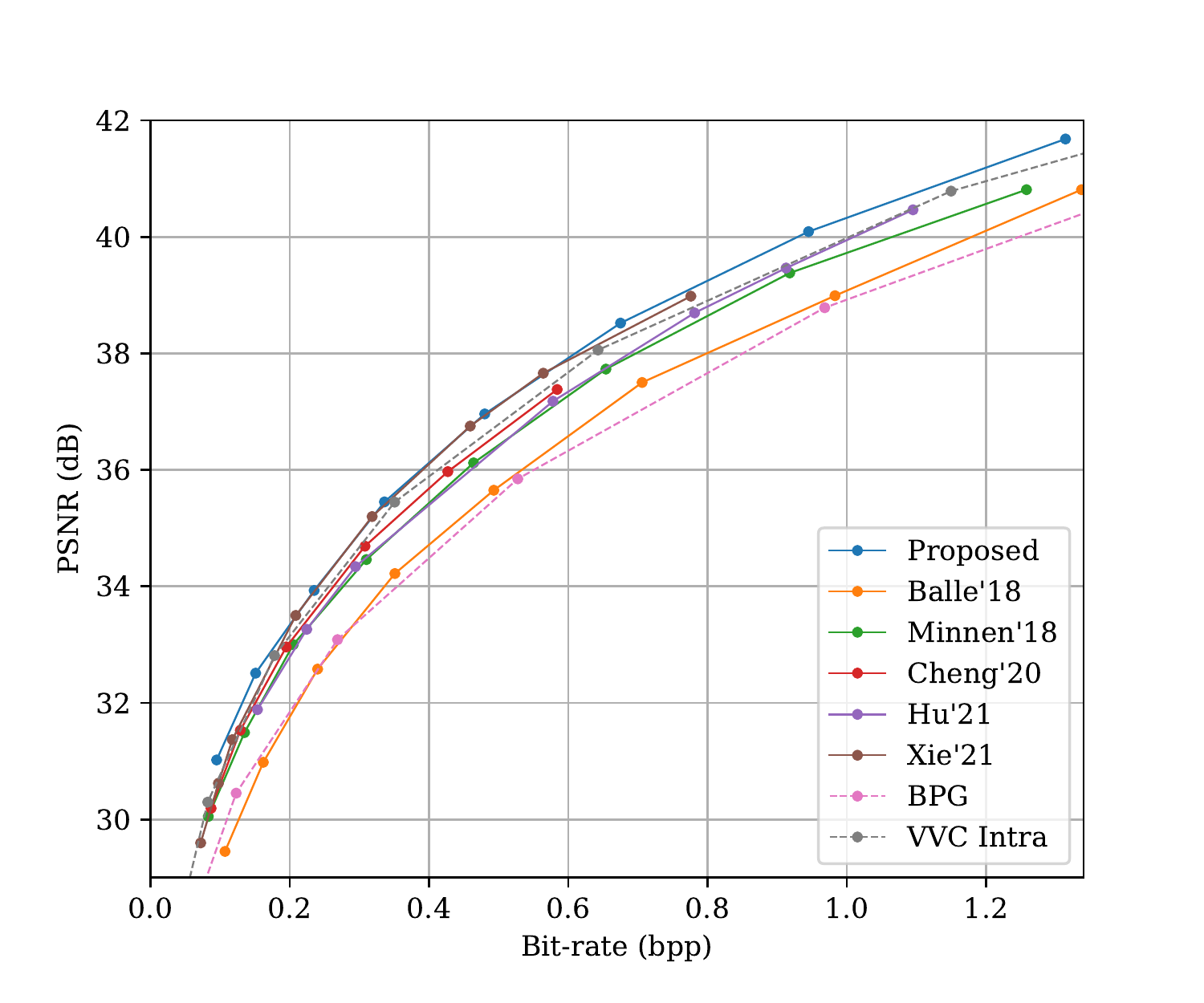}
}
\end{center}
\caption{R-D performance averaged on (a) Kodak (b)Tecnick (c) CLIC dataset with MSE optimized. {\it Please zoom in for more details.}} 
\end{figure*}

% \begin{figure}[!t]
% \centering
% \includegraphics[width=0.86\linewidth]{Figures/rd_curve_tecnick.pdf}
% % where an .eps filename suffix will be assumed under latex, 
% % and a .pdf suffix will be assumed for pdflatex; or what has been declared
% % via \DeclareGraphicsExtensions.
% \caption{R-D performance averaged on Tecnick dataset.}
% \label{fig:rd_tecnick}
% \end{figure}

% \begin{figure}[!t]
% \centering
% \includegraphics[width=0.86\linewidth]{Figures/rd_curve_clic.pdf}
% % where an .eps filename suffix will be assumed under latex, 
% % and a .pdf suffix will be assumed for pdflatex; or what has been declared
% % via \DeclareGraphicsExtensions.
% \caption{R-D performance averaged on CLIC dataset.}
% \label{fig:rd_clic}
% \end{figure}

{\bf Model Settings.} Our {\it TinyLIC}, shown in Fig.~\ref{fig:network}, is implemented on top of the open-source CompressAI PyTorch library~\cite{begaint2020compressai}, by which we can easily share our models and materials for reproducible research. Eight models are trained from the scratch to match different bitrates (or quality levels) by adapting $\lambda$ in \eqref{eq:vae_rd}. {All of them share the same network architecture and channel numbers, which can be found in our supplemental materials.} 
As for MSE loss optimized model, $\lambda$ is chosen from \{0.0018, 0.0035, 0.0067, 0.013, 0.025, 0.0483, 0.0932, 0.18\}\footnote{For MS-SSIM loss optimized model, $\lambda$ is from \{2.40, 4.58, 8.73, 16.64, 31.73, 60.50, 115.37, 220.00\}. Detailed performance evaluations of MS-SSIM optimized model are given in supplemental material.}. 
%while for MS-SSIM loss $\lambda$ is from \{2.40, 4.58, 8.73, 16.64, 31.73, 60.50, 115.37, 220.00\}\footnote{In implementation the actual loss functions are $\lambda \times 255^2 \times D + R$ for MSE oriented optimization and $\lambda \times (1 - D) + R$ for the optimization using MS-SSIM as suggested in \cite{begaint2020compressai}.}. 
{We use mixed quantization estimator for training by encoding the $\lceil \bf y-\mu \rfloor$ to the
bitstream instead of $\lceil \bf y \rfloor$ and restore the symbol using
$\lceil \bf y-\mu \rfloor + \mu$ as in \cite{minnen2020channel}.} % We later show at our supplemental materials, very few models can cover the whole bitrate range without noticeable performance degradation. 

For convolutional feature embedding, we use small-scale $3\times3$ or $5\times5$ convolutional kernels for lightweight computation, and simply enforce the resampling by a stride of 2 at each spatial dimension. The number of RNABs for four ICSA units in main coder are 2, 2, 6, and 2 respectively, i.e., $d_1=2$, $d_2=2$, $d_3=6$, $d_4=2$, having different $d_i$ follows the suggestions presented in~\cite{liu2021swin}; And they are 2 and 2 in hyper coder, i.e., $d_5=2$ and $d_6=2$. 
The numbers of heads used in the self-attention layer for RNABs at four stages of the main coder are 8, 12, 16, 20, and 12; while they are fixed at 12 in hyper. The window size is 7$\times$7 for RNABs in the main coder, while it is  3$\times$3 in the hyper coder; And the hidden channels are expanded by a factor of 2 for all MLP layers used in our work.

{\bf Testing.} We use three popular datasets that contain diverse images for evaluation, i.e., the Kodak dataset\footnote{\url{https://r0k.us/graphics/kodak/}} with image resolution at 768$\times$512, Tecnick dataset\footnote{\url{https://tecnick.com/?aiocp_dp=testimages}} with image size of $1200\times1200$ and CLIC professional validation dataset\footnote{\url{http://compression.cc/tasks/\#image}} which contains 41 images at {2k spatial resolution approximately}.  These datasets are widely used for image coding competitions. Both peak signal-to-noise ratio (PSNR) and MS-SSIM are used to quantify the decoded image quality, and the bpp (bits per pixel) measures the compressed bitrate.

\subsection{Evaluation}

{\bf Anchor \& Alternatives.} We set HEVC Intra compliant BPG method as the anchor, a.k.a., BPG, to derive BD-rate gains. We also offer the results of VVC Intra using its latest reference software~\cite{vvc_intra}. Meanwhile a broad collection of prominent LIC solutions are included for comparison with their best-produced results, including the Ball\'e'18~\cite{balle2018variational}, Minnen'18~\cite{minnen2018joint}, Cheng'20~\cite{cheng2020learned}, Minnen'20~\cite{minnen2020channel}, Ma'20~\cite{9204799}, Hu'21~\cite{9376651}, and Xie'21~\cite{xie2021enhanced}. These methods are representative examples that plot the technical development history for the past years as discussed in Sec.~\ref{sec:related_work}. Among them, Minnen'18 is the seminal framework that forms the foundation modules (e.g., nonlinear transform, context modeling using hyperpriors \& spatial-channel neighbors) for future improvements.

\subsubsection{Quantitative Performance} 
Rate-distortion (R-D) curves are plotted in Fig.~\ref{fig:rd_kodak},~\ref{fig:rd_tecnick},~\ref{fig:rd_clic} while the BD-rate gains against the BPG anchor are given in Table~\ref{tbl:bd-rate} and Fig.~\ref{fig:complexity}. For a fair comparison, we try our best to ensure a similar bitrate range across different approaches for BD-rate computation~\cite{BDData}.

\begin{table}[t]
\caption{Averaged BD-rate (\%) improvement against the BPG anchor for different datasets. \textcolor{red}{\bf Red} and \textcolor{blue}{\underline{blue}} indicates the best and the second best performance, respectively.}
\label{tbl:bd-rate}
\centering
\begin{tabular}{cccc}
\toprule
Method & Kodak & Tecnick & CLIC \\ 
\midrule
Ball\'e'18~\cite{balle2018variational} & 3.93 & 9.42 & 0.67 \\
Minnen'18~\cite{minnen2018joint} & -11.3 & -11.66 & -18.22 \\
Cheng'20~\cite{cheng2020learned} & -17.71 & -17.69 & -23.53 \\
Minnen'20~\cite{minnen2020channel} & -16.81 & -22.48 & - \\
Ma'20~\cite{9204799} & -16.3 & -22.6 & -20.3 \\
Hu'21~\cite{9376651} & -9.42 & -14.15 & -18.89 \\  
Xie'21~\cite{xie2021enhanced} & \textcolor{blue}{\underline{-21.55}} & \textcolor{blue}{\underline{-23.85}} & \textcolor{blue}{\underline{-28.13}} \\
VVC & -20.53 & -20.38 & -25.76 \\
{\bf Ours} - {\it TinyLIC} & \textcolor{red}{\bf -21.77} & \textcolor{red}{\bf -24.30} & \textcolor{red}{\bf -28.60} \\
% {\underline{0.216}} & \textcolor{blue}{\underline{0.167}}
\bottomrule
\end{tabular}
\end{table}

\begin{figure*}[t]
\begin{center}
\includegraphics[width=0.9\linewidth]{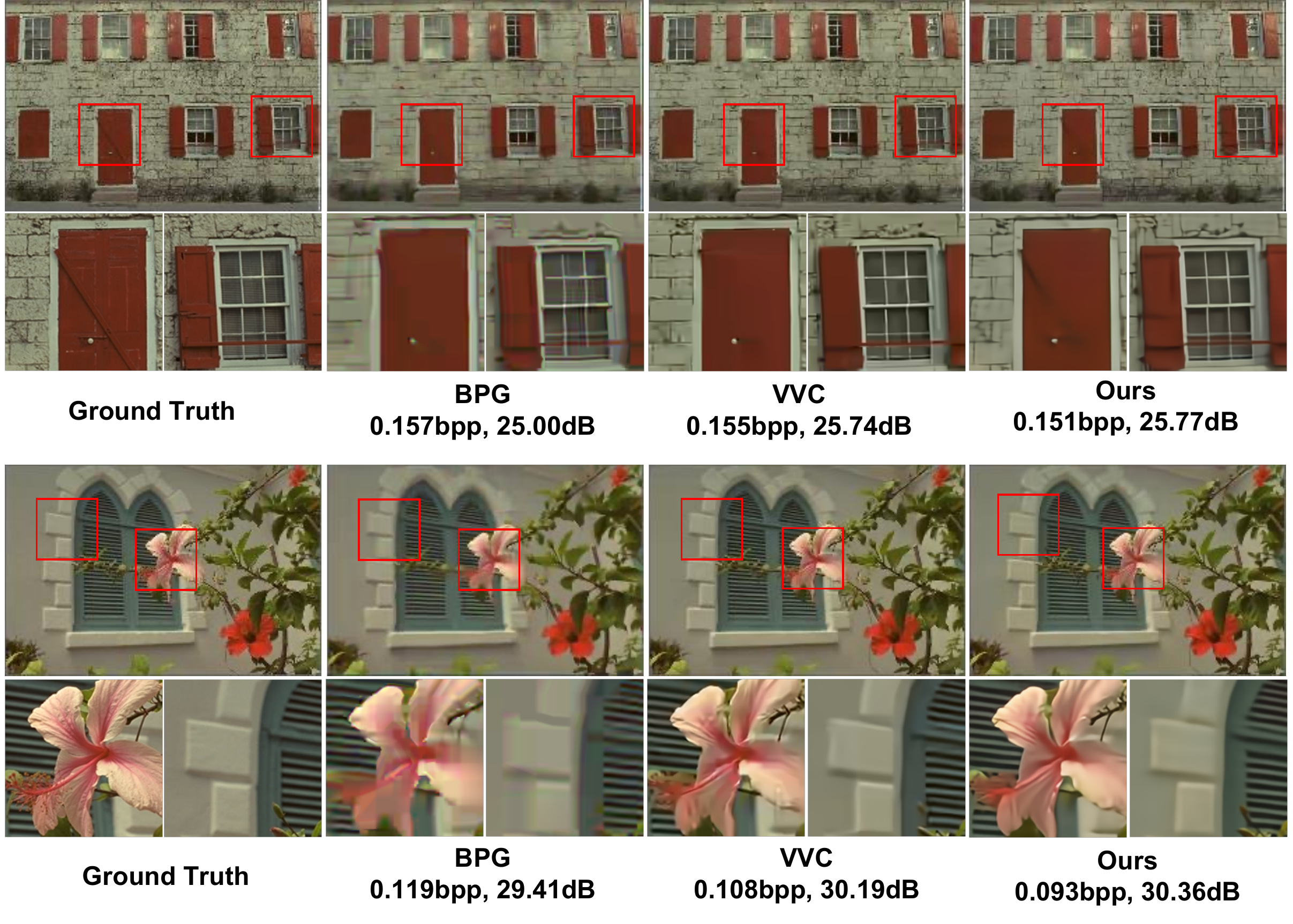}
\end{center}
\caption{{\bf Qualitative Visualization.}
Reconstructions and close-ups of the BPG, VVC and our method. Corresponding bpp and PSNR are marked.} 
\label{fig:visual_compare}
\end{figure*}

{\bf Overall BD-rate Gain.} As for the results tested on Kodak dataset, the proposed method outperforms all other solutions for the distortion  measured by PSNR shown in Fig.~\ref{fig:rd_kodak}. In Table~\ref{tbl:bd-rate}, our method provides {21.77\%} BD-rate improvement against the anchor BPG, while the VVC just offers a 20.53\% gain. Coding gains are further enlarged for Tecnick and CLIC datasets as illustrated in Fig.~\ref{fig:rd_tecnick} and~\ref{fig:rd_clic}. For instance, $\approx$4\% relative gain is captured for CLIC images that are widely used for image compression competitions, e.g., 24.30\% BD-rate improvement to BPG for ``Ours'' versus 20.38\% of it for ``VVC'' in Table~\ref{tbl:bd-rate}. 
In the meantime, our method also shows consistent performance lead in comparison to other notable LICs.

{\bf Discussion.} Almost all solutions report the increase of coding efficiency for test images with larger resolution in Table~\ref{tbl:bd-rate}. For instance, the average BD-rate gain for CLIC dataset is larger than it is for the Kodak dataset. This is because: an image sample with a larger spatial resolution would exhibit higher local coherency, for which it is easier  to exploit neighborhood correlation for better compression efficiency. 

It also evidences that the VAE architecture with nonlinear transform and entropy context conditioned on joint hyperpriors and spatial-channel neighbors which was first proposed in Minnen'18 is a well-generalized solution regardless of the different techniques used in its modular components like ReLU or GDN-based activation, simple convolution or ICSA, etc~\cite{cheng2020learned,chen2021end,minnen2018joint,lu2021transformer}. When comparing to coding efficiency offered by the Minnen'18, almost 10 absolute percentage points improvement is reported against the same BPG anchor.

% A slight performance inconsistency  is spotted for Xie \etal ~2021~\cite{xie2021enhanced} among testing datasets, e.g., ranked as the second best for Kodak dataset, but only fourth place for CLIC. This is largely because  a different structure using invertible neural network and feature enhancement network is applied in Xie \etal~2021~\cite{xie2021enhanced}.

The attention mechanism further reveals its outstanding effectiveness to adaptively weigh and aggregate highly correlated information from the results of Cheng'20~\cite{cheng2020learned} and our {\it TinyLIC}. Compared with Cheng'20~\cite{cheng2020learned}, our method not only extends the attention embedding to all stages (but not just the bottleneck layer) in main and hyper coders but also replaces the convolution-based attention computation with the self-attention to flexibly characterize  any dynamic input.

The VVC Intra is expected to succeed its predecessor HEVC Intra (BPG) because of its outstanding performance~\cite{vvc_intra}. We then switch the anchor from BPG to VVC Intra to derive coding gains of the proposed {\it TinyLIC}, for which $\approx$3\% BD-rate gains are captured on average across all three datasets.

\subsubsection{Qualitative Visualization}
Figure~\ref{fig:visual_compare} visualizes the reconstructions and closeups generated by the BPG, VVC Intra, and our method. Ground-truth labels are also provided for side-by-side illustration. We particularly use the VVC Intra compressed image for qualitative comparison because  a set of normative in-loop filters are enabled in the VVC intra profile which promises outstanding subjective quality of decoded image~\cite{VVC_ALF}. As seen, the proposed method noticeably improves the subjective quality with more sharp textures and less noise. Reconstruction snapshots of other test images from different datasets will be shown in our supplemental material.

\subsubsection{Complexity}
% Two metrics are used for complexity measures, including the MACs/pixel and model parameters.

{We report the size of model parameters and MACs/pixel of each  LIC solution in Fig.~\ref{fig:complexity}. As seen, the proposed method provides the most competitive performance-complexity tradeoff. In comparison to the Minnen'18~\cite{minnen2018joint}, our method uses a comparable-size model (e.g., 28.34M vs. 25.5M) and slightly more  MACs/pixel for $>10$ absolute percentage points improvement against the same BPG anchor. 
%but improves the coding performance significantly 
Compared with the Cheng'20~\cite{cheng2020learned} - another representative work succeeding the Minnen'18~\cite{minnen2018joint}, the proposed {\it TinyLIC} uses a smaller model, e.g., $\approx$4.6\% model size reduction from 29.63M to 28.34M, much less MACs/pixel with $>$50\% reduction from {1077.12 to 516.8} for more than 5\% relative BD-rate improvement to the same BPG anchor. Although Xie'21~\cite{xie2021enhanced} offers the closest BD-rate gain (e.g., -21.55\% versus -21.77\% in Table~\ref{tbl:bd-rate}), its model size is enlarged about 2$\times$, and its MACs/pixel is also close to 2$\times$ of that of {\it TinyLIC}.}

Besides the  MACs/pixel, decoding latency is another important metric for computational complexity which is highly related to the parallelism of entropy engine. 
%In practice, we generally desire for the fast
Because of the implementation differences (especially for the arithmetic coding engine), it is hard to directly compare the decoding runtime in a fair manner. We then choose to implement typical methods  on the same platform (e.g., CompressAI) for detailed comparison (see Sec. \ref{sec:latency}). %In general, sequential autoregressive decoding is applied in Minnen'18, Cheng'20 and Xie'21. Channel-wise speedup is used in Minnen'20 and

%\lm{Due to the different implementations of the methods mentioned above (especially the implementation of the arithmetic coding engine), it is difficult to directly compare the encoding/decoding times in a fair manner. We solve this problem by using the same implementation platform, with only the context model changed, as will be discussed in \ref{sec:latency}.}

% \begin{table}[t]
% \caption{Averaged encoding / decoding latency for Kodak dataset. 2CC is 2-stage Checkerboard Context modeling. Latency is measured by ms (millisecond) collected on Python platform.}
% \label{table:latency}
% \centering
% \begin{tabular}{c|c|c|c}
% \hline
% \multirow{2}{*}{Method} & \multirow{2}{*}{BD-rate (\%) $\downarrow$} & \multicolumn{2}{c}{Latency (s) $\downarrow$} \\
% \cline{3-4}
% & & Encoder & Decoder \\
% \hline
% %Ball\'e et al. 2018~\cite{balle2016end} & 3.70 & 40 & 29 \\
% ICLR'18~\cite{balle2016end}[S] & 3.93 & & \\
% NIPS'18~\cite{minnen2018joint}[S] & -11.30 & 2.640 & 5.566 \\
% CVPR'20~\cite{cheng2020learned}[S] & -17.71 & 2.626 & 5.564 \\
% ICIP'20~\cite{minnen2020channel}[CW] & -16.81 & 0.524 & 0.665 \\
% ICLR'21~\cite{qian2020learning}[SG] & -14.49 & 341 & 352 \\
% ACMMM'21~\cite{xie2021enhanced}[S] & -21.55 & 2.780 & 6.343 \\
% \textbf{Ours} & \textbf{-22.23} & \textbf{} & \textbf{} \\
% \hline
% \end{tabular}
% \end{table}

\begin{figure}[t]
\begin{center}
\subfloat[]{
% \label{fig:win_size_low}
\includegraphics[width=0.45\linewidth]{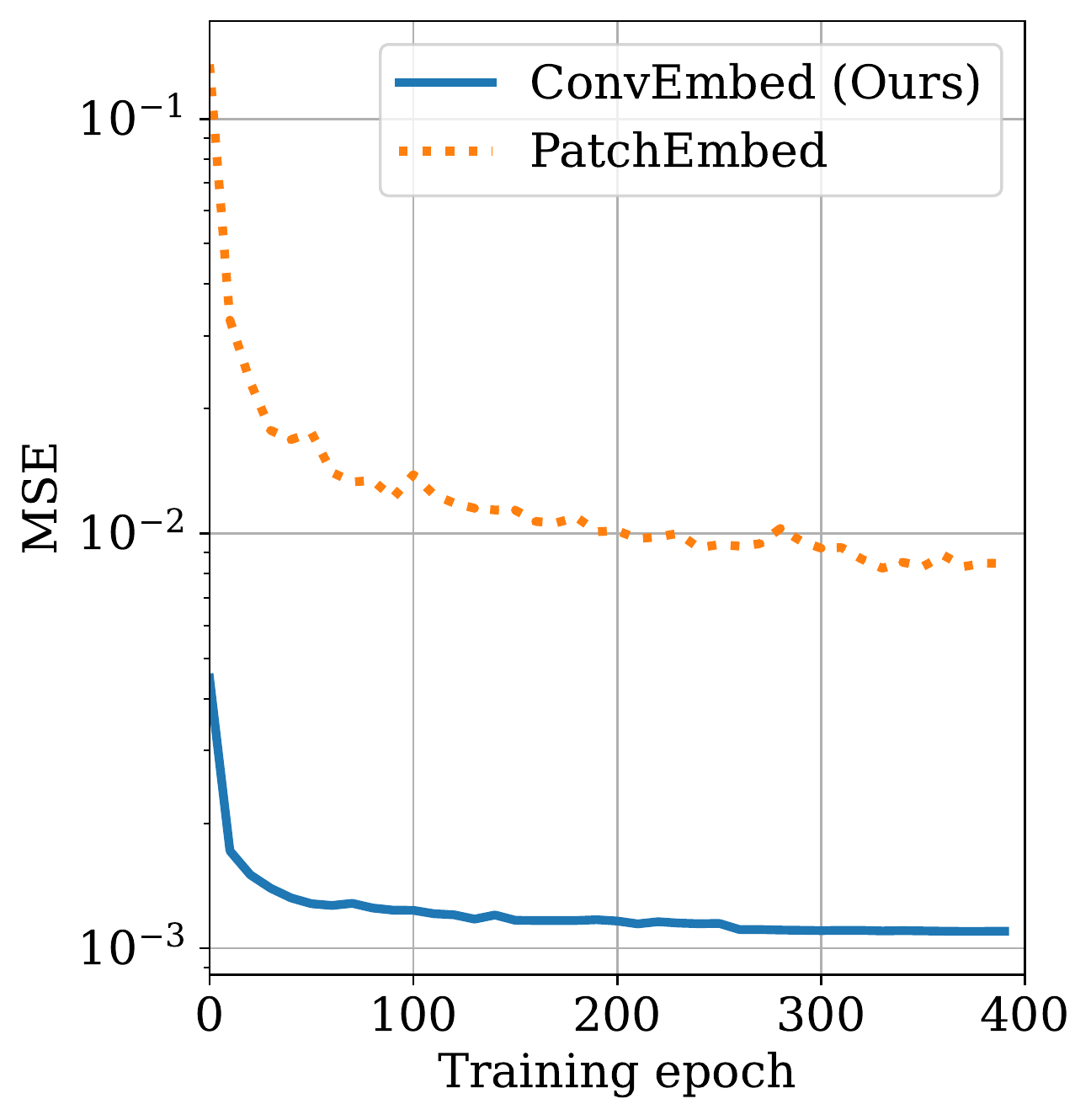}
}
\subfloat[]{
% \label{fig:win_size_high}
\includegraphics[width=0.45\linewidth]{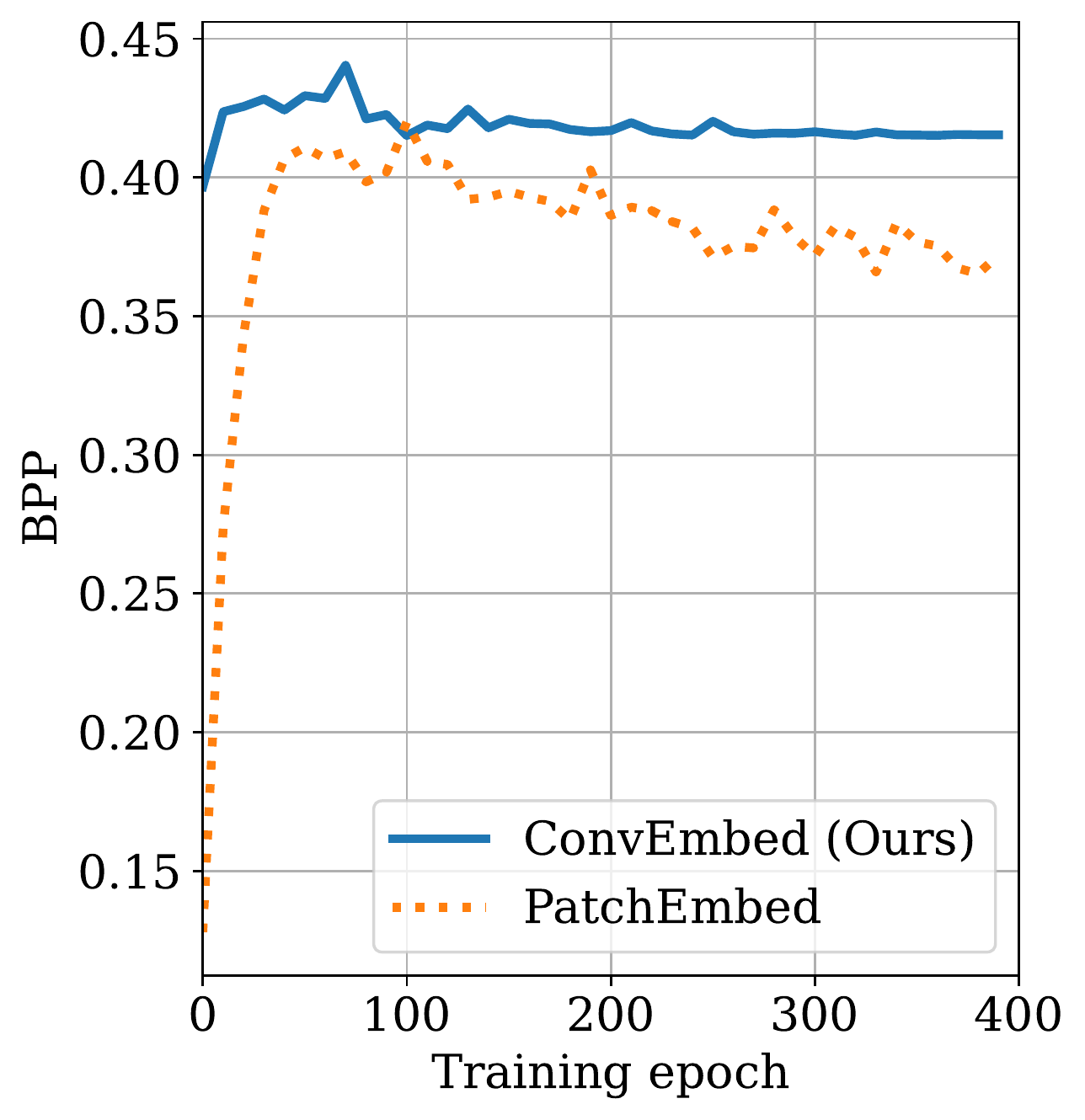}
}
\end{center}
\caption{{\bf Token Embedding.} (a) $\log(\rm MSE)$ versus epoch (b) BPP versus epoch.}
\label{fig:token_embedding} 
\end{figure}

\section{Ablation Studies}
In this section, we dissect the {\it TinyLIC} to offer more insightful discussions on its modular components like the transform block backbone and entropy context models.

\subsection{Transform Block Backbone} \label{sec:backbone}
The {\it TinyLIC} stacks ICSA units to form the transform block backbone in default.

\subsubsection{Feature Embedding}

Our ICSA applies a convolutional layer to do token embedding, noted as ``ConvEmbed''. Apparently, there are many other solutions for the same purpose. One prevailing method can directly extract  non-overlapping patches from input feature tensor for token embedding. We then follow the implementation in~\cite{zhu2021transformer} to perform patch-based token embedding, a.k.a., ``PatchEmbed'', to replace default ``ConvEmbed'' in {\it TinyLIC}. As shown in Fig.~\ref{fig:token_embedding}, we can clearly notice that default ``ConvEmbed'' provides much faster convergence rate than the ``PatchEmbed'' in model training. This coincides with the claim in~\cite{CvT}, e.g., convolutional token embedding can improve {the convergence stability of model}. Other advantages of the use of ``ConvEmbed'' like implicit position embedding, flexible token size support, etc, can be referred in~\cite{CvT} as well.

\begin{figure}[t]
\begin{center}
\subfloat[]{
\label{fig:win_size_low}
\includegraphics[width=0.45\linewidth]{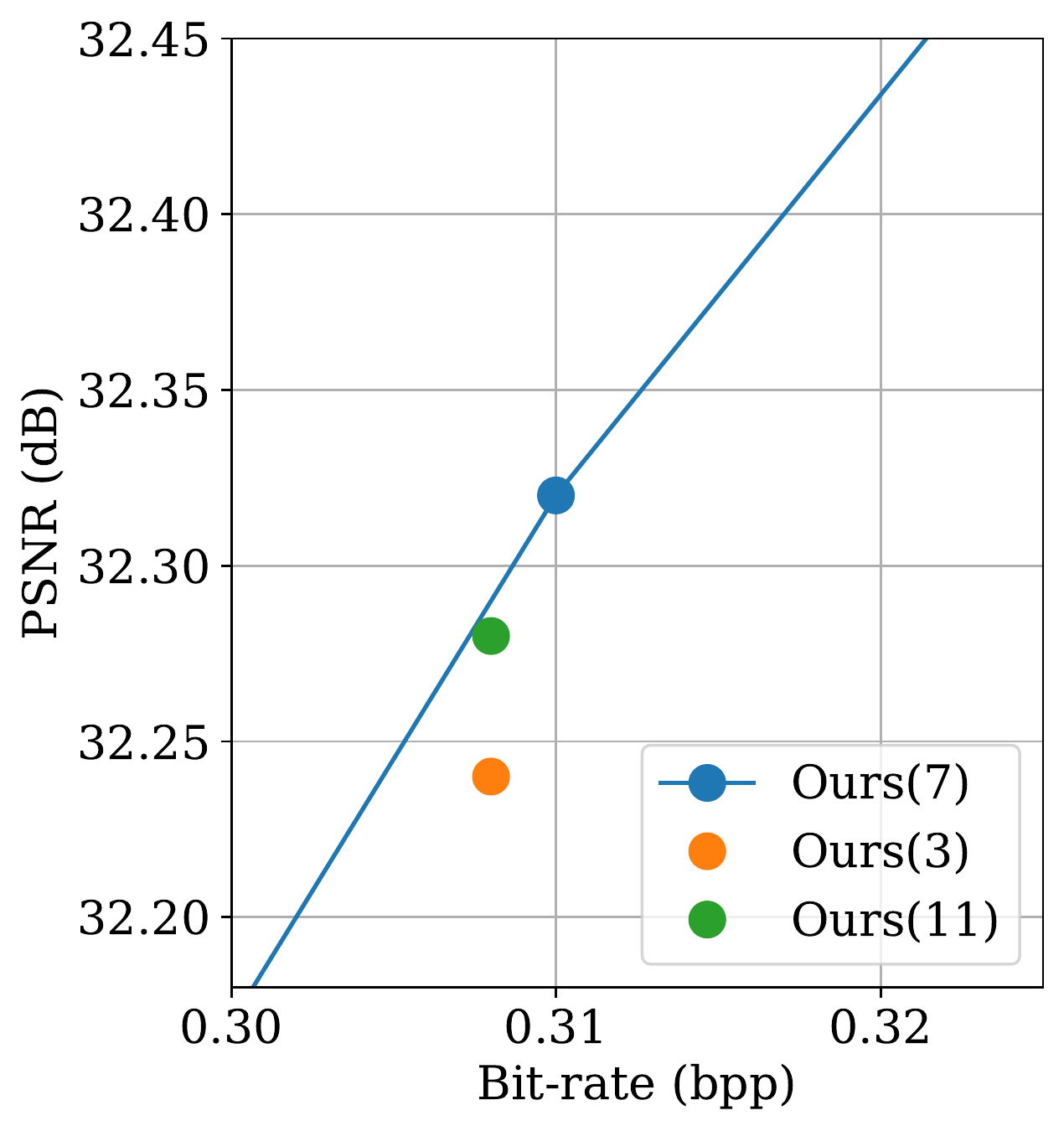}
}
\subfloat[]{
\label{fig:win_size_high}
\includegraphics[width=0.45\linewidth]{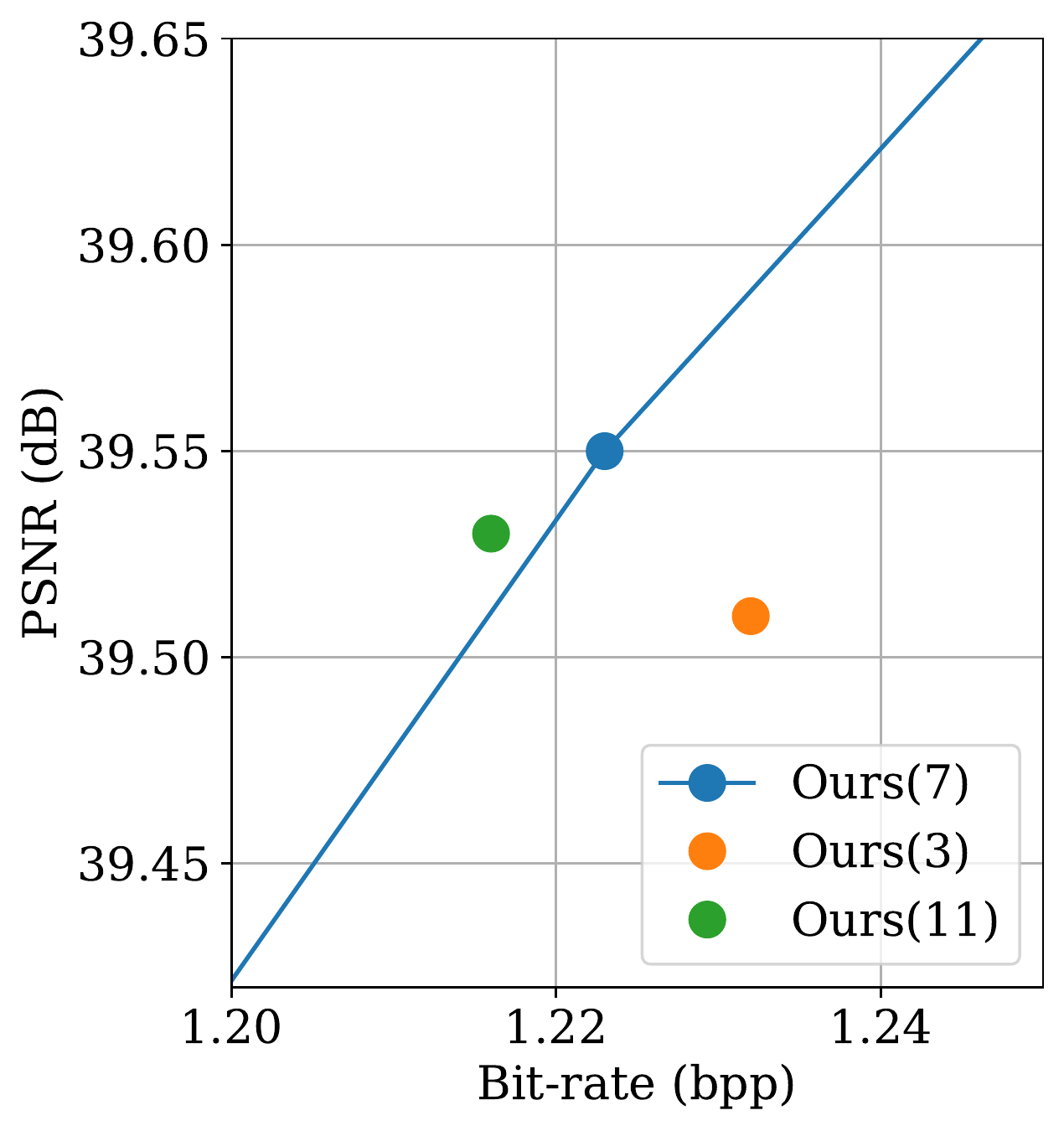}
}
\end{center}
\caption{{\bf Attentive Window Size.} (a) Low bitrate with $\lambda = 0.0067$, (b) High bitrate with $\lambda = 0.0932.$ Solid line is R-D curve of default {\it TinyLIC}.} 
\label{fig:win_size_compare} 
\end{figure}

\begin{figure}[t]
\begin{center}
\subfloat[SwinT]{
\label{fig:swint}
\includegraphics[width=0.4\linewidth]{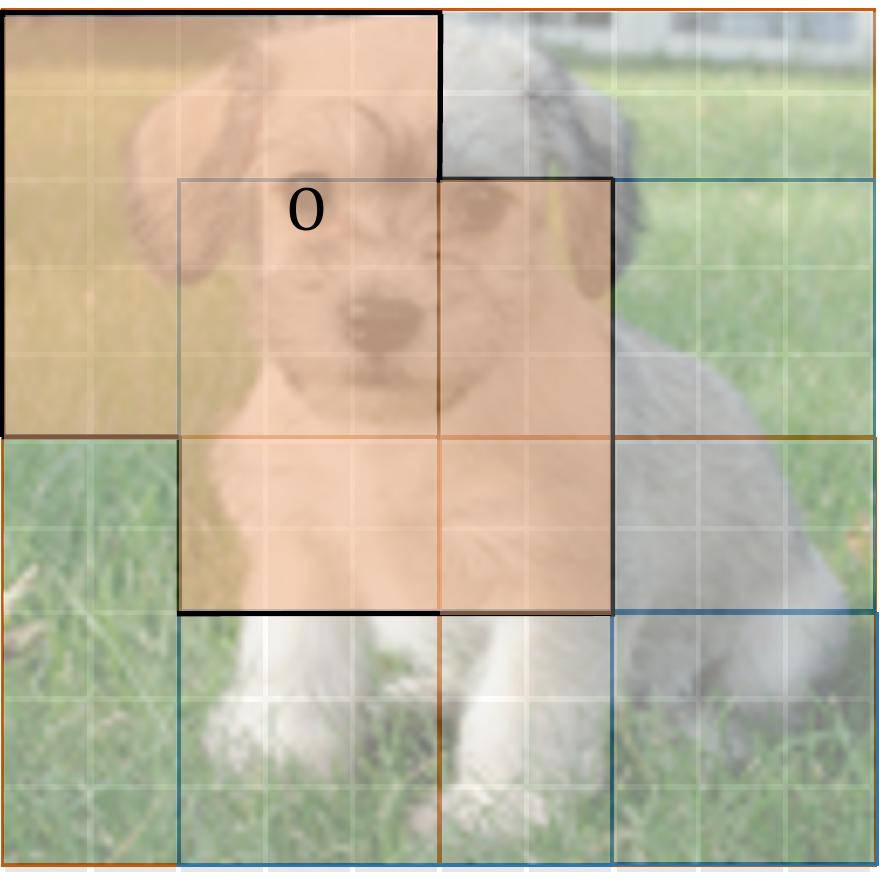}
}
\hspace{0.8cm}
\subfloat[NAT]{
\label{fig:nat}
\includegraphics[width=0.4\linewidth]{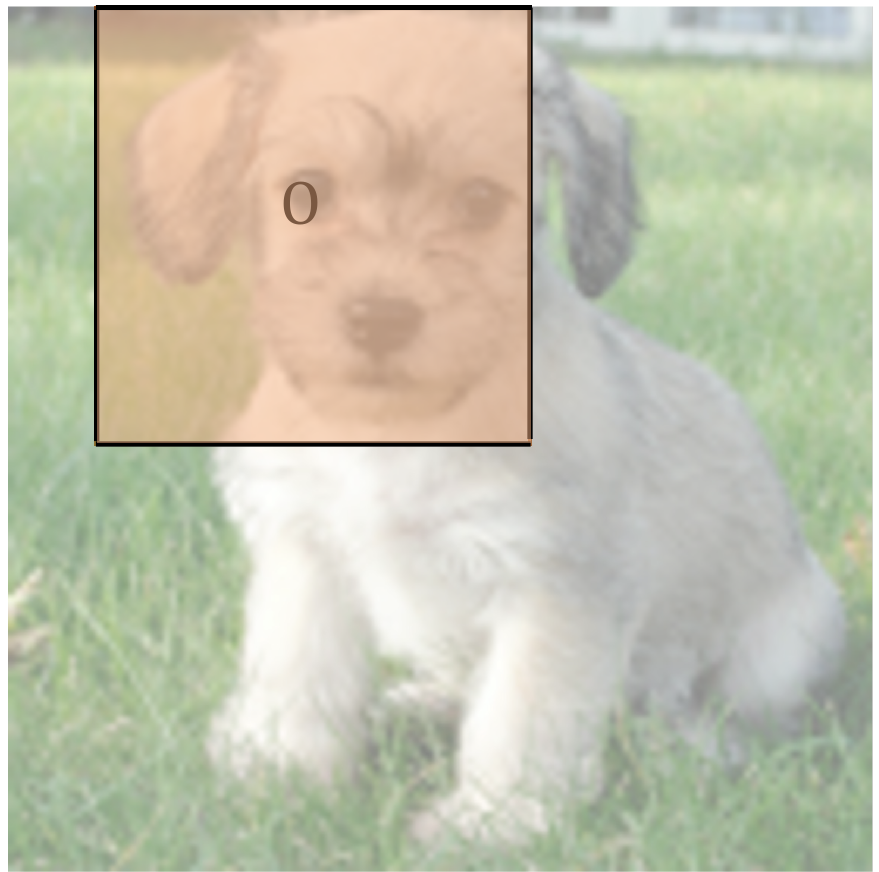}
}
\end{center}
\caption{{\bf Successive Windows for Self-Attention Computation.} (a) SwinT (b) NAT. Although the receptive field of SwinT is enlarged to that of NAT due to the shifted window mechanism, less correlated elements are included for information aggregation.} 
\label{fig:swin_vs_na} 
\end{figure}

\subsubsection{SwinT vs. NAT}

{Previous works~\cite{zhu2021transformer,lu2021transformer} have illustrated the efficiency of Swin Transformer (SwinT)~\cite{liu2021swin} for content-adaptive aggregation. Thus, we replace the NAT with the SwinT in RNABs for evaluation. All other settings of training and testing are the same for a fair comparison.  Upon the Kodak dataset, the {\it TinyLIC} with the SwinT also provides competitive compression performance.}

{On the other hand, a slight 0.1 dB PSNR loss at the same bitrate is reported compared to the coder using default NAT.  As seen, although the introduction of the shifted-window mechanism can essentially enlarge the receptive field for each pixel in the SwinT, experimental results suggest that a smaller local window used in the NAT sufficiently aggregates 
 local neighbors for compact representation. Having a larger window size may also include uncorrelated pixels which instead hurts the efficiency of the self-attention mechanism with coding performance loss. }

%. Howbors evserfor , such rule-based block partition strategy may introduce a lot of irrelevant locations into attention calculation, which will constraint the performance. In addition, the principle of expanding the receptive field should not be at the expense of the closer pixels. It may lead to a certain tendency for information aggregation with a number of layers stacked as shown in Fig.~\ref{fig:swint}.}

%\lm{Stacking RNABs not only ensures enough receptive field for information aggregation through local neighborhood but also offers the content-adaptive weighting and embedding of neighboring windows by self-attention computation as in Fig.~\ref{fig:nat}. It is reported an averaged 0.1dB improvement on the Kodak dataset can be obtained when campared with the SwinT.}

\subsubsection{Self-Attention Window Size}

To understand the impact of window size in the window-based self-attention layer of RNAB, we further examine 3$\times$3 and 11$\times$11 window settings in addition to the default 7$\times$7. Since the square window is used, we label them using ``Ours(3)'', ``Ours(7)'' and ``Ours(11)'' respectively as in Fig.~\ref{fig:win_size_compare}. Two different $\lambda$s at 0.0067 and 0.0932 are experimented for typical low and high bitrate scenarios.

When using the 3$\times$3 window, the performance drops slightly ($\approx0.05$ dB) while the performance keeps unchanged for the 11$\times$11 attentive window, e.g., overlapped with the default R-D curve. 
When compared with the default 7$\times$7 window, proportionally more patches are used for 3$\times$3 window, leading to the sharp increase of model parameters and potential throughput bottleneck; although fewer patches are used for 11$\times$11 attentive window, per window computation is increased and it also imposes more strict resolution limitations of input content\footnote{The input image size should be a multiple of $2^6 \times w^2$ since resampling by a stride of 2 at each dimension is enforced 6 times as pictured in Fig.~\ref{fig:network}. Here $w$ is the side length of a square.}. All of these suggest a 7$\times$7 window is a justified option for a balanced tradeoff.

\subsection{Entropy Context Model} \label{sec:ablation_entropy_context}

This section comprehensively examines entropy context models for insightful comparison. 

\subsubsection{Various Context Modeling Methods}

The use of joint hyperpriors and spatial-channel neighbors was first proposed in~\cite{minnen2018joint} where Minnen et al. applied the autoregressive manner to exploit statistical correlations. As aforementioned, data dependency in such an autoregressive model forces sequential processing, leading to unbearable decoding latency. Since then, serial refinements have been developed to alleviate the autoregressive dependence for high-throughput processing, such as the channel-wise grouping~\cite{minnen2020channel}, checkerboard patterning~\cite{he2021checkerboard}, the combination of channel-wise grouping and checkerboard patterning~\cite{he2022elic}, and the proposed MCM, etc.

\begin{figure}[t]
\begin{center}
\subfloat[]{
\includegraphics[width=0.45\linewidth]{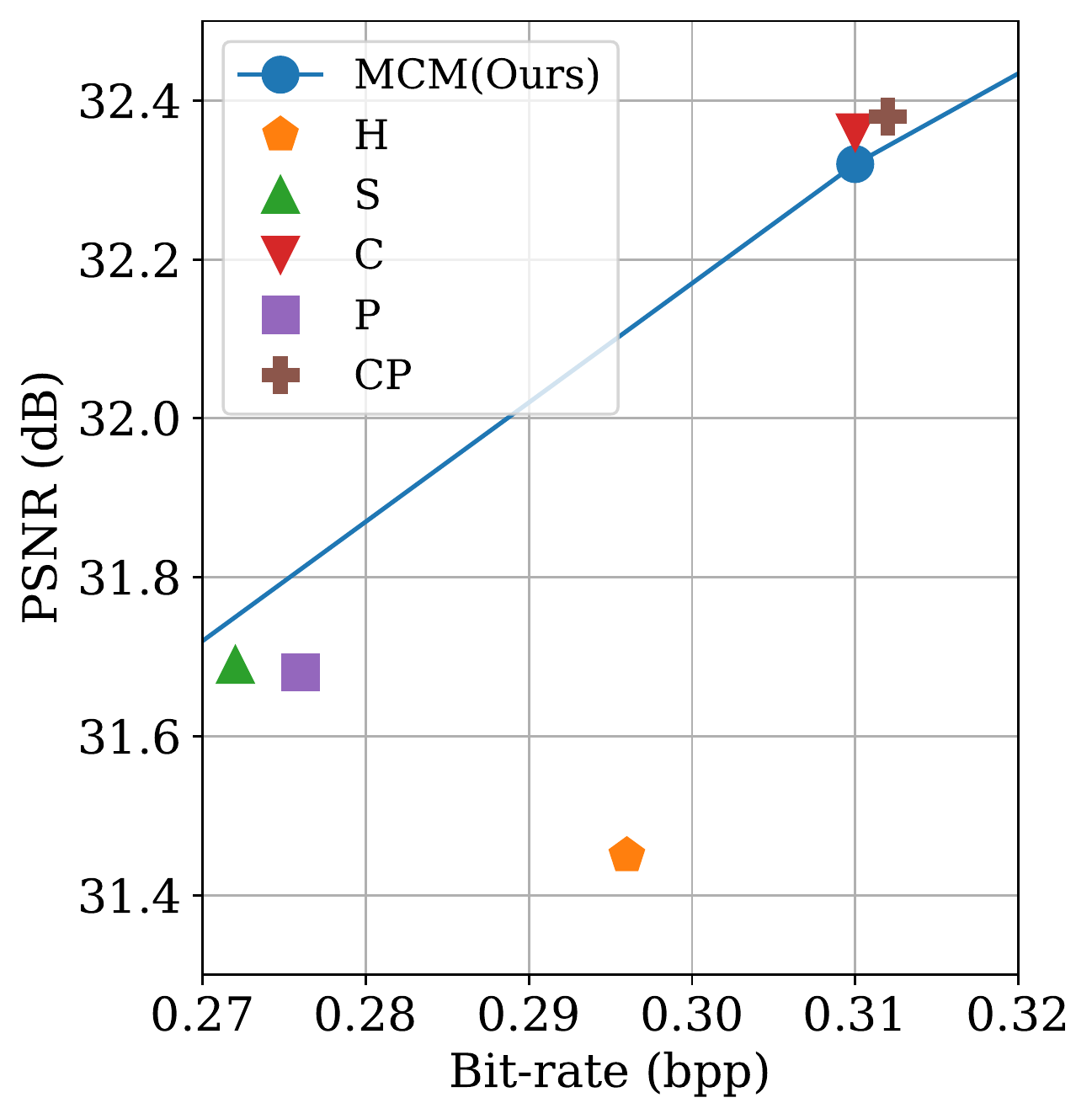}
}
\subfloat[]{
\includegraphics[width=0.45\linewidth]{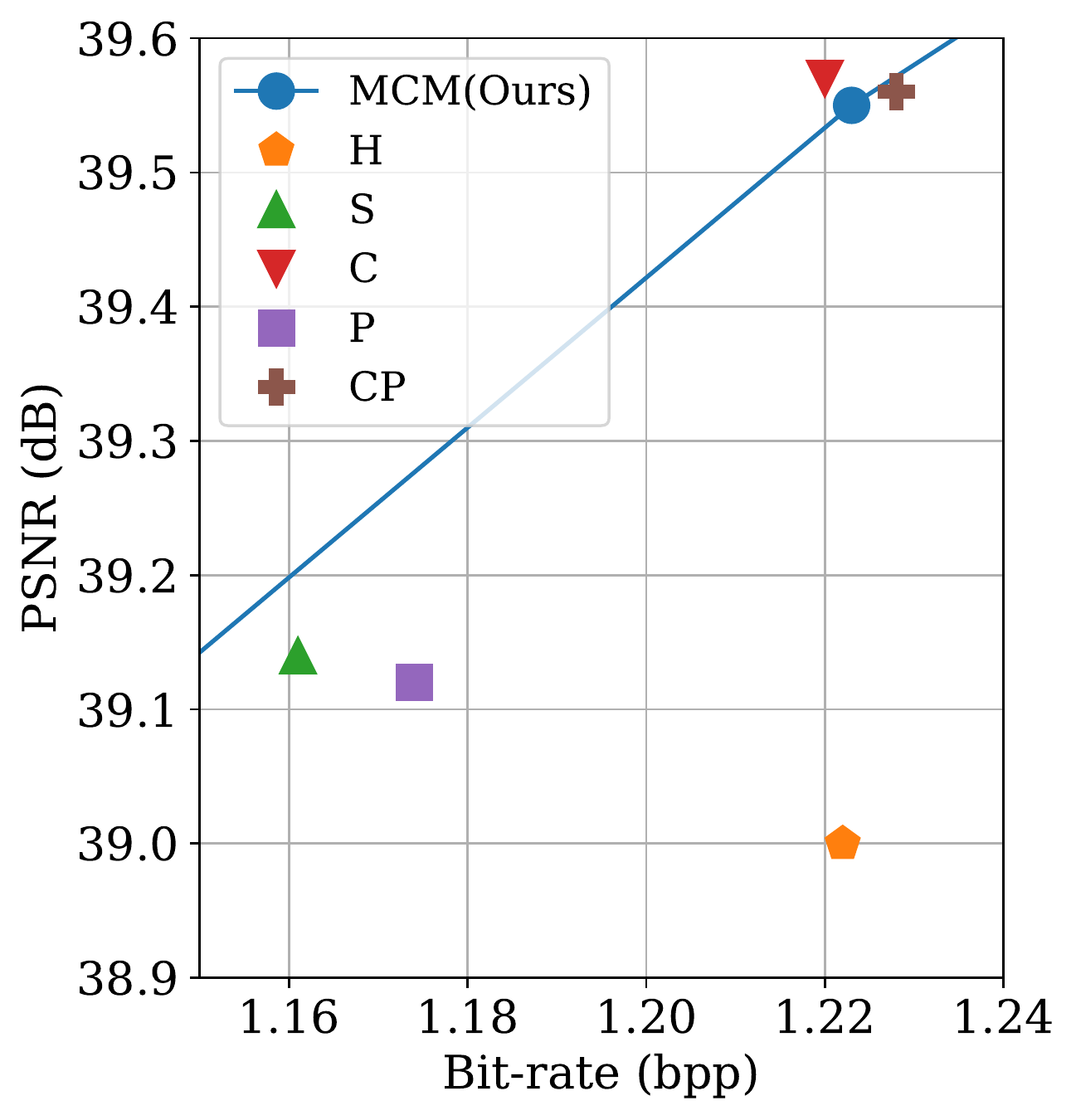}
}
\end{center}
\caption{{\bf Entropy Context Models.} (a) Low bitrate with $\lambda = 0.0067$, (b) High bitrate with $\lambda = 0.0932.$ Solid line is R-D curve of default {\it TinyLIC}.} 
\label{fig:entropy_model} 
\end{figure}

\begin{table}[htbp]
\caption{Computational Complexity and Decoding Latency}
\label{tbl:latency}
\centering
\begin{tabular}{cccc}
\toprule
Entropy Model & $\Delta$ Parameters (MB) & $\Delta$ MACs/pixel & Latency (s) \\ 
\midrule
H~\cite{balle2018variational} & -9.53 & -43.84 & 0.044 \\
S~\cite{minnen2018joint} & -1.03 & -29.92 & 11.079 \\
C~\cite{minnen2020channel} & +88.01 & +355.84 & 0.657 \\
P~\cite{he2021checkerboard} & -1.04 & -29.92 & 0.088 \\
CP~\cite{he2022elic} & +19.18 & +67.84 & 0.201 \\
MCM(Ours) & - & - & 0.167 \\
\bottomrule
\end{tabular}\\
\vspace{0.3mm}
The parameters and MACs/pixel are evaluated only for the entropy model.
\end{table}

{For a fair comparison, we implement typical context modeling methods on the same platform as the proposed MCM. For instance, the ``H'' model only uses hyperpriors~\cite{balle2018variational} for entropy context modeling while the ``S'' model applies the  context modeling following the autoregressive manner to utilize the hyperpriors and  spatial-channel neighbors jointly~\cite{minnen2018joint}. The channel-wise grouping used in~\cite{minnen2020channel}, the two-stage spatial checkerboard arrangement in~\cite{he2021checkerboard}, and the combination of two-stage spatial checkerboard arrangement and uneven channel-wise grouping in~\cite{he2022elic} are referred to as the ``C'', ``P'' and ``CP'' models respectively. }

\subsubsection{Performance vs. Decoding Latency} \label{sec:latency}
Results are given in Fig.~\ref{fig:entropy_model} and Table~\ref{tbl:latency}. As seen, the proposed MCM offers the best performance-complexity tradeoff. For example, the MCM model, the ``C'' model, and the ``CP'' model report the leading coding performance at both low and high bitrates, even outperforming the spatial autoregressive model (e.g., the ``S'' model). The leading performance obtained by the ``C'' and ``CP'' models comes from the use of an excessive amount of extra model parameters and increased MACs/pixels. More importantly, the MCM reports $>60\times$, $\approx4\times$, and $\approx1.2\times$ decoding speedup when compared with the respective ``S'', ``C'' and ``CP'' models. Although the ``H'' and ``P'' models are faster than the MCM, their coding performance suffers. {Note that the performance of our method with only hyperpriors (e.g., TinyLIC with the ``H'' model in entropy coding) is also significantly better than BPG ({$\approx$19.86\% BD-rate improvement on average}), suggesting the effectiveness of our transform backbone.} {The latency measurement is conducted on the Python platform. Numbers may be different on other platforms but the trend would remain.}

\subsubsection{Channel Slicing Strategy} \label{sec:channel_split}

\begin{figure}[t]
\begin{center}
\subfloat[]{
\includegraphics[width=0.45\linewidth]{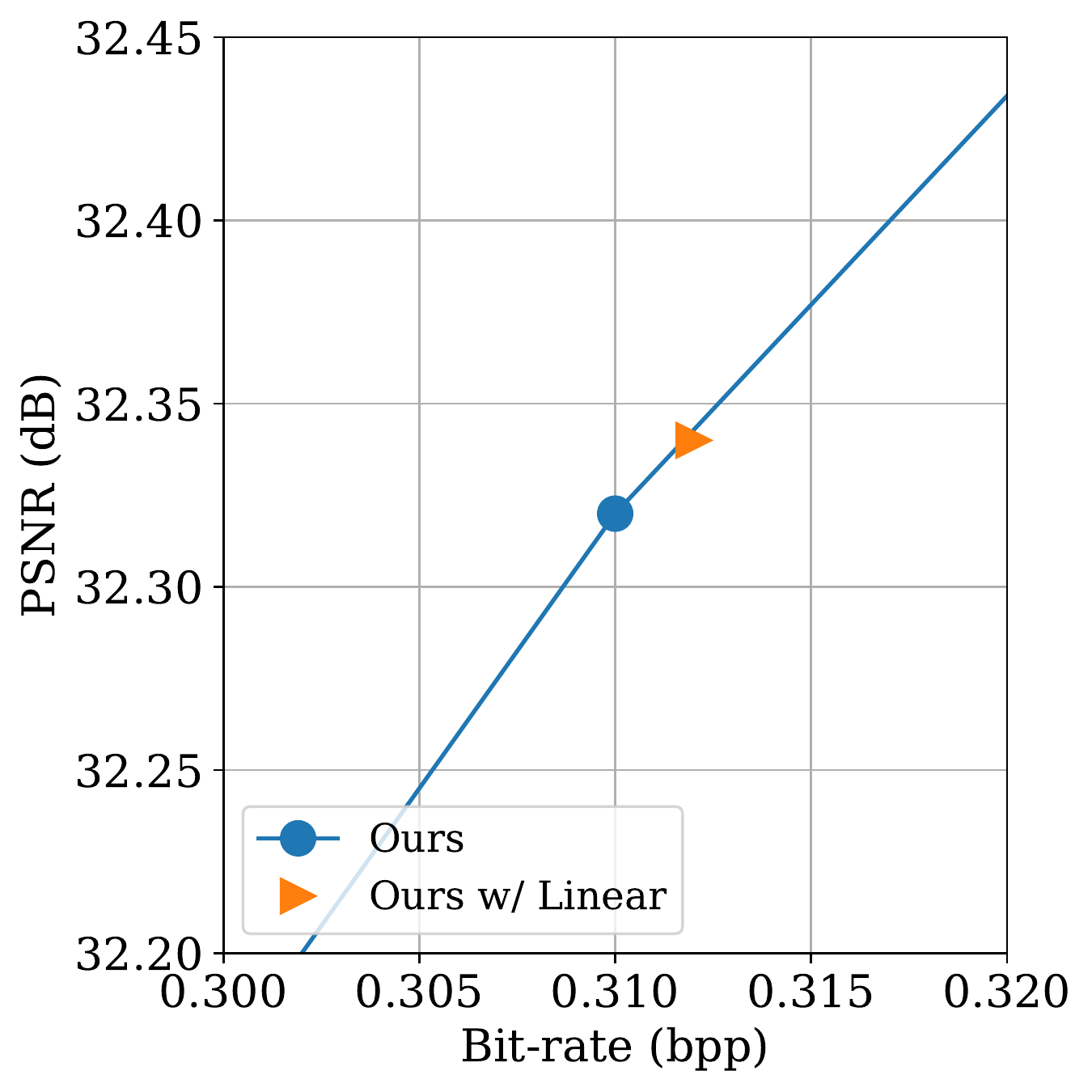}
}
\subfloat[]{
\includegraphics[width=0.45\linewidth]{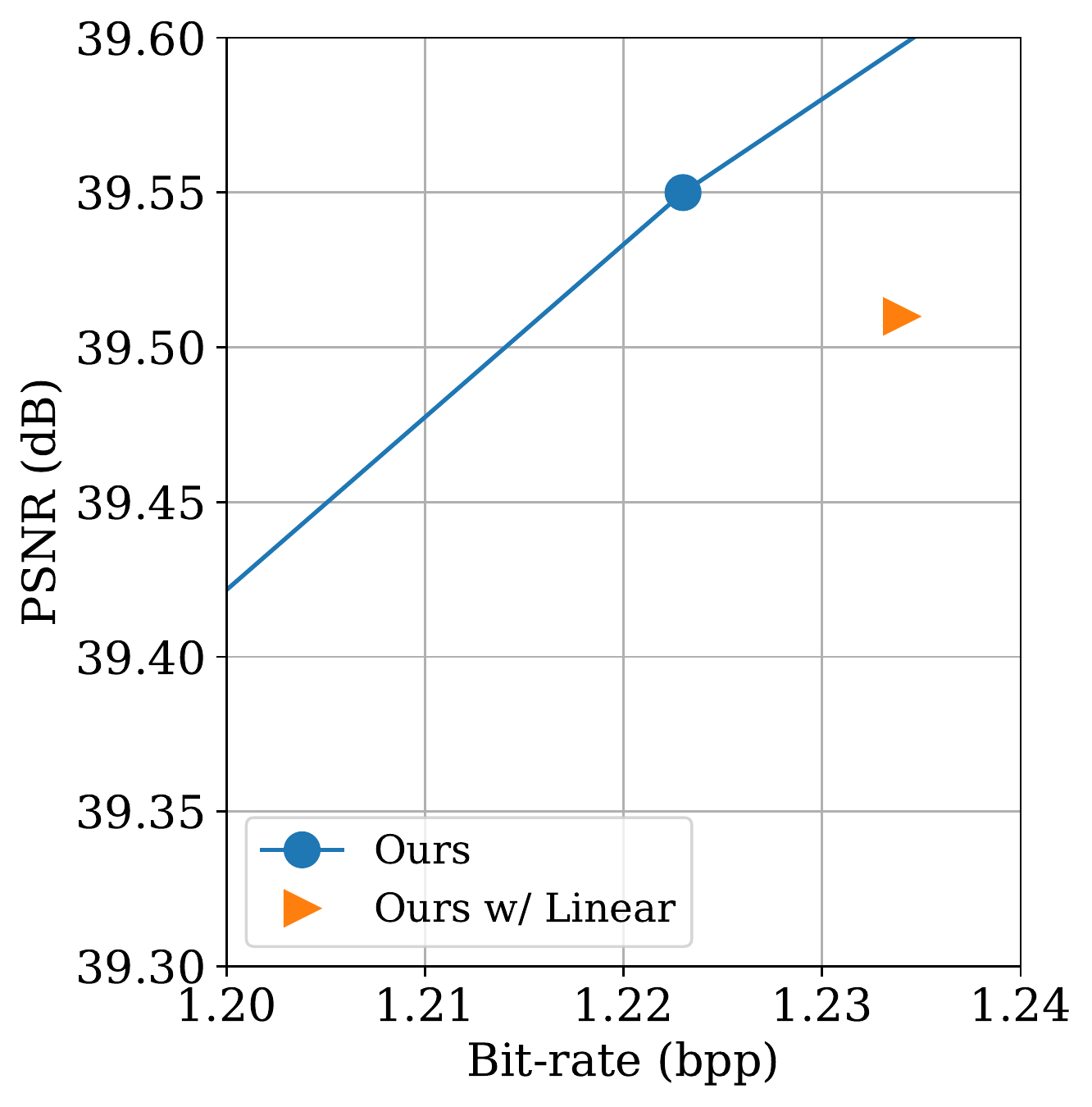}
}
\end{center}
\caption{{\bf Channel Slicing Strategy.} (a) Low bitrate with $\lambda = 0.0067$, (b) High bitrate with $\lambda = 0.0932.$ Solid line is R-D curve of default {\it TinyLIC} with the Cosine Strategy.} 
\label{fig:channel_split} 
\end{figure}

We compare several channel slicing strategies where the Linear scheme is used in \cite{minnen2020channel} to evenly group channels, and the Cosine scheme sets variable-size channel groups as handcrafted rules used in~\cite{he2022elic}. As for Linear slicing, the latent feature tensor with a total of 320 channels is evenly sliced into four 80-channel groups while, on the contrary, non-uniform channel groups with variable channels, e.g., 24-, 69-, 104- and 123-channel, are produced following the Cosine scheme. The stage-wise GCP is specifically applied for grouped channels accordingly as in Sec.~\ref{sec:entropy_mcc}. As shown in Fig.~\ref{fig:channel_split}, our default Cosine slicing strategy is better than the Linear slicing method.

\subsubsection{Progressive Decoding}

\begin{figure}[t]
\begin{center}
\includegraphics[width=1\linewidth]{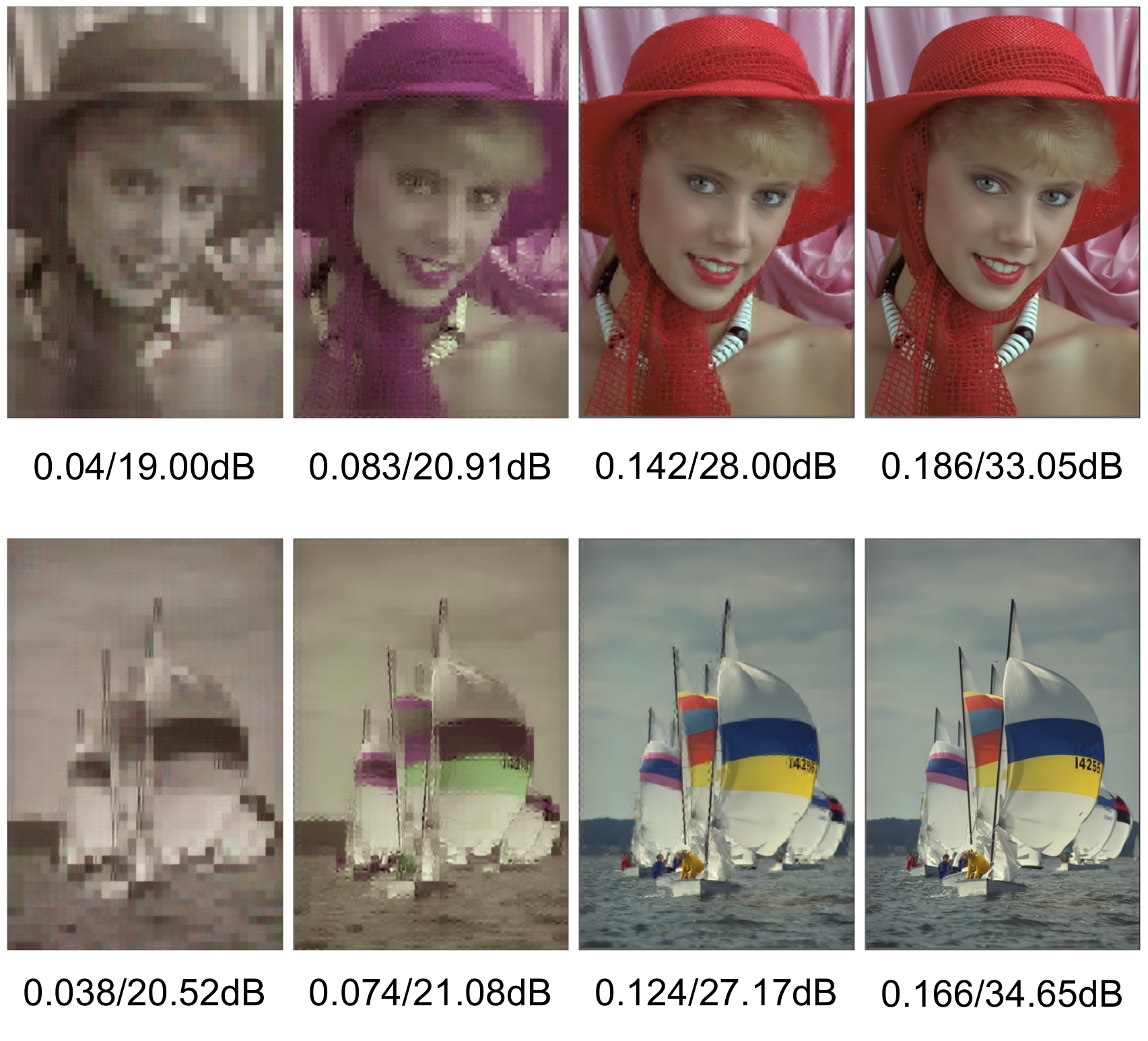}
\end{center}
\caption{{\bf Progressive Decoding.} The reconstruction results of four stages by our proposed MCM.}
\label{fig:progressive}
\end{figure}

%As a version of channel conditional model, 
A take-away point offered by the MCM model is the support of the progressive decoding as in Fig.~\ref{fig:progressive}. By decoding the first group of channels, general structural information are  reconstructed with only a quarter bpp consumption. By further decoding and augmenting the following groups of channels, we can observe the restoration of chrominance  and high-frequency components progressively. Apparently, having such a progressive decoding capability would benefit networked applications that often comes with unreliable connections.

%Since the progressive decoding is usually adopted for preview, it is also a good choice for transmission as a subcontract in networking environment.

% In addition, we also report the rate reduction contributed by multistage contexts in Table~\ref{tbl:ctx_contribution}. As seen, by only enabling the hyperpriors (by completely ignoring stage-wise contexts), the bpp value increases sharply with much higher value than the default setting using both hyperpriors and multistage context models in {\it TinyLIC}. By  enabling the context prediction for latent elements (e.g., ctx\#$k$ represents the use of context model in stage\#$k$, $k=1,2,3$) from stage\#1 to stage\#3, more bit saving is reported, clearly evidencing the effectiveness of our proposed MCM to gradually aggregate neighborhood information.

% \begin{table}[t]
% \caption{Progressive Rate Reduction Contributed by Context Models Used in Multiple Stages. (Results Are Averaged on Kodak Set with $\lambda = 0.0067$; Similar Outcomes Are Observed for Other $\lambda$s).}
% \label{tbl:ctx_contribution}
% \centering
% \begin{tabular}{ccccc}
% \hline
% hyperpriors & ctx\#1 & ctx\#2 & ctx\#3 & bpp $\downarrow$ \\
% \hline
% \checkmark & & & & 0.480 \\
% \checkmark & & & \checkmark & 0.396 \\
% \checkmark & & \checkmark & \checkmark & 0.313 \\
% \checkmark & \checkmark & \checkmark & \checkmark & \textbf{0.279} \\
% \hline
% \end{tabular}
% \end{table}

\IEEEpubidadjcol

\section{Conclusion}
A novel learned image coding method - {\it TinyLIC} was developed in this work, presenting the superior compression performance, e.g., averaged 3\% BD-rate gains against the VVC Intra anchor, and high-throughput computation, e.g., almost 60$\times$ decoding speedup compared with prominent learned image coding approach. Joint high-performance compression and high throughput computation of the proposed {\it TinyLIC} comes from the intelligent use of adaptive neighborhood information aggregation. To this end,  we integrate the convolution and self-attention to form the content-adaptive transform by which we can dynamically characterize and embed the neighborhood information conditioned on the input content; we further propose the multistage context model using local spatial-channel neighbors in a managed order for entropy coding to unknit the autoregressive dependency for parallel processing while still retaining the efficiency as the autoregressive model. Companion material is also offered to demonstrate that the proposed {\it TinyLIC} is a well-generalized method with promising prospects for practical applications.

% \section*{Acknowledgments}
% We particularly thank the authors of ICLR'18~\cite{balle2018variational}, NIPS'18~\cite{minnen2018joint}, CVPR'20~\cite{cheng2020learned}, ICLR'21~\cite{qian2020learning}, TPAMI'20~\cite{9204799}, TPAMI'21~\cite{9376651}, and ACMMM'21~\cite{xie2021enhanced} for their publicly accessible models and materials  used in our comparative studies. 

\bibliographystyle{IEEEtran}
\bibliography{reference}

\begin{thebibliography}{10}
\providecommand{\url}[1]{#1}
\csname url@samestyle\endcsname
\providecommand{\newblock}{\relax}
\providecommand{\bibinfo}[2]{#2}
\providecommand{\BIBentrySTDinterwordspacing}{\spaceskip=0pt\relax}
\providecommand{\BIBentryALTinterwordstretchfactor}{4}
\providecommand{\BIBentryALTinterwordspacing}{\spaceskip=\fontdimen2\font plus
\BIBentryALTinterwordstretchfactor\fontdimen3\font minus
  \fontdimen4\font\relax}
\providecommand{\BIBforeignlanguage}[2]{{%
\expandafter\ifx\csname l@#1\endcsname\relax
\typeout{** WARNING: IEEEtran.bst: No hyphenation pattern has been}%
\typeout{** loaded for the language `#1'. Using the pattern for}%
\typeout{** the default language instead.}%
\else
\language=\csname l@#1\endcsname
\fi
#2}}
\providecommand{\BIBdecl}{\relax}
\BIBdecl

\bibitem{Gibson2017RateDF}
J.~D. Gibson, ``Rate distortion functions and rate distortion function lower
  bounds for real-world sources,'' \emph{Entropy}, vol.~19, p. 604, 2017.

\bibitem{wang2003multiscale}
Z.~Wang, E.~P. Simoncelli, and A.~C. Bovik, ``Multiscale structural similarity
  for image quality assessment,'' in \emph{The Thrity-Seventh Asilomar
  Conference on Signals, Systems \& Computers}, 2003, pp. 1398--1402.

\bibitem{balle2020nonlinear}
J.~Ball{\'e}, P.~A. Chou, D.~Minnen, S.~Singh, N.~Johnston, E.~Agustsson, S.~J.
  Hwang, and G.~Toderici, ``Nonlinear transform coding,'' \emph{IEEE Journal of
  Selected Topics in Signal Processing}, vol.~15, no.~2, pp. 339--353, 2020.

\bibitem{952802}
V.~Goyal, ``Theoretical foundations of transform coding,'' \emph{IEEE Signal
  Processing Magazine}, vol.~18, no.~5, pp. 9--21, 2001.

\bibitem{BDData}
G.~Bjontegaard, ``Calculation of average {PSNR} differences between {R-D}
  curves,'' in \emph{Doc. VCEG-M33, ITU-T VCEG 13th Meeting}, 2001.

\bibitem{HEVC_overview}
G.~J. Sullivan, J.-R. Ohm, W.-J. Han, and T.~Wiegand, ``Overview of the high
  efficiency video coding (hevc) standard,'' \emph{IEEE Transactions on
  Circuits and Systems for Video Technology}, vol.~22, no.~12, pp. 1649--1668,
  2012.

\bibitem{balle2018variational}
J.~Ball{\'e}, D.~Minnen, S.~Singh, S.~J. Hwang, and N.~Johnston, ``Variational
  image compression with a scale hyperprior,'' in \emph{International
  Conference on Learning Representations}, 2018.

\bibitem{minnen2018joint}
D.~Minnen, J.~Ball{\'{e}}, and G.~Toderici, ``Joint autoregressive and
  hierarchical priors for learned image compression,'' in \emph{Advances in
  Neural Information Processing Systems}, 2018, pp. 10\,794--10\,803.

\bibitem{cheng2020learned}
Z.~Cheng, H.~Sun, M.~Takeuchi, and J.~Katto, ``Learned image compression with
  discretized gaussian mixture likelihoods and attention modules,'' in
  \emph{Proceedings of the IEEE/CVF Conference on Computer Vision and Pattern
  Recognition}, 2020, pp. 7939--7948.

\bibitem{minnen2020channel}
D.~Minnen and S.~Singh, ``Channel-wise autoregressive entropy models for
  learned image compression,'' in \emph{2020 IEEE International Conference on
  Image Processing (ICIP)}.\hskip 1em plus 0.5em minus 0.4em\relax IEEE, 2020,
  pp. 3339--3343.

\bibitem{xie2021enhanced}
Y.~Xie, K.~L. Cheng, and Q.~Chen, ``Enhanced invertible encoding for learned
  image compression,'' in \emph{Proceedings of the ACM International Conference
  on Multimedia}, 2021.

\bibitem{Qian2021Entroformer}
Y.~Qian, X.~Sun, M.~Lin, Z.~Tan, and R.~Jin, ``{Entroformer: A
  Transformer-based Entropy Model for Learned Image Compression},'' in
  \emph{International Conference on Learning Representations}, 2022.

\bibitem{bross2021overview}
B.~Bross, Y.-K. Wang, Y.~Ye, S.~Liu, J.~Chen, G.~J. Sullivan, and J.-R. Ohm,
  ``Overview of the versatile video coding (vvc) standard and its
  applications,'' \emph{IEEE Transactions on Circuits and Systems for Video
  Technology}, vol.~31, no.~10, pp. 3736--3764, 2021.

\bibitem{GrayQuantization}
R.~M. Gray and D.~L. Neuhoff, ``Quantization,'' \emph{IEEE Transactions on
  Information Theory}, vol.~44, no.~6, pp. 2325--2383, Oct 1998.

\bibitem{cabac}
D.~Marpe, H.~Schwarz, and T.~Wiegand, ``Context-based adaptive binary
  arithmetic coding in the h.264/avc video compression standard,'' \emph{IEEE
  Transactions on Circuits and Systems for Video Technology}, vol.~13, no.~7,
  pp. 620--636, 2003.

\bibitem{sullivan_videoConcepts}
G.~J. Sullivan and T.~Wiegand, ``Video compression---from concepts to the
  {H.264/AVC} standard,'' \emph{Proceedings of the IEEE}, vol.~93, no.~1, pp.
  18--31, 2005.

\bibitem{JPEG}
{Overview of JPEG}, ``{\url{https://jpeg.org/jpeg/}},'' 2018.

\bibitem{JPEG2K}
D.~T. Lee, ``Jpeg 2000: Retrospective and new developments,'' \emph{Proceedings
  of the IEEE}, vol.~93, no.~1, pp. 32--41, Jan 2005.

\bibitem{DCT}
N.~Ahmed, T.~Natarajan, and K.~R. Rao, ``Discrete cosine transform,''
  \emph{IEEE transactions on Computers}, vol. 100, no.~1, pp. 90--93, 1974.

\bibitem{zhang2020image}
X.~Zhang, C.~Yang, X.~Li, S.~Liu, H.~Yang, I.~Katsavounidis, S.-M. Lei, and
  C.-C.~J. Kuo, ``Image coding with data-driven transforms: Methodology,
  performance and potential,'' \emph{IEEE Transactions on Image Processing},
  vol.~29, pp. 9292--9304, 2020.

\bibitem{hevc_intra}
J.~Lainema, F.~Bossen, W.-J. Han, J.~Min, and K.~Ugur, ``Intra coding of the
  hevc standard,'' \emph{IEEE Trans. Circuits Syst. Video Technol.}, vol.~22,
  no.~12, pp. 1792--1801, Dec. 2012.

\bibitem{vvc_intra}
J.~Pfaff, A.~Filippov, S.~Liu, X.~Zhao, J.~Chen, S.~De-Luxán-Hernández,
  T.~Wiegand, V.~Rufitskiy, A.~K. Ramasubramonian, and G.~Van~der Auwera,
  ``Intra prediction and mode coding in vvc,'' \emph{IEEE Transactions on
  Circuits and Systems for Video Technology}, vol.~31, no.~10, pp. 3834--3847,
  2021.

\bibitem{chen2021end}
T.~Chen, H.~Liu, Z.~Ma, Q.~Shen, X.~Cao, and Y.~Wang, ``End-to-end learnt image
  compression via non-local attention optimization and improved context
  modeling,'' \emph{IEEE Transactions on Image Processing}, vol.~30, pp.
  3179--3191, 2021.

\bibitem{VVC_parallelism}
\BIBentryALTinterwordspacing
P.~Belememis, N.~Panagou, T.~Loukopoulos, and M.~Koziri, ``{Review and
  comparative analysis of parallel video encoding techniques for VVC},'' in
  \emph{Applications of Digital Image Processing XLIII}, A.~G. Tescher and
  T.~Ebrahimi, Eds., vol. 11510, International Society for Optics and
  Photonics.\hskip 1em plus 0.5em minus 0.4em\relax SPIE, 2020, pp. 258 -- 276.
  [Online]. Available: \url{https://doi.org/10.1117/12.2569283}
\BIBentrySTDinterwordspacing

\bibitem{HT-CABAC}
V.~Sze and M.~Budagavi, ``High throughput cabac entropy coding in hevc,''
  \emph{IEEE Transactions on Circuits and Systems for Video Technology},
  vol.~22, no.~12, pp. 1778--1791, 2012.

\bibitem{liang2021swinir}
J.~Liang, J.~Cao, G.~Sun, K.~Zhang, L.~Van~Gool, and R.~Timofte, ``Swinir:
  Image restoration using swin transformer,'' in \emph{IEEE International
  Conference on Computer Vision Workshops}, 2021.

\bibitem{jia2016dynamic}
X.~Jia, B.~De~Brabandere, T.~Tuytelaars, and L.~V. Gool, ``Dynamic filter
  networks,'' \emph{Advances in neural information processing systems},
  vol.~29, 2016.

\bibitem{hassani2022neighborhood}
A.~Hassani, S.~Walton, J.~Li, S.~Li, and H.~Shi, ``Neighborhood attention
  transformer,'' \emph{arXiv preprint arXiv:2204.07143}, 2022.

\bibitem{lu2021transformer}
M.~Lu, P.~Guo, H.~Shi, C.~Cao, and Z.~Ma, ``Transformer-based image
  compression,'' in \emph{IEEE Data Compression Conference}, 2022.

\bibitem{Liu_CSTR}
H.~Liu, M.~Lu, Z.~Chen, X.~Cao, Z.~Ma, and Y.~Wang, ``End-to-end neural video
  coding using a compound spatiotemporal representation,'' \emph{IEEE
  Transactions on Circuits and Systems for Video Technology}, pp. 1--1, 2022.

\bibitem{he2022elic}
D.~He, Z.~Yang, W.~Peng, R.~Ma, H.~Qin, and Y.~Wang, ``Elic: Efficient learned
  image compression with unevenly grouped space-channel contextual adaptive
  coding,'' in \emph{Proceedings of the IEEE/CVF Conference on Computer Vision
  and Pattern Recognition}, 2022, pp. 5718--5727.

\bibitem{J2K_Wavelet}
B.-F. Wu and C.-F. Lin, ``A high-performance and memory-efficient pipeline
  architecture for the 5/3 and 9/7 discrete wavelet transform of jpeg2000
  codec,'' \emph{IEEE Transactions on Circuits and Systems for Video
  Technology}, vol.~15, no.~12, pp. 1615--1628, 2005.

\bibitem{wallace1992jpeg}
G.~K. Wallace, ``The jpeg still picture compression standard,'' \emph{IEEE
  transactions on consumer electronics}, vol.~38, no.~1, pp. xviii--xxxiv,
  1992.

\bibitem{rabbani2002jpeg2000}
M.~Rabbani, ``Jpeg2000: Image compression fundamentals, standards and
  practice,'' \emph{Journal of Electronic Imaging}, vol.~11, no.~2, p. 286,
  2002.

\bibitem{DCT_drift}
\BIBentryALTinterwordspacing
A.~T. Hinds, Y.~A. Reznik, L.~Yu, Z.~Ni, and C.~Zhang, ``{Drift analysis for
  integer IDCT},'' in \emph{Applications of Digital Image Processing XXX},
  A.~G. Tescher, Ed., vol. 6696, International Society for Optics and
  Photonics.\hskip 1em plus 0.5em minus 0.4em\relax SPIE, 2007, pp. 432 -- 447.
  [Online]. Available: \url{https://doi.org/10.1117/12.740220}
\BIBentrySTDinterwordspacing

\bibitem{AVC_transform}
H.~Malvar, A.~Hallapuro, M.~Karczewicz, and L.~Kerofsky, ``Low complexity
  transform and quantization in {H.264/AVC},'' \emph{IEEE Trans. Circuits Syst.
  Video Technol.}, vol.~13, no.~7, pp. 598--603, Jul. 2003.

\bibitem{yao_selfDict}
Y.~Xue and Y.~Wang, ``Video coding using a self-adaptive redundant dictionary
  consisting of spatial and temporal prediction candidates,'' in \emph{Proc.
  IEEE Int. Conf. Multimedia and Expo (ICME)}, 2014.

\bibitem{Xu_CompressDict}
M.~Xu, S.~Li, J.~Lu, and W.~Zhu, ``Compressibility constrained sparse
  representation with learnt dictionary for low bit-rate image compression,''
  \emph{IEEE Transactions on Circuits and Systems for Video Technology},
  vol.~24, no.~10, pp. 1743--1757, 2014.

\bibitem{RLS-DLA}
K.~Skretting and K.~Engan, ``Image compression using learned dictionaries by
  rls-dla and compared with k-svd,'' in \emph{2011 IEEE International
  Conference on Acoustics, Speech and Signal Processing (ICASSP)}, 2011, pp.
  1517--1520.

\bibitem{li2001edge}
X.~Li and M.~T. Orchard, ``Edge-directed prediction for lossless compression of
  natural images,'' \emph{IEEE Transactions on image processing}, vol.~10,
  no.~6, pp. 813--817, 2001.

\bibitem{VVC_Q_Entropy}
H.~Schwarz, M.~Coban, M.~Karczewicz, T.-D. Chuang, F.~Bossen, A.~Alshin,
  J.~Lainema, C.~R. Helmrich, and T.~Wiegand, ``Quantization and entropy coding
  in the versatile video coding (vvc) standard,'' \emph{IEEE Transactions on
  Circuits and Systems for Video Technology}, vol.~31, no.~10, pp. 3891--3906,
  2021.

\bibitem{balle2016end}
J.~Ball{\'e}, V.~Laparra, and E.~P. Simoncelli, ``End-to-end optimized image
  compression,'' in \emph{International Conference on Learning
  Representations}, 2017.

\bibitem{qian2020learning}
Y.~Qian, Z.~Tan, X.~Sun, M.~Lin, D.~Li, Z.~Sun, L.~Hao, and R.~Jin, ``Learning
  accurate entropy model with global reference for image compression,'' in
  \emph{International Conference on Learning Representations}, 2020.

\bibitem{kim2021joint}
J.-H. Kim, B.~Heo, and J.-S. Lee, ``Joint global and local hierarchical priors
  for learned image compression,'' 2021.

\bibitem{ViT_Survey}
K.~Han, Y.~Wang, H.~Chen, X.~Chen, J.~Guo, Z.~Liu, Y.~Tang, A.~Xiao, C.~Xu,
  Y.~Xu, Z.~Yang, Y.~Zhang, and D.~Tao, ``A survey on vision transformer,''
  \emph{IEEE Transactions on Pattern Analysis \& Machine Intelligence}, no.~01,
  pp. 1--1, feb 5555.

\bibitem{carion2020end}
N.~Carion, F.~Massa, G.~Synnaeve, N.~Usunier, A.~Kirillov, and S.~Zagoruyko,
  ``End-to-end object detection with transformers,'' in \emph{European
  Conference on Computer Vision}.\hskip 1em plus 0.5em minus 0.4em\relax
  Springer, 2020, pp. 213--229.

\bibitem{dosovitskiy2020image}
A.~Dosovitskiy, L.~Beyer, A.~Kolesnikov, D.~Weissenborn, X.~Zhai,
  T.~Unterthiner, M.~Dehghani, M.~Minderer, G.~Heigold, S.~Gelly \emph{et~al.},
  ``An image is worth 16x16 words: Transformers for image recognition at
  scale,'' \emph{arXiv preprint arXiv:2010.11929}, 2020.

\bibitem{zhou2021deepvit}
D.~Zhou, B.~Kang, X.~Jin, L.~Yang, X.~Lian, Z.~Jiang, Q.~Hou, and J.~Feng,
  ``Deepvit: Towards deeper vision transformer,'' \emph{arXiv:2103.11886},
  2021.

\bibitem{wang2021uformer}
Z.~Wang, X.~Cun, J.~Bao, and J.~Liu, ``Uformer: A general u-shaped transformer
  for image restoration,'' \emph{arXiv:2106.03106}, 2021.

\bibitem{liu2021swin}
Z.~Liu, Y.~Lin, Y.~Cao, H.~Hu, Y.~Wei, Z.~Zhang, S.~Lin, and B.~Guo, ``Swin
  transformer: Hierarchical vision transformer using shifted windows,''
  \emph{International Conference on Computer Vision (ICCV)}, 2021.

\bibitem{zhu2021transformer}
Y.~Zhu, Y.~Yang, and T.~Cohen, ``Transformer-based transform coding,'' in
  \emph{International Conference on Learning Representations}, 2022.

\bibitem{liu2022convnet}
Z.~Liu, H.~Mao, C.-Y. Wu, C.~Feichtenhofer, T.~Darrell, and S.~Xie, ``A convnet
  for the 2020s,'' 2022.

\bibitem{he2021checkerboard}
D.~He, Y.~Zheng, B.~Sun, Y.~Wang, and H.~Qin, ``Checkerboard context model for
  efficient learned image compression,'' in \emph{Proceedings of the IEEE/CVF
  Conference on Computer Vision and Pattern Recognition}, 2021, pp.
  14\,771--14\,780.

\bibitem{islam2019much}
M.~A. Islam, S.~Jia, and N.~D. Bruce, ``How much position information do
  convolutional neural networks encode?'' in \emph{International Conference on
  Learning Representations}, 2019.

\bibitem{gulati2020conformer}
A.~Gulati, J.~Qin, C.-C. Chiu, N.~Parmar, Y.~Zhang, J.~Yu, W.~Han, S.~Wang,
  Z.~Zhang, Y.~Wu \emph{et~al.}, ``Conformer: Convolution-augmented transformer
  for speech recognition,'' \emph{arXiv preprint arXiv:2005.08100}, 2020.

\bibitem{xiao2021early}
T.~Xiao, M.~Singh, E.~Mintun, T.~Darrell, P.~Doll{\'a}r, and R.~Girshick,
  ``Early convolutions help transformers see better,'' \emph{arXiv preprint
  arXiv:2106.14881}, 2021.

\bibitem{He_2016_CVPR}
K.~He, X.~Zhang, S.~Ren, and J.~Sun, ``Deep residual learning for image
  recognition,'' in \emph{Proceedings of the IEEE Conference on Computer Vision
  and Pattern Recognition (CVPR)}, June 2016.

\bibitem{hendrycks2016gaussian}
D.~Hendrycks and K.~Gimpel, ``Gaussian error linear units (gelus),''
  \emph{arXiv preprint arXiv:1606.08415}, 2016.

\bibitem{liu2020unified}
J.~Liu, G.~Lu, Z.~Hu, and D.~Xu, ``A unified end-to-end framework for efficient
  deep image compression,'' \emph{arXiv preprint arXiv:2002.03370}, 2020.

\bibitem{kingma2014adam}
D.~P. Kingma and J.~Ba, ``Adam: A method for stochastic optimization,''
  \emph{arXiv preprint arXiv:1412.6980}, 2014.

\bibitem{begaint2020compressai}
J.~B{\'e}gaint, F.~Racap{\'e}, S.~Feltman, and A.~Pushparaja, ``Compressai: a
  pytorch library and evaluation platform for end-to-end compression
  research,'' \emph{arXiv preprint arXiv:2011.03029}, 2020.

\bibitem{9204799}
H.~Ma, D.~Liu, N.~Yan, H.~Li, and F.~Wu, ``End-to-end optimized versatile image
  compression with wavelet-like transform,'' \emph{IEEE Transactions on Pattern
  Analysis and Machine Intelligence}, vol.~44, no.~3, pp. 1247--1263, 2022.

\bibitem{9376651}
Y.~Hu, W.~Yang, Z.~Ma, and J.~Liu, ``Learning end-to-end lossy image
  compression: A benchmark,'' \emph{IEEE Transactions on Pattern Analysis and
  Machine Intelligence}, vol.~44, no.~8, pp. 4194--4211, 2022.

\bibitem{VVC_ALF}
M.~Karczewicz, N.~Hu, J.~Taquet, C.-Y. Chen, K.~Misra, K.~Andersson, P.~Yin,
  T.~Lu, E.~François, and J.~Chen, ``Vvc in-loop filters,'' \emph{IEEE
  Transactions on Circuits and Systems for Video Technology}, vol.~31, no.~10,
  pp. 3907--3925, 2021.

\bibitem{CvT}
H.~Wu, B.~Xiao, N.~Codella, M.~Liu, X.~Dai, L.~Yuan, and L.~Zhang, ``Cvt:
  Introducing convolutions to vision transformers,'' in \emph{2021 IEEE/CVF
  International Conference on Computer Vision (ICCV)}, 2021, pp. 22--31.

\end{thebibliography}


\begin{thebibliography}{1}
\providecommand{\url}[1]{#1}
\csname url@samestyle\endcsname
\providecommand{\newblock}{\relax}
\providecommand{\bibinfo}[2]{#2}
\providecommand{\BIBentrySTDinterwordspacing}{\spaceskip=0pt\relax}
\providecommand{\BIBentryALTinterwordstretchfactor}{4}
\providecommand{\BIBentryALTinterwordspacing}{\spaceskip=\fontdimen2\font plus
\BIBentryALTinterwordstretchfactor\fontdimen3\font minus
  \fontdimen4\font\relax}
\providecommand{\BIBforeignlanguage}[2]{{%
\expandafter\ifx\csname l@#1\endcsname\relax
\typeout{** WARNING: IEEEtran.bst: No hyphenation pattern has been}%
\typeout{** loaded for the language `#1'. Using the pattern for}%
\typeout{** the default language instead.}%
\else
\language=\csname l@#1\endcsname
\fi
#2}}
\providecommand{\BIBdecl}{\relax}
\BIBdecl

\bibitem{minnen2020channel}
D.~Minnen and S.~Singh, ``Channel-wise autoregressive entropy models for
  learned image compression,'' in \emph{2020 IEEE International Conference on
  Image Processing (ICIP)}.\hskip 1em plus 0.5em minus 0.4em\relax IEEE, 2020,
  pp. 3339--3343.

\bibitem{he2022elic}
D.~He, Z.~Yang, W.~Peng, R.~Ma, H.~Qin, and Y.~Wang, ``Elic: Efficient learned
  image compression with unevenly grouped space-channel contextual adaptive
  coding,'' in \emph{Proceedings of the IEEE/CVF Conference on Computer Vision
  and Pattern Recognition}, 2022, pp. 5718--5727.

\bibitem{he2021checkerboard}
D.~He, Y.~Zheng, B.~Sun, Y.~Wang, and H.~Qin, ``Checkerboard context model for
  efficient learned image compression,'' in \emph{Proceedings of the IEEE/CVF
  Conference on Computer Vision and Pattern Recognition}, 2021, pp.
  14\,771--14\,780.

\bibitem{cheng2020learned}
Z.~Cheng, H.~Sun, M.~Takeuchi, and J.~Katto, ``Learned image compression with
  discretized gaussian mixture likelihoods and attention modules,'' in
  \emph{Proceedings of the IEEE/CVF Conference on Computer Vision and Pattern
  Recognition}, 2020, pp. 7939--7948.

\bibitem{xie2021enhanced}
Y.~Xie, K.~L. Cheng, and Q.~Chen, ``Enhanced invertible encoding for learned
  image compression,'' in \emph{Proceedings of the ACM International Conference
  on Multimedia}, 2021.

\bibitem{balle2018variational}
J.~Ball{\'e}, D.~Minnen, S.~Singh, S.~J. Hwang, and N.~Johnston, ``Variational
  image compression with a scale hyperprior,'' in \emph{International
  Conference on Learning Representations}, 2018.

\bibitem{minnen2018joint}
D.~Minnen, J.~Ball{\'{e}}, and G.~Toderici, ``Joint autoregressive and
  hierarchical priors for learned image compression,'' in \emph{Advances in
  Neural Information Processing Systems}, 2018, pp. 10\,794--10\,803.

\bibitem{ScalingNet}
J.~Lin, M.~Akbari, H.~Fu, Q.~Zhang, S.~Wang, J.~Liang, D.~Liu, F.~Liang,
  G.~Zhang, and C.~Tu, ``Variable-rate multi-frequency image compression using
  modulated generalized octave convolution,'' in \emph{2020 IEEE 22nd
  International Workshop on Multimedia Signal Processing (MMSP)}, 2020, pp.
  1--6.

\bibitem{yao_book}
Y.~Wang and Y.-Q. Zhang, \emph{Video processing and communications}.\hskip 1em
  plus 0.5em minus 0.4em\relax Prentice hall Upper Saddle River, NJ, 2002,
  vol.~1.

\end{thebibliography}

% \clearpage
% % \newpage
% \appendix
% \subfile{Sections/6-supplement.tex}

% \newpage

 

% \end{IEEEbiography}



\vfill

\end{document}

\title{Supplementary Materials - \it High-Efficiency Lossy Image Coding Through Adaptive Neighborhood Information Aggregation}

\author{Ming Lu,~\IEEEmembership{Student Member,~IEEE,}
        Fangdong Chen,
        Shiliang Pu,
        and 
        Zhan Ma,~\IEEEmembership{Senior Member,~IEEE}
        % <-this % stops a space
\thanks{M. Lu and Z. Ma are with Nanjing University, Nanjing, Jiangsu, China. E-mails: luming@smail.nju.edu.cn, mazhan@nju.edu.cn.}
\thanks{F. Chen and S. Pu are with Hikvision Inc., Hangzhou, Zhejiang, China. E-mails: chenfangdong@hikvision.com, pushiliang.hri@hikvision.com.}% <-this % stops a space
%\thanks{Manuscript received April 19, 2021; revised August 16, 2021.}
}



\maketitle

\begin{abstract}
This companion document provides additional information to further evidence the generalization of the proposed {\it TinyLIC} for the support of extra functionalities.
\end{abstract}

\begin{IEEEkeywords}
Learned image coding, adaptive neighborhood information aggregation, convolution, self-attention, multistage context model.
\end{IEEEkeywords}

\section{Architecture Details of {\it TinyLIC}}

\subsection{Architecture of Transform Networks}
Table~\ref{tab:main_arch} and Table~\ref{tab:hyper_arch} detail the architecture of transform networks used in {\it TinyLIC}. The example of ``Conv: k5c128s2'' stands for a convolutional layer having convolutions with spatial kernel size at 5$\times$5 (k5), 128 channels (c128), and a stride of 2 based spatial downsampling (s2) at both dimensions. The same convention is applied to other convolutional settings. It is worth to point out that in transposed convolutions (TConv) at decoder, ``s2'' stands for the spatial upsampling at a stride of 2. Stacked RNAB blocks are used for nonlinear transform without changing the channel numbers. Interested parties can either follow these settings to implement the {\it TinyLIC} from the scratch or clone our project from \url{https://njuvision.github.io/TinyLIC} directly for reproducible research. 

\begin{table}[htbp]
\caption{Network Settings of Main Transform Networks.}
\label{tab:main_arch}
\centering
\begin{tabular}{cc}
\toprule
Main Encoder ($g_a$) & Main Decoder ($g_s$) \\
\hline
Conv: k5c128s2 & RNAB$\times$2 \\
RNAB$\times$2 & TConv: k3c256s2 \\
Conv: k3c192s2 & RNAB$\times$6 \\
RNAB$\times$2 & TConv: k3c192s2 \\
Conv: k3c256s2 & RNAB$\times$2 \\
RNAB$\times$6 & TConv: k3c128s2 \\
Conv: k3c320s2 & RNAB$\times$2 \\
RNAB$\times$2 & TConv: k5c3s2 \\
\bottomrule
\end{tabular}
\end{table}

\begin{table}[htbp]
\caption{Network Settings of Hyper Transform Networks.}
\label{tab:hyper_arch}
\centering
\begin{tabular}{cc}
\toprule
Hyper Encoder ($h_a$) & Hyper Decoder ($h_s$) \\
\hline
Conv: k3c192s2 & RNAB$\times$2 \\
RNAB$\times$2 & TConv: k3c192s2 \\
Conv: k3c192s2 & RNAB$\times$2 \\
RNAB$\times$2 & TConv: k3c384s2 \\
\bottomrule
\end{tabular}
\end{table}

\subsection{Architecture of Multistage Context Model}

\begin{figure}[t]
\begin{center}
\subfloat[]{
\includegraphics[width=0.15\linewidth]{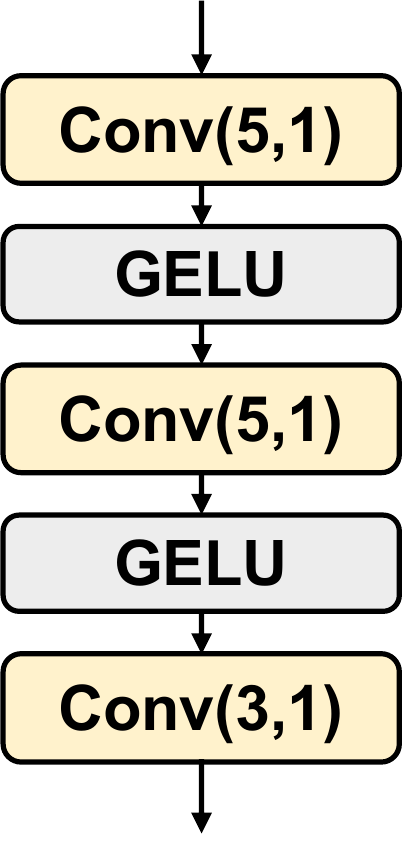}
\label{fig:g_cc}
}
\hspace{1.8cm}
\subfloat[]{
\includegraphics[width=0.15\linewidth]{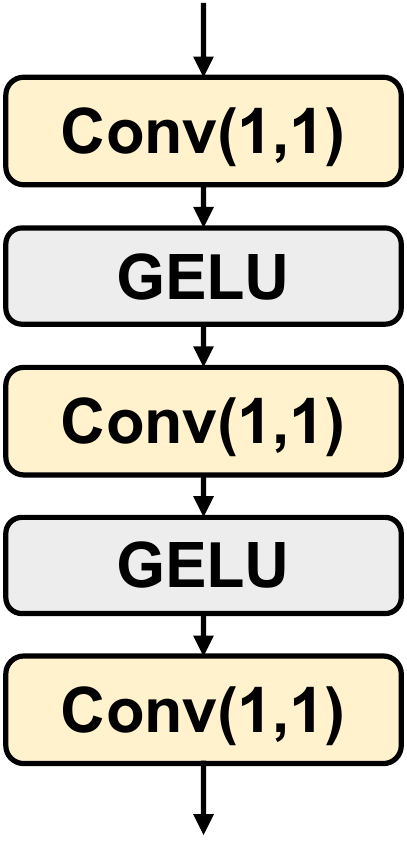}
\label{fig:g_ep}
}
\end{center}
\caption{{\bf Network Compositions} of (a) $g_{cc}(\cdot)$ and (b) $g_{ep}(\cdot)$.}
\end{figure}

Following Minnen et al.~\cite{minnen2020channel}, we stack convolutions to form the $g_{cc}(\cdot)$ as in Fig.~\ref{fig:g_cc} to analyze and embed cross-channel information. The entropy parameter networks $g_{ep}(\cdot)$ used for mean and scale deviation are shown in Fig. \ref{fig:g_ep}.

\begin{figure}[t]
\begin{center}
\subfloat[]{
\includegraphics[width=0.8\linewidth]{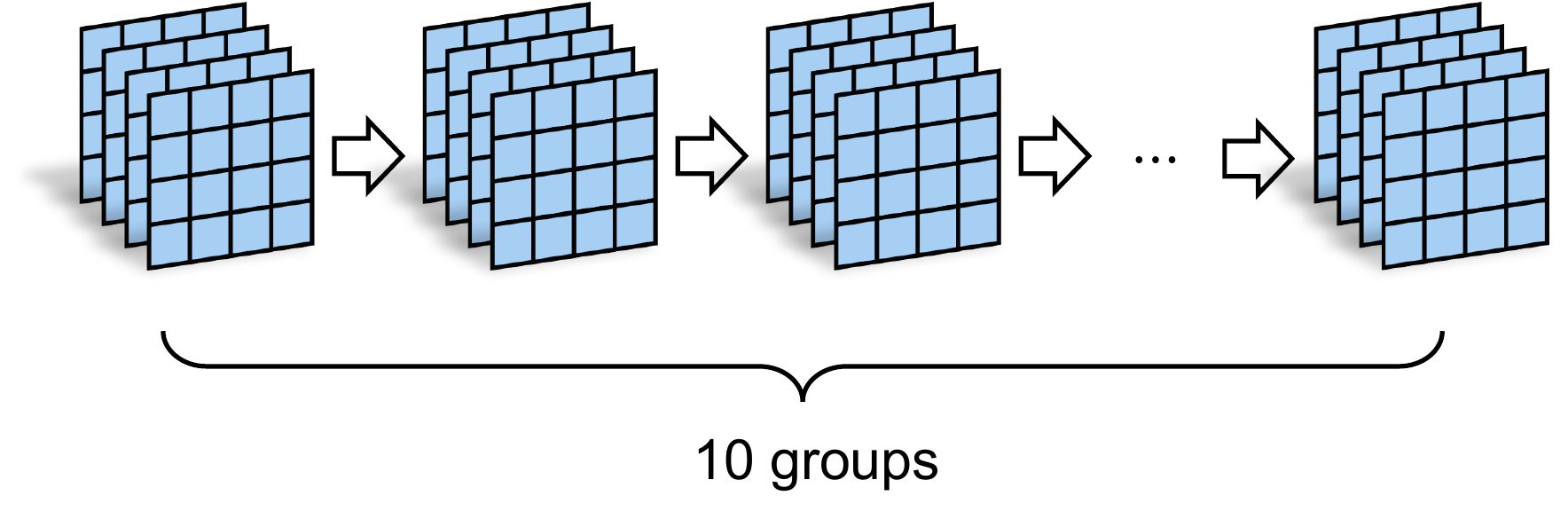}
} \\
\subfloat[]{
\includegraphics[width=0.8\linewidth]{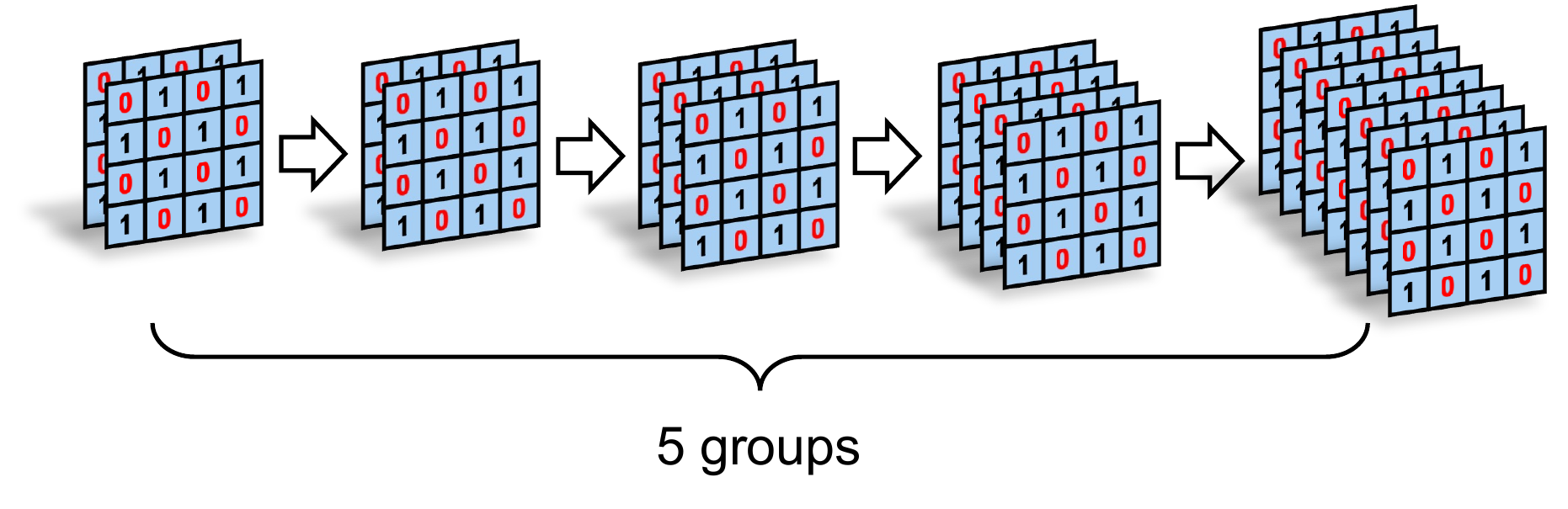}
} \\
\subfloat[]{
\includegraphics[width=0.7\linewidth]{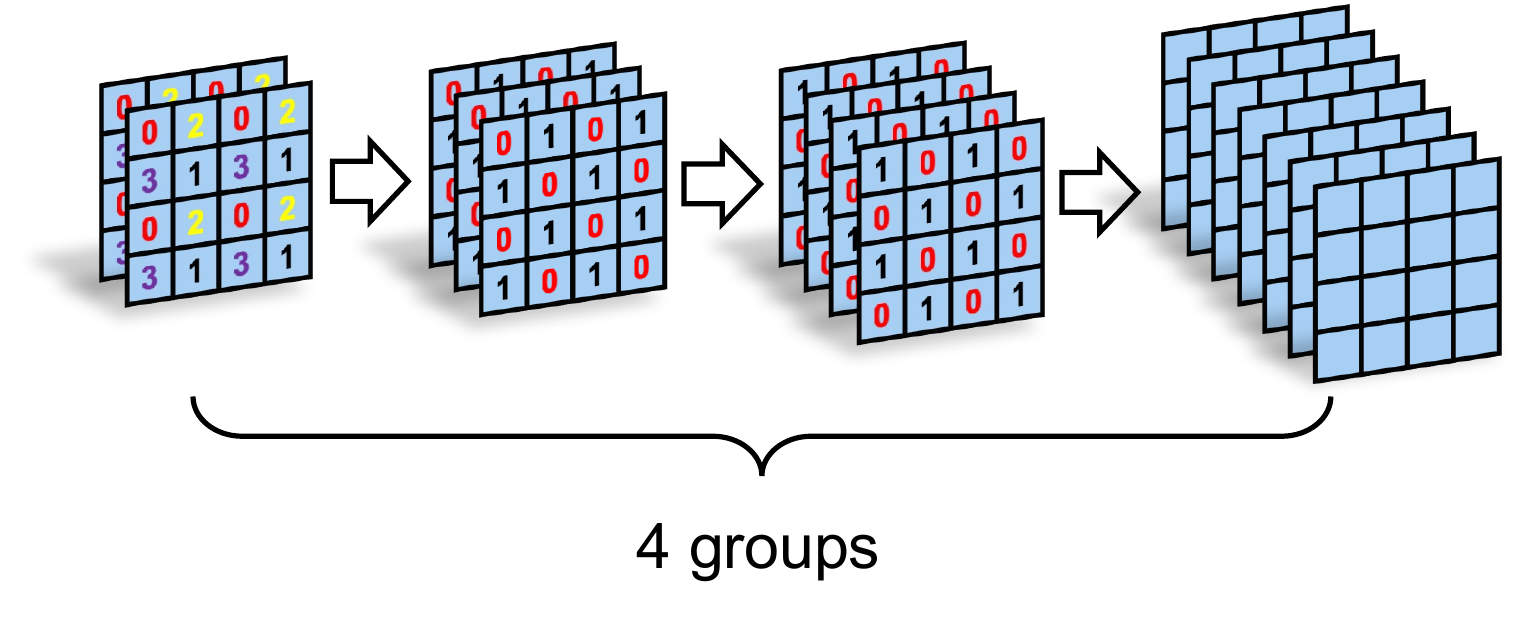}
}
\end{center}
\caption{{\bf Latent Grouping Strategy.} (a) Solely Uniform Channel Grouping as in Minnen'20~\cite{minnen2020channel}; (b) Non-uniform Channel Grouping and Uniform Spatial Checkerboard Grouping as in~\cite{he2022elic} (c) Non-uniform Channel Grouping and Non-uniform Spatial GCP Grouping in the proposed MCM.}
\label{fig:latent_slicing_strategy}
\end{figure}

\subsection{Additional Experiment Settings}

\subsubsection{Latent Grouping Strategies}
As in the main text, our MCM is a generalized approach by extending the existing methods in~\cite{minnen2020channel,he2021checkerboard,he2022elic}. Figure~\ref{fig:latent_slicing_strategy} plots the latent feature grouping for various well-known methods. Recalling the  evaluations in the main text, our MCM offers the best performance-complexity tradeoff by grouping features non-uniformly from both spatial and channel dimensions. Although both Minnen'20~\cite{minnen2020channel} and He et al.~\cite{he2022elic} present a close performance to our method, they require a much larger-size model and more MACs/pixel for computation.  This is because our MCM uses the least stages by intelligently allocating computation gradually to leverage the spatial-channel dependency. 
For example, the least number of channels are grouped in the first stage of the proposed MCM, upon which the fine-grained 4-Step GCP is applied for spatial grouping and context modeling. On the contrary, the last stage presents the largest number of channels, in which feature elements at different spatial locations are processed concurrently using channel-wise aggregation only.  %Such a non-uniform spatial-channel grouping suggest

\subsubsection{Bitrate Range}
As shown in Table~\ref{tab:bitrange}, our proposed TinyLIC covers a much wider bit range and proved better performance than Cheng'20~\cite{cheng2020learned} and Xie'21~\cite{xie2021enhanced} using a unified model. 

\begin{table}[htbp]
\centering
\caption{Bitrate ranges (in bpp) and BD-rate for different methods evaluated in Kodak Dataset}
\begin{tabular}{@{}ccc@{}}
\toprule
\multicolumn{1}{c}{\textbf{Method}} & \multicolumn{1}{c}{\textbf{Birate Range}} & \multicolumn{1}{c}{\textbf{BD-rate}} \\ 
\midrule
BPG & $0.07\sim1.93$ & - \\
VVC & $0.05\sim1.43$ & -20.53\% \\
Ball\'e'18~\cite{balle2018variational} & $0.13\sim1.66$ & 3.93\% \\
Minnen'18~\cite{minnen2018joint} & $0.11\sim1.59$ & -11.30\% \\ 
Cheng'20~\cite{cheng2020learned} & $0.12\sim0.81$ & -17.71\% \\ 
Xie'21~\cite{xie2021enhanced} & $0.10\sim1.06$ & -21.55\% \\ 
TinyLIC (Ours) & $0.126\sim1.63$ & -21.77\% \\
\bottomrule
\label{tab:bitrange}
\end{tabular}
\end{table}

\subsection{R-D Performance for MS-SSIM Optimized Model}

\begin{figure}[htbp]
\begin{center}
\includegraphics[width=0.8\linewidth]{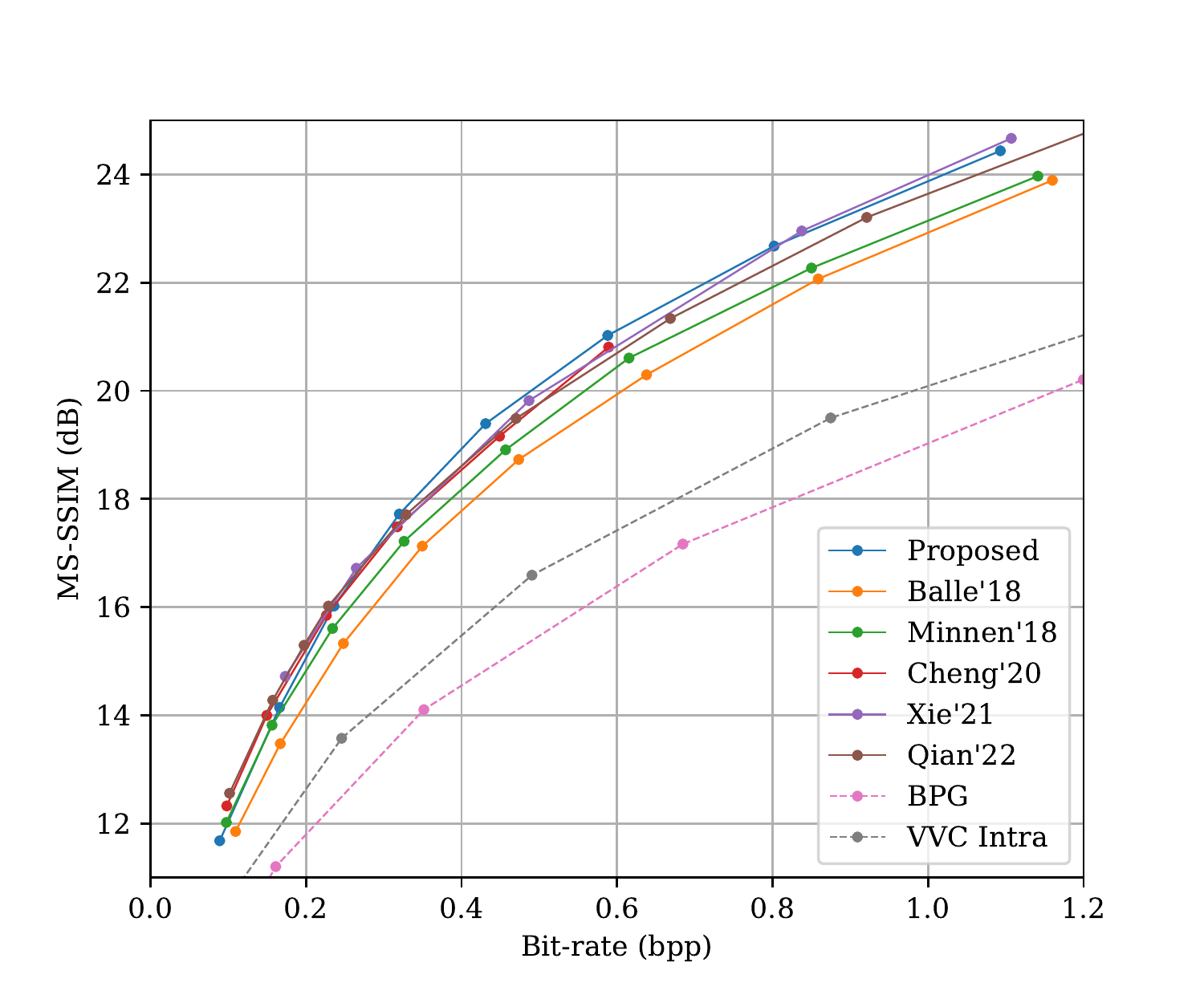}
\caption{R-D performance averaged on Kodak dataset with MS-SSIM optimized model.}
\label{fig:msssim_rd}
\end{center}

\end{figure}

We plot the R-D curve with MS-SSIM optimized model  Fig.~\ref{fig:msssim_rd}. Similarly, our method also offers the most competitive gains to the BPG anchor.

\subsection{BD-rate Performance on Extra Dataset}
In addition to the Kodak, CLIC and Tecnick datasets, we further evaluate the {\it TinyLIC} on common test dataset suggested by the IEEE 1857.11 Learning-based Image Coding committee. This dataset is referred to as the NIC$\_$Dataset:
\begin{itemize}
    \item  	The NIC$\_$Dataset is a public dataset at \url{https://pan.baidu.com/s/1dPTg9JRh4PS748zxdCUUtA} with access code {\tt p76h}.
    \item Test set contains 24$\times$4 = 96 images with 4 different resolutions (ClassA$\_$6K, ClassB$\_$4K, ClassC$\_$2K, ClassD$\_$Kodak). %2) The test set in the folder of /test_crop is used to boost the test speed, which is cropped to a resolution of no more than 720p from the original test images at the center position. Note that the final submitted techniques should be tested using the original test images in the folder of /test.
   % \item 
\end{itemize}

As quantitatively measured in Table~\ref{tab:NIC_dataset_performance}, we can still observe the lead of BD-rate gains of {\it TinyLIC} to the most recent VVC Intra, for the compression of RGB images at various resolutions and bitrates.  Note that NIC$\_$Dataset also provides training and validation images. However, to evidence the model generalization, we directly reuse pretrained {\it TinyLIC} to compress image samples from the test set of NIC$\_$Dataset.

\begin{table}[t]
    \centering
    \caption{BD-rate Performance of VVC Intra and {\it TinyLIC} on NIC$\_$ Dataset. Anchor is the BPG. Distortion is measured by PSNR.}
    \label{tab:NIC_dataset_performance}
    \begin{tabular}{c|c|c|c|c}
    \hline
    \multirow{2}{*}{Class} & \multicolumn{2}{c|}{VVC Intra} & \multicolumn{2}{c}{\it TinyLIC}\\
    \cline{2-5}
    & High Bitrate & Low Bitrate & High Bitrate & Low Bitrate\\
    \hline
    A & -15.1\% & -23.6\% & -22.5\% & -26.6\%\\
    B & -15.3\% & -23.7\% & -19.3\% & -23.6\%\\
    C & -22.4\% & -28.8\% & -28.7\% & -31.3\%\\
    D* & -19.0\% & -23.5\% & -20.5\% & -26.4\%\\
    \hline
    Ave. & -17.9\% & -24.9\% & -22.8\% & -27.0\%\\
\hline
    \end{tabular}\\
    * Class D images are from Kodak testing samples.
\end{table}

% \begin{comment}
% \subsection{Variable-rate Model}

\begin{table}[t]
    \centering
        \caption{BD-rate Performance of Variable-Rate Model Enabled by the ScalingNet~\cite{ScalingNet} Against the Anchor Using Multiple Rate-Specific Models for the proposed {\it TinyLIC}.     Numbers Are Averaged for each Dataset. {\it The smaller number the better.}}
    \label{tab:variable_rate_nic_model}
    \begin{tabular}{c|c|c}
    \hline
   \multirow{2}{*}{ dataset } & \multicolumn{2}{c}{BD-rate}\\
   \cline{2-3}
    & High Bitrate & Low Bitrate\\
    \hline
    Kodak & -1.35\% & -0.9\%\\
    CLIC & -1.78\% &  +0.46\% \\
    Tecnick & -1.87\%& +0.36\%\\
    \hline
         %&  \\
         %& 
    \end{tabular}\\
\end{table}

\subsection{Extra Visualizations}

\begin{figure*}[t]
\begin{center}
\subfloat[]{
\includegraphics[width=0.9\linewidth]{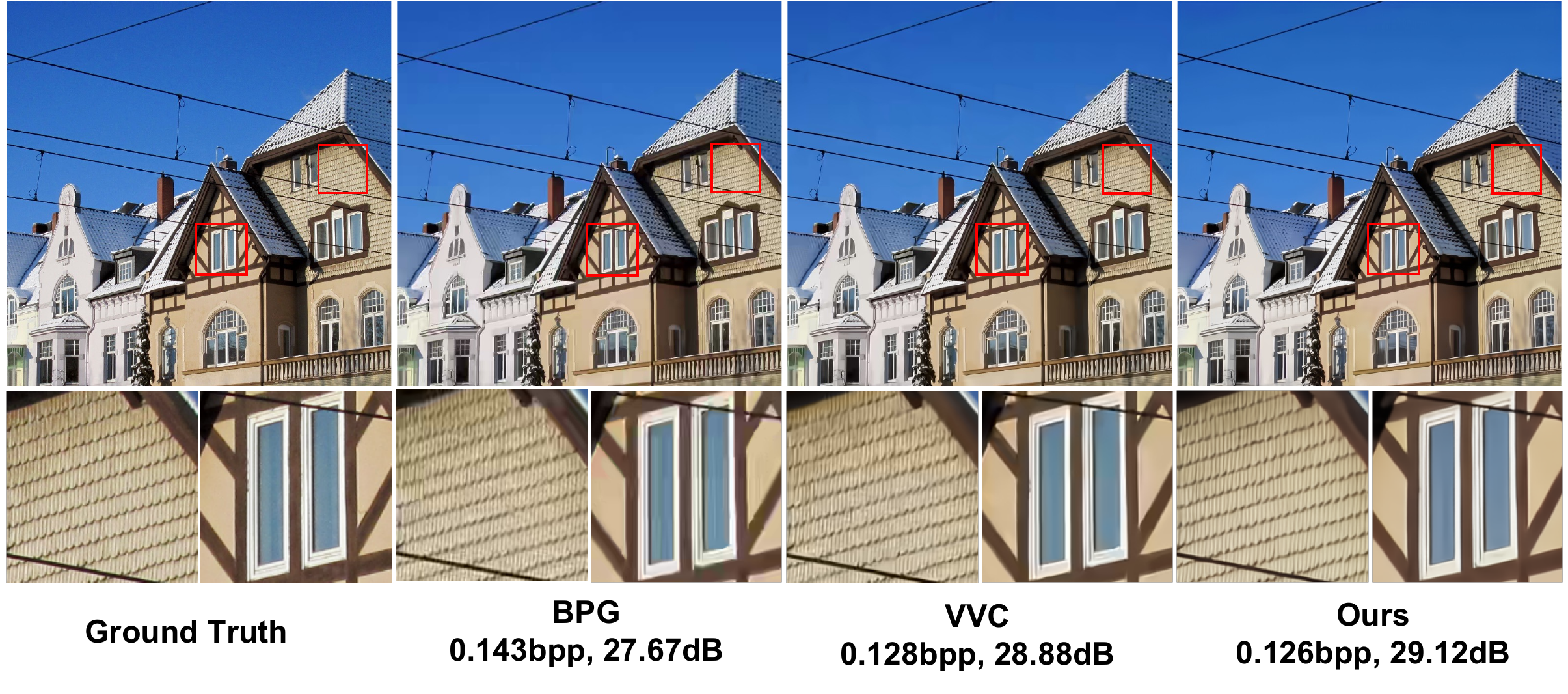}\label{fig:tecnick_visual_wall}
}

\subfloat[]{
\includegraphics[width=0.9\linewidth]{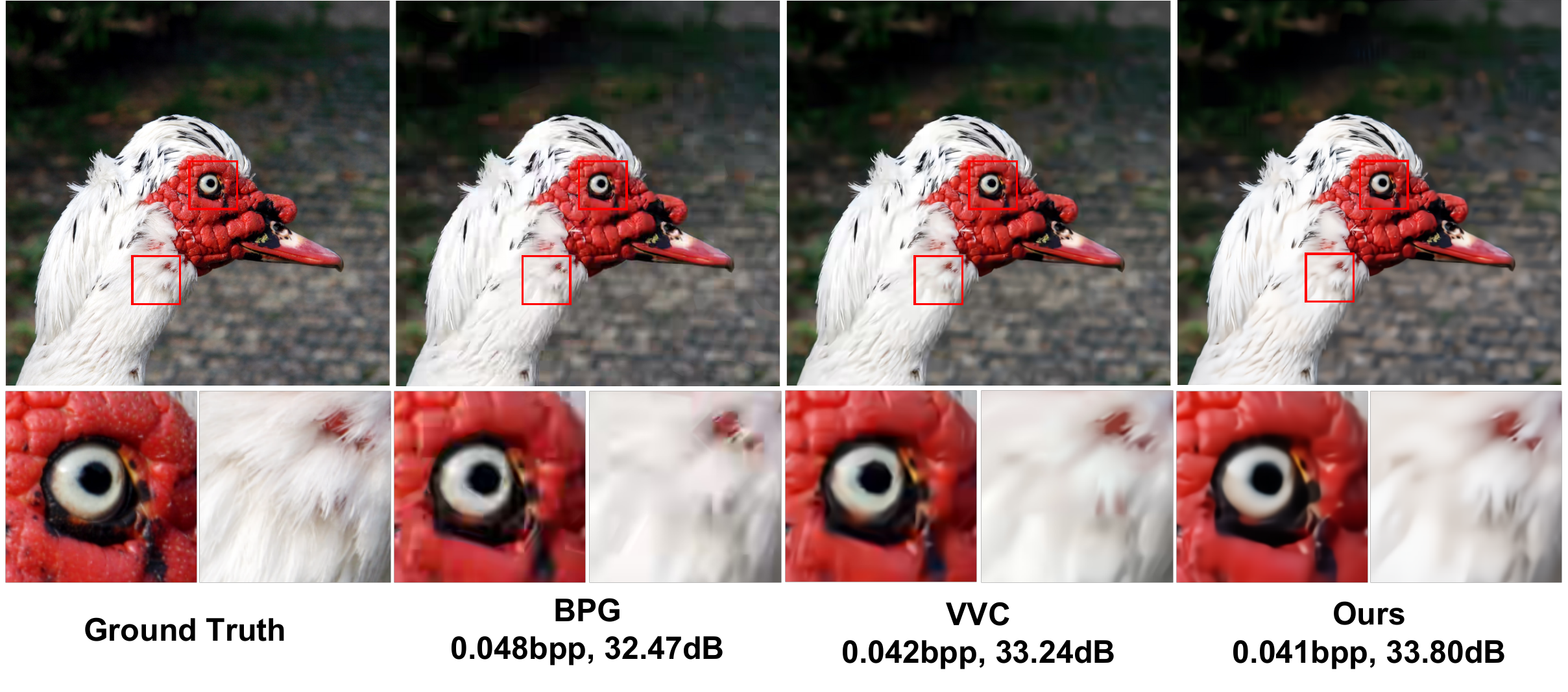}
}
\end{center}
\caption{{\bf Qualitative Visualization on Tecnick Dataset.}
Reconstructions and close-ups of the BPG, VVC and our {\it TinyLIC}. Both bpp and PSNR are marked. (a) RGB\_OR\_1200x1200\_023, (b) RGB\_OR\_1200x1200\_056.} 
\label{fig:tecnick_visual}
\end{figure*}

\begin{figure*}[t]
\begin{center}
\subfloat[]{
\includegraphics[width=0.9\linewidth]{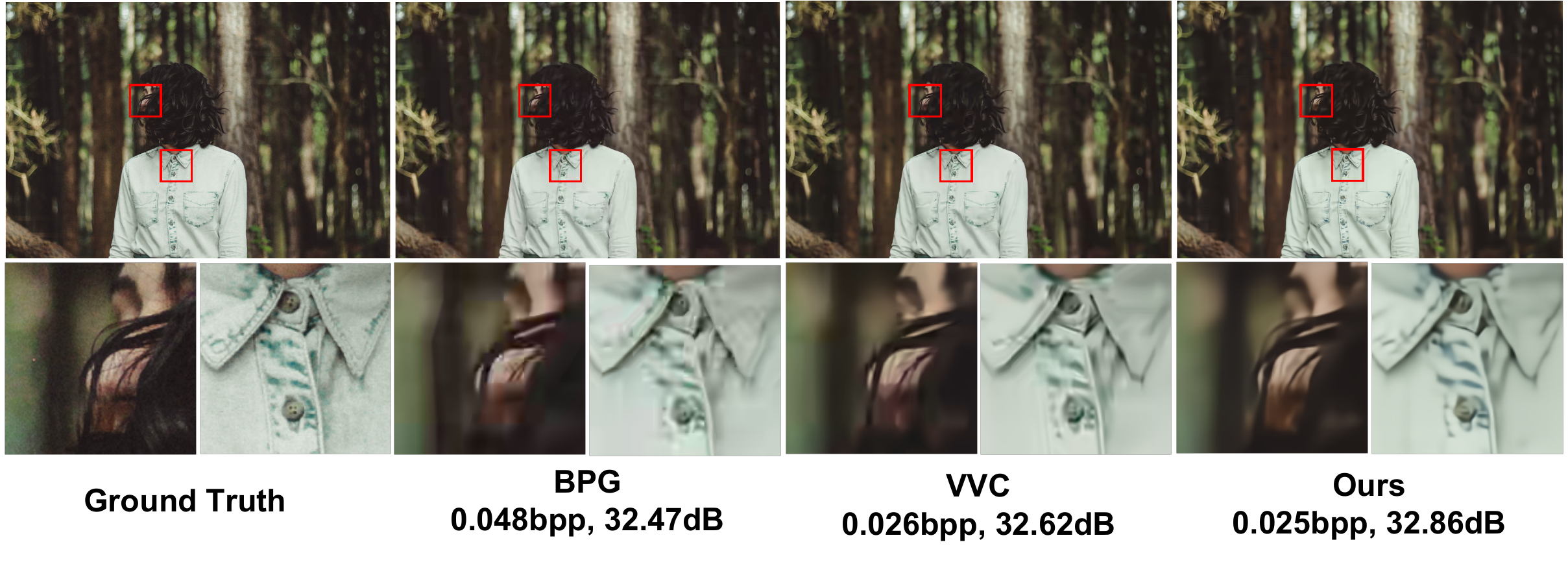} \label{fig:clic_visual_hair}
}

\subfloat[]{
\includegraphics[width=0.9\linewidth]{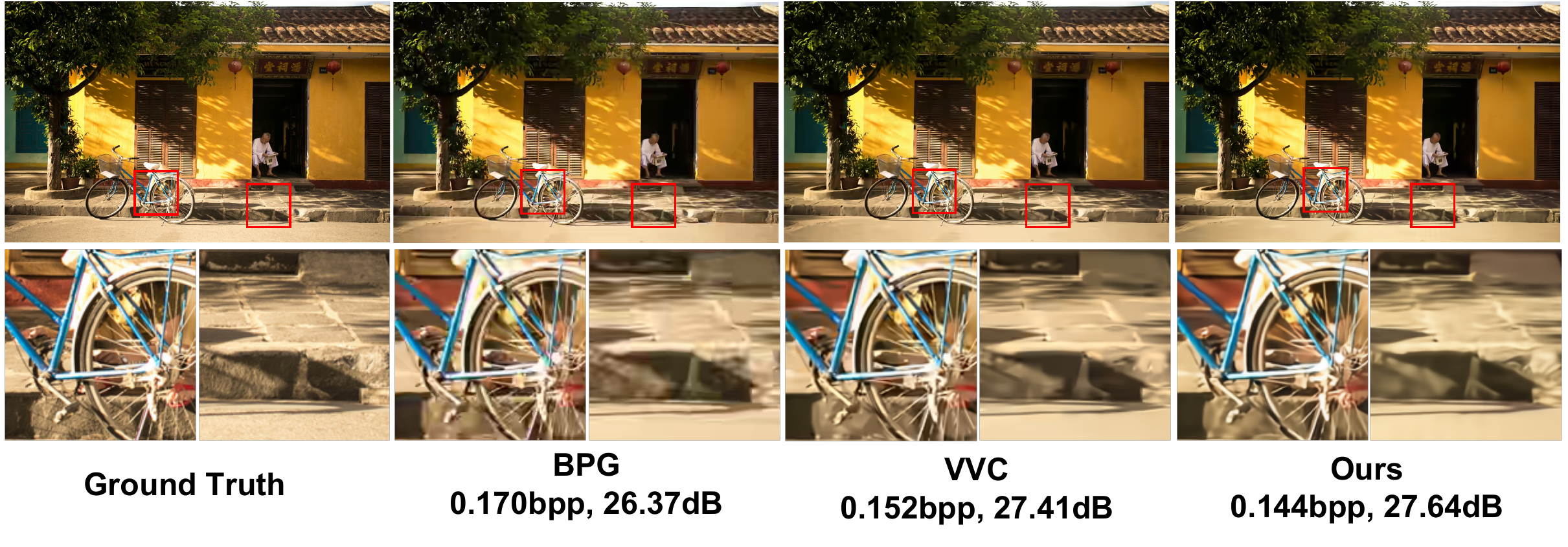}
}
\end{center}
\caption{{\bf Qualitative Visualization on CLIC Dataset.}
Reconstructions and close-ups of the BPG, VVC and our {\it TinyLIC}. Both bpp and PSNR are marked. (a) allef-vinicius-109434, (b) thong-vo-428.} 
\label{fig:clic_visual}
\end{figure*}

We also offer more qualitative visualizations using Tecnick and CLIC image samples in Fig.~\ref{fig:tecnick_visual} and Fig.~\ref{fig:clic_visual} respectively. Similar to the results in the main content of this work, we can clearly observe the subjective improvements of the proposed {\it TinyLIC} in comparison to the BPG and VVC. For wall tile textures and flying hair in closeups of respective Fig.~\ref{fig:tecnick_visual_wall} and~\ref{fig:clic_visual_hair}, our {\it TinyLIC} provides sharper and less noisy reconstructions which are closer to the ground truth samples. 

\subsection{Support of Various Image Sources}
\begin{figure}[t]
    \centering
    \includegraphics[scale=0.65]{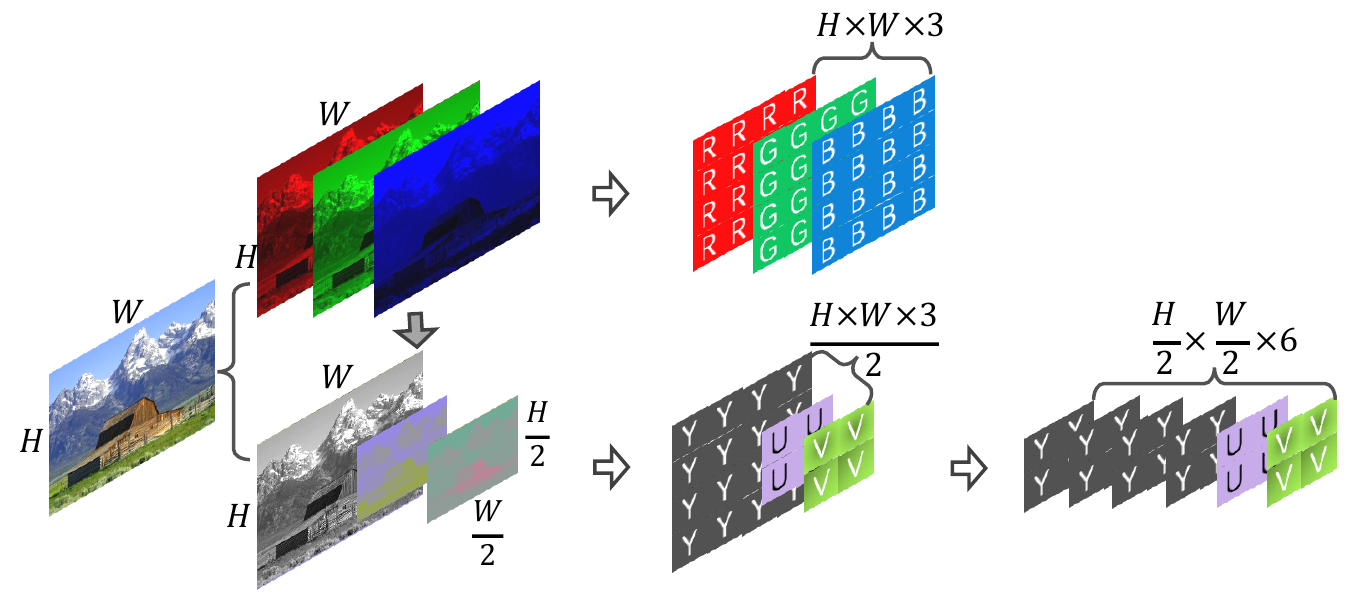}
    \caption{{\bf Pixel Arrangement} for {\it TinyLIC} to process various image sources in both training and inference stages.}
    \label{fig:rbg_YUV420}
\end{figure}

To ensure broader adoption of the proposed {\it TinyLIC} in vast scenarios, one key feature is to support different image formats as the input. In addition to the RGB sources, here we exemplify the use cases of the support of YUV420\footnote{The use of YUV420 allows us to use low-resolution chrominance for data saving without noticeable perceptual distortion~\cite{yao_book} because the human visual system is more sensitive to luminance components.}  and Y (monochrome) images. As illustrated in Fig.~\ref{fig:rbg_YUV420}, a native RGB image at a size of $H\times W\times 3$ is processed directly by stacking R, G, B attributes of each pixel; while for an image in YUV420 format, it is first converted from the native RGB representation, and then rearranged to a pile of YYYYUV at a size of $\frac{H}{2}\times\frac{W}{2}\times 6$ for compression. Besides, if we want to compress a monochrome image, we can just need to process the luminance component of the native RGB content, a.k.a, Y attribute as in Fig.~\ref{fig:rbg_YUV420} if using YUV color space. 

\begin{table}[htbp]
    \centering
        \caption{BD-rate Performance of {\it TinyLIC} Upon YUV420 Images. Anchor is the VVC Intra.  Numbers are Averaged for each Dataset. {\it The smaller number the better}}
    \label{tab:YUV420_BD}
    \begin{tabular}{c|c|c|c|c}
    \hline
   \multirow{2}{*}{ dataset } & \multicolumn{2}{c}{Y BD-rate $\downarrow$} & \multicolumn{2}{|c}{YUV BD-rate $\downarrow$}\\
   \cline{2-5}
    & High Bitrate & Low Bitrate & High Bitrate & Low Bitrate \\
    \hline
    Kodak & -20.72\% & -16.77\% & -18.74\% & -13.57\%\\
   % CLIC & -2.28\% &-2.25\%  \\
    %Tecnick & -2.17\% & -2.17\%\\
    \hline
         %&  \\
         %& 
    \end{tabular}\\
\end{table}

\bibliographystyle{IEEEtran}
\bibliography{reference}

\vfill